\documentclass{article}
\usepackage[utf8]{inputenc}
\usepackage{cite}
\usepackage{amsmath,amssymb,amsfonts}
\usepackage{balance}
\usepackage{tipa}
\usepackage{subfig}
\usepackage{float}
\usepackage{algorithm}
\usepackage{algpseudocode}
\usepackage{graphicx}
\usepackage{textcomp}
\usepackage{placeins}
\usepackage{letltxmacro}
\usepackage[colorlinks=true,linkcolor=blue,urlcolor=blue,citecolor=blue,anchorcolor=blue]{hyperref}

\DeclareTextCommand{\DJ}{OT1}{%
  \raisebox{-0.1ex}{\scalebox{0.75}[1.4]{--}}\kern-.4em D%
}

\DeclareTextCommand{\dj}{OT1}{%
  \raisebox{-0.1ex}{\scalebox{0.75}[2.2]{--}}\kern-.6em d%
}

\begin{document}

\date{}

\title{Reply to Comment on ``TVOR: Finding Discrete Total Variation Outliers among Histograms''}
\author{Nikola Bani{\'{c}} and Neven Elezovi{\'{c}}}

\maketitle

\begin{abstract}
In this paper, we respond to a critique of one of our papers previously published in this journal, entitled ``TVOR: Finding Discrete Total Variation Outliers among Histograms''. Our paper proposes a method for smoothness outliers detection among histograms by using the relation between their discrete total variations (DTV) and their respective sample sizes. In this response, we demonstrate point by point that, contrary to its claims, the critique has not found any mistakes or problems in our paper, either in the used datasets, methodology, or in the obtained top outlier candidates. On the contrary, the critique's claims can easily be shown to be mathematically unfounded, to directly contradict the common statistical theorems, and to go against well established demographic terms. Exactly this is done in the reply here by providing both theoretical and experimental evidence. Additionally, due to critique's compalint, a more extensive research on top outlier candidate, i.e. the Jasenovac list is conducted and in order to clear any of the critique's doubts, new evidence of its problematic nature unseen in other lists are presented. This reply is accompanied by additional theoretical explanations, simulations, and experimental results that not only confirm the earlier findings, but also present new data. The source code is at https://github.com/DiscreteTotalVariation/TVOR.
\end{abstract}


\section{Introduction}
\label{sec:intro}

\newtheorem{theorem}{Theorem}
\newtheorem{hypothesis}{Hypothesis}
\newcommand{\E}[1]{\mathbb{E}\left[ #1 \right]}
\newcommand{\Var}[1]{\mathrm{Var}\left[ #1 \right]}
\newcommand{\V}[1]{\left\lVert #1 \right\rVert_{V}}
\newcommand{\Norm}[2]{\left\lVert #1 \right\rVert_{#2}}
\newcommand{\N}[2]{\mathcal{N}\left(#1,#2\right)}

In the recent critique~\cite{ornik2021comment} of one of our papers~\cite{banic2021tvor}, it is claimed that the method Total Variation Outlier Recognizer~(TVOR) proposed in our paper is inapplicable to the United States Holocaust Memorial Museum~(USHMM) dataset. The critique also gives significant attention to the top outlier candidate, i.e. the Jasenovac inmates list, claiming that too few errors were identified and that they were not historically analyzed. In this response we show that the critique is mathematically unfounded, that it directly contradicts the common statistical theorems, and that it goes against well established demographic terms. As for the Jasenovac list, additional effort has been put into a more detailed investigation of its data with the resulting conclusion that the errors there are counted at least in thousands, which is significantly more than the dozens that are ``reasonable to expect'' as the critique puts it. This result not only helps to explain the highly expressed age heaping problem of the Jasenovac list, but it also uncovers potential problems with its data veracity.

The reply is structured as follows: Section~\ref{sec:background} gives a brief description of TVOR, Sections~\ref{sec:preliminaries}, \ref{sec:assumptions}, \ref{sec:bias}, \ref{sec:dearth}, \ref{sec:processing}, \ref{sec:objections} are direct replies to the corresponding sections in the critique, Section~\ref{sec:jasenovac} investigates the data of the Jasenovac list as required by the critique, and Section~\ref{sec:conclusions} concludes the paper.

\begin{figure*}[htb]
    \centering
    
    \subfloat[]{
    \includegraphics[width=0.48\linewidth]{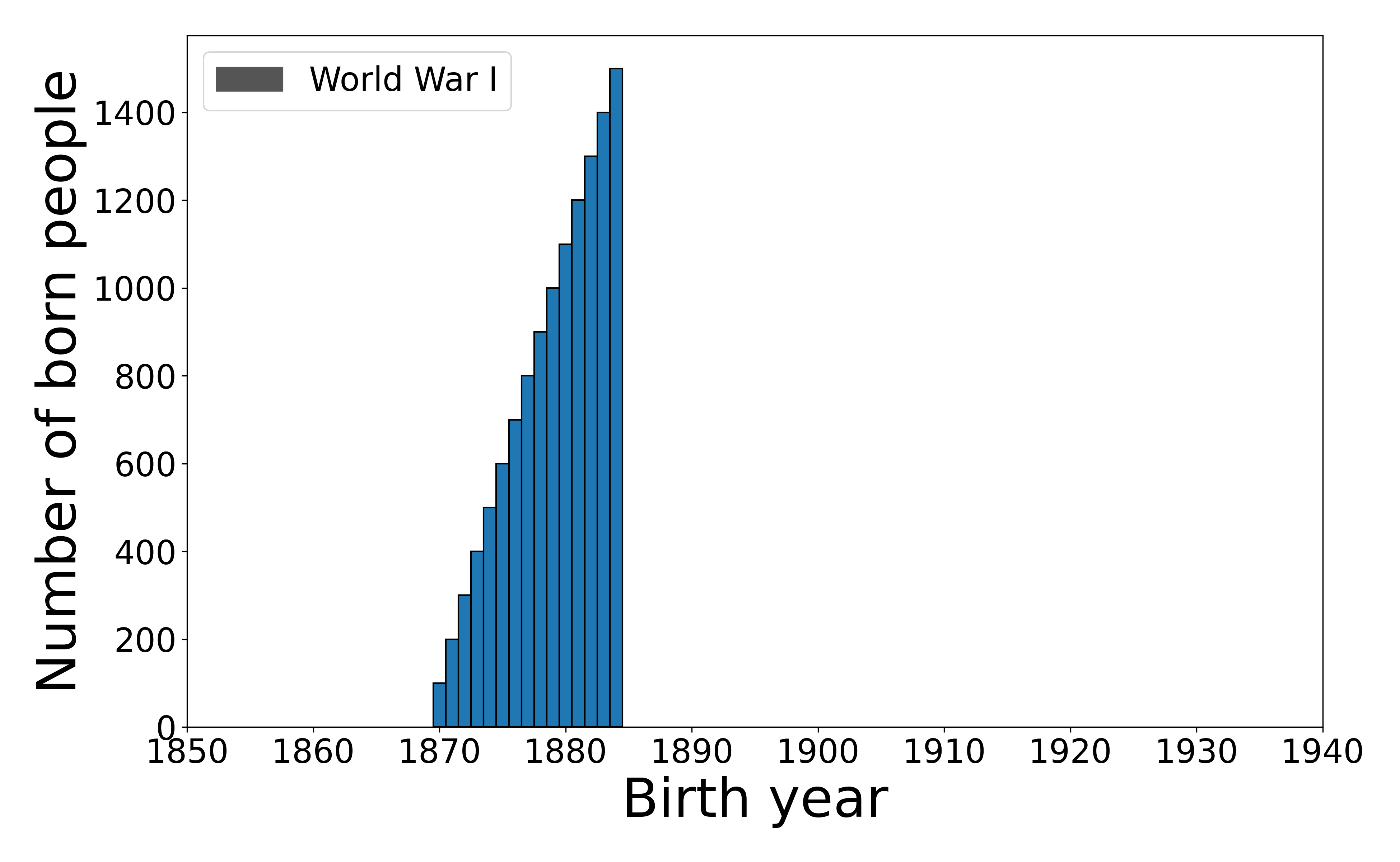}
    \label{fig:same_dtv_1}
    }%
    \subfloat[]{
    \includegraphics[width=0.48\linewidth]{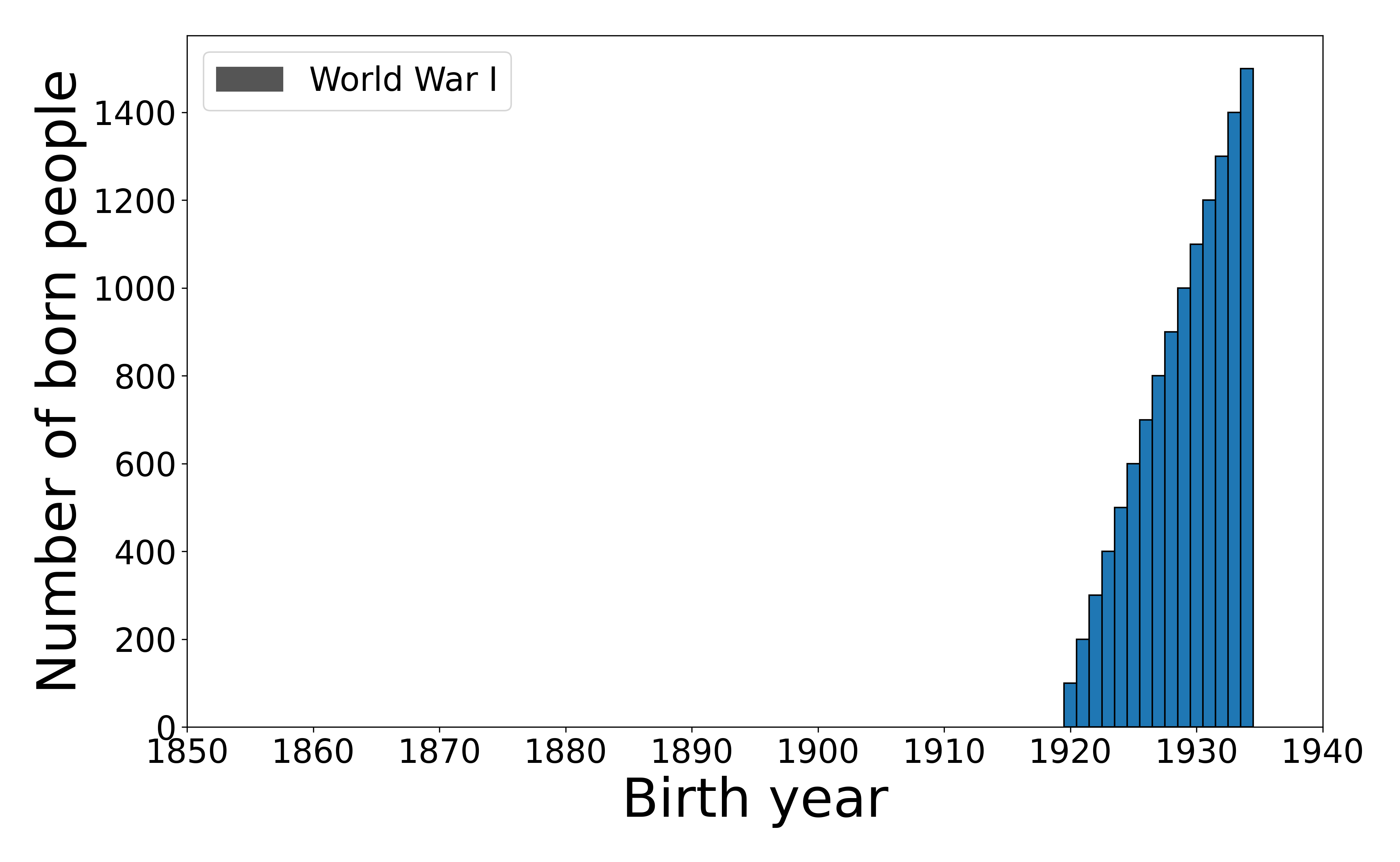}
    \label{fig:same_dtv_2}
    }
    
    \caption{The histograms of two different samples with the same DTV: (a)~an older population and (b)~a younger population.}
  \label{fig:same_dtv}
    
\end{figure*}

\begin{figure}[htb]
    \centering
    
    \includegraphics[width=\linewidth]{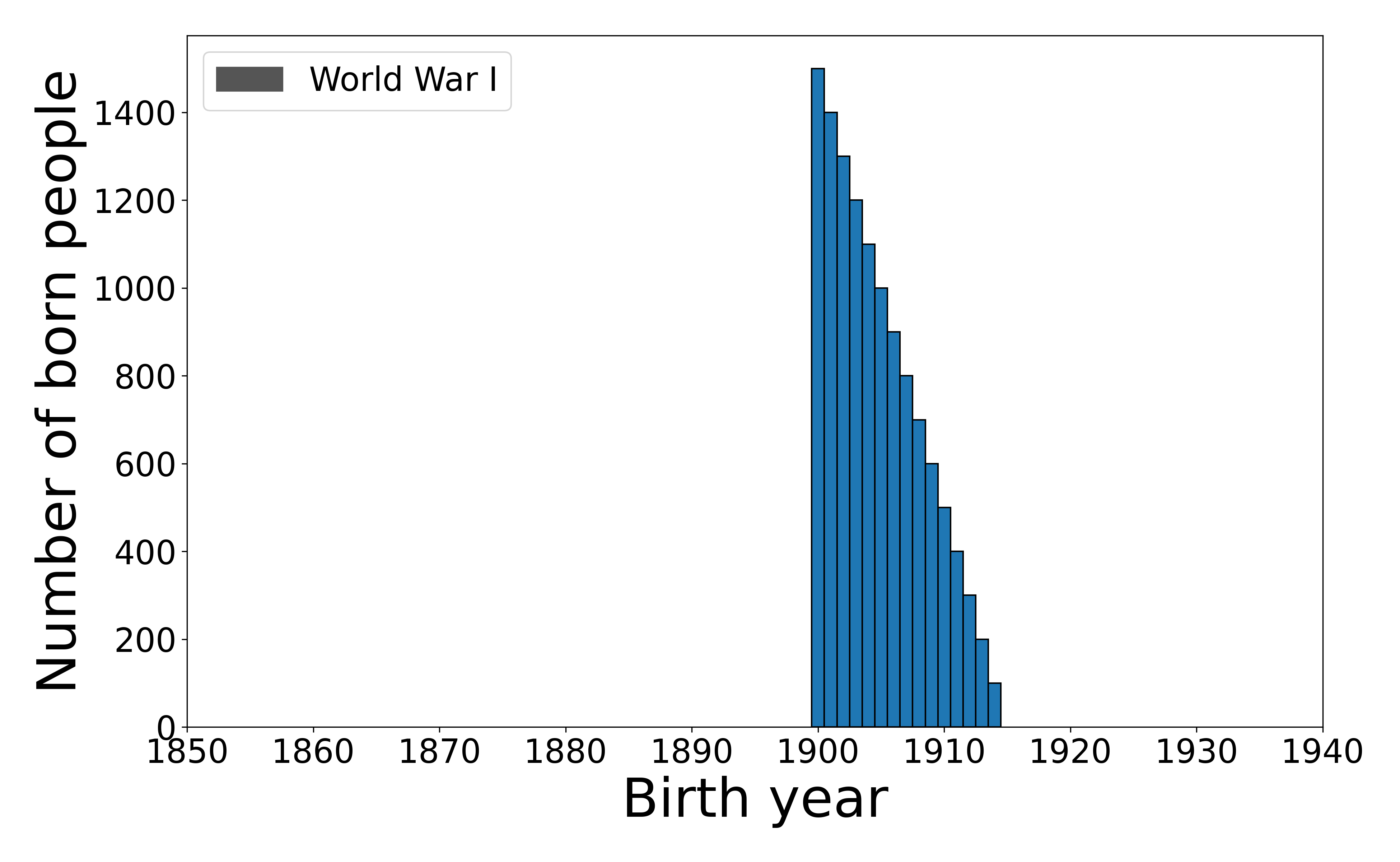}
    
    \caption{The histogram of a sample with different bin probabilities than the ones in Fig.~\ref{fig:same_dtv}, but with the same DTV.}
  \label{fig:same_dtv_3}
    
\end{figure}

\section{Background}
\label{sec:background}

\subsection{Scoring}
\label{subsec:scoring}

If $\mathbf{x}_n$ is a histogram with $N$ elements arranged in $n$ bins so that $\sum\limits_{i=1}^{n}x_i=N$, then its discrete total variation~(DTV) is
\begin{equation}
    \label{eq:dtv}
    \V{\mathbf{x}_{n}}=\sum\limits_{i=2}^{n}\lvert x_{i}-x_{i-1}\rvert.
\end{equation}
In~\cite{banic2021tvor} for histograms of similar smoothness the value of expected DTV denoted as  $\E{\V{\mathbf{x}_n}}$ has been modelled as
\begin{equation}
\label{eq:non-uniform_model}
m=aN+b\sqrt{N}
\end{equation}
where the values of $a$ and $b$ are obtained through fitting on the values of $N$ and $\V{\mathbf{x}_{n}}$ of all given histograms. To recognize the potential outlier candidates among these histograms in terms of the DTV, the score $d'$ was introduced and defined as
\begin{equation}
\label{eq:d2}
\begin{gathered}
d'=\frac{\left|\V{\mathbf{x}_n}-m\right|}{\sqrt{N}}=\frac{\left|\V{\mathbf{x}_n}-aN+b\sqrt{N}\right|}{\sqrt{N}}.
\end{gathered}
\end{equation}
This score is essentially based on the difference between the observed and expected histogram's DTV. Here the denominator is used to properly handle the effects of randomness. The histograms with higher score $d'$ values are considered to be more likely candidates for outliers in terms of the DTV. This ranking is the core of the Total Variation Outlier Recognizer~(TVOR) method. It must be stressed here that TVOR has only few assumptions and that it only expects the analyzed histograms to share a relatively similar smoothness, with exception of the outliers. The $d'$ score is also not supposed to be used in some probabilistic interpretations since its only purpose is to enable ranking. Because of that, it is their interrelation that matters, not their individual value.

\subsection{The importance of the sample size}
\label{subsec:importance}

In~\cite{banic2021tvor}, the upper bound for $\E{\V{\mathbf{x}_n}}$ was also given as
\begin{equation}
\label{eq:non-uniform2}
	\E{\V{\mathbf{x}_n}}\le\V{{\cal D}}\cdot N+\E{\V{{\cal R}}}\sqrt{N}
\end{equation}
where ${\cal R}$ is a deviation from the theoretical distribution due to the randomness. As the sample size $N$ grows, the influence of randomness gets ever smaller and any overall deviation is much more likely to be caused by the difference in the theoretical smoothness. Because of that, the deviations of larger samples can effectively be scored more strictly due to the higher certainty of a lower influence of the randomness. A good illustration of how the role of randomness vanishes with sample size is given in the TVOR paper~\cite{banic2021tvor} in Figure~22.

\begin{figure*}[htbp]
    \centering
    
  \subfloat[]{
  \includegraphics[width=0.48\linewidth]{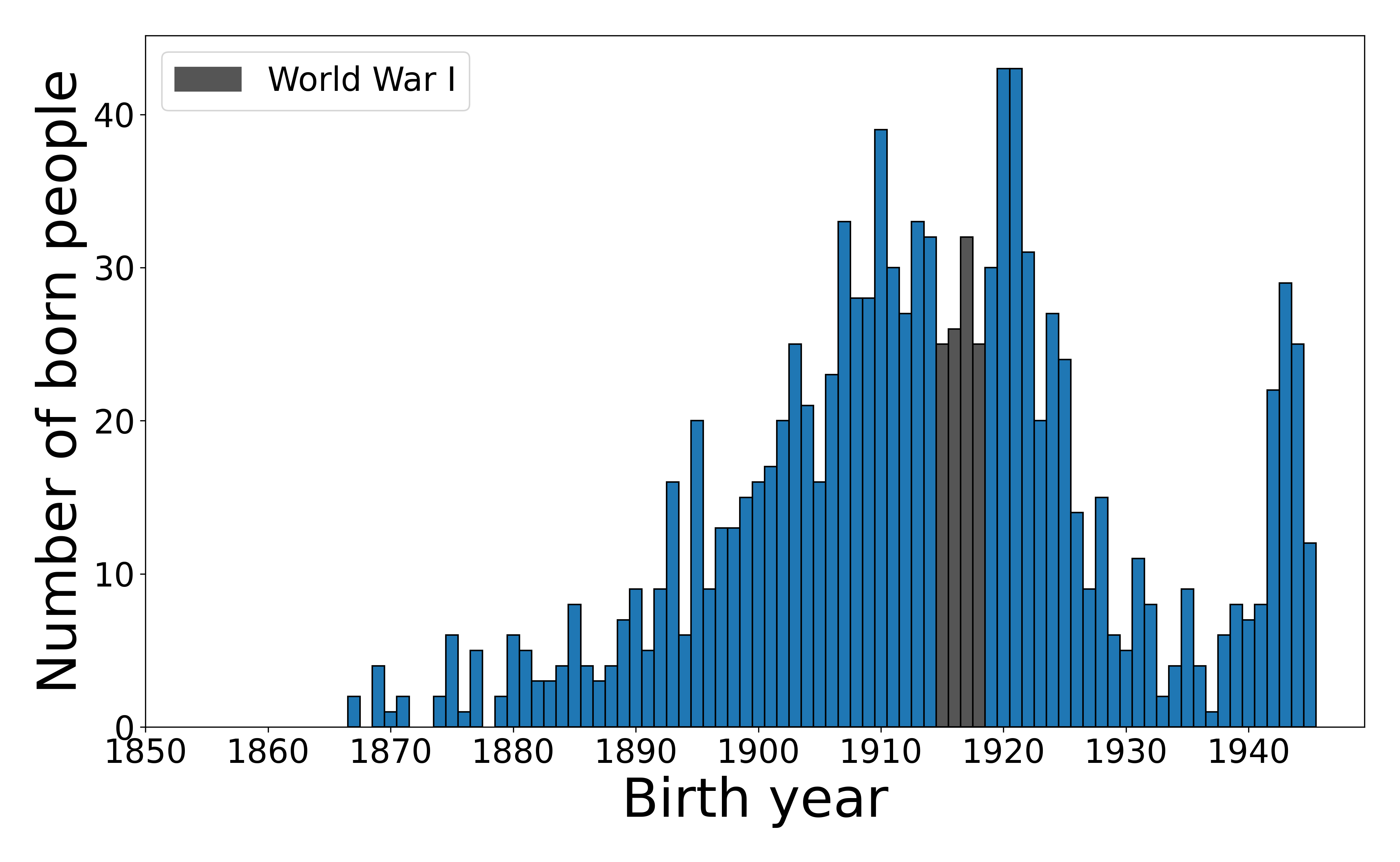}
  \label{fig:24835}
  }%
  \subfloat[]{
  \includegraphics[width=0.48\linewidth]{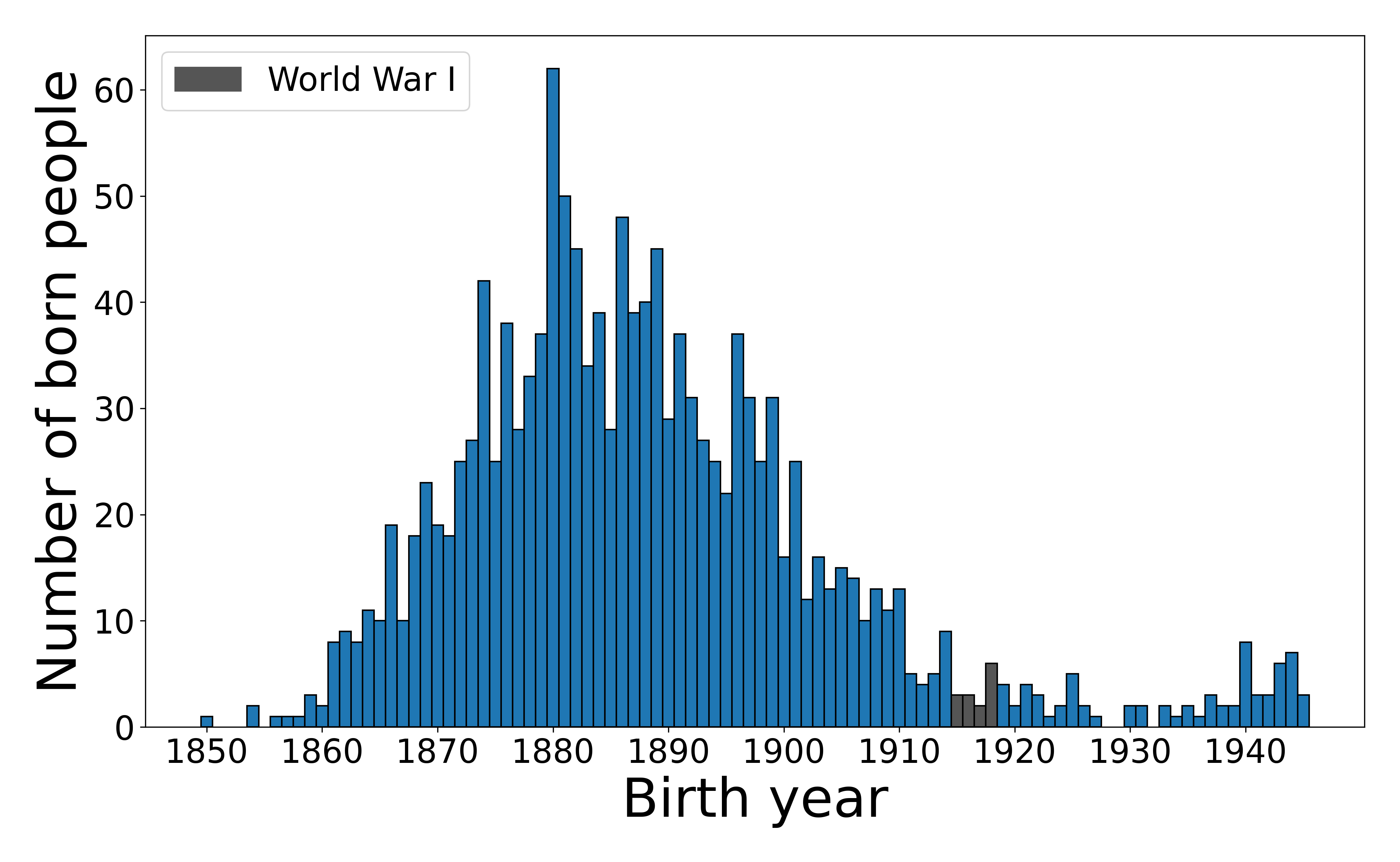}
  \label{fig:20780}
  }
  \\
  \subfloat[]{
  \includegraphics[width=0.48\linewidth]{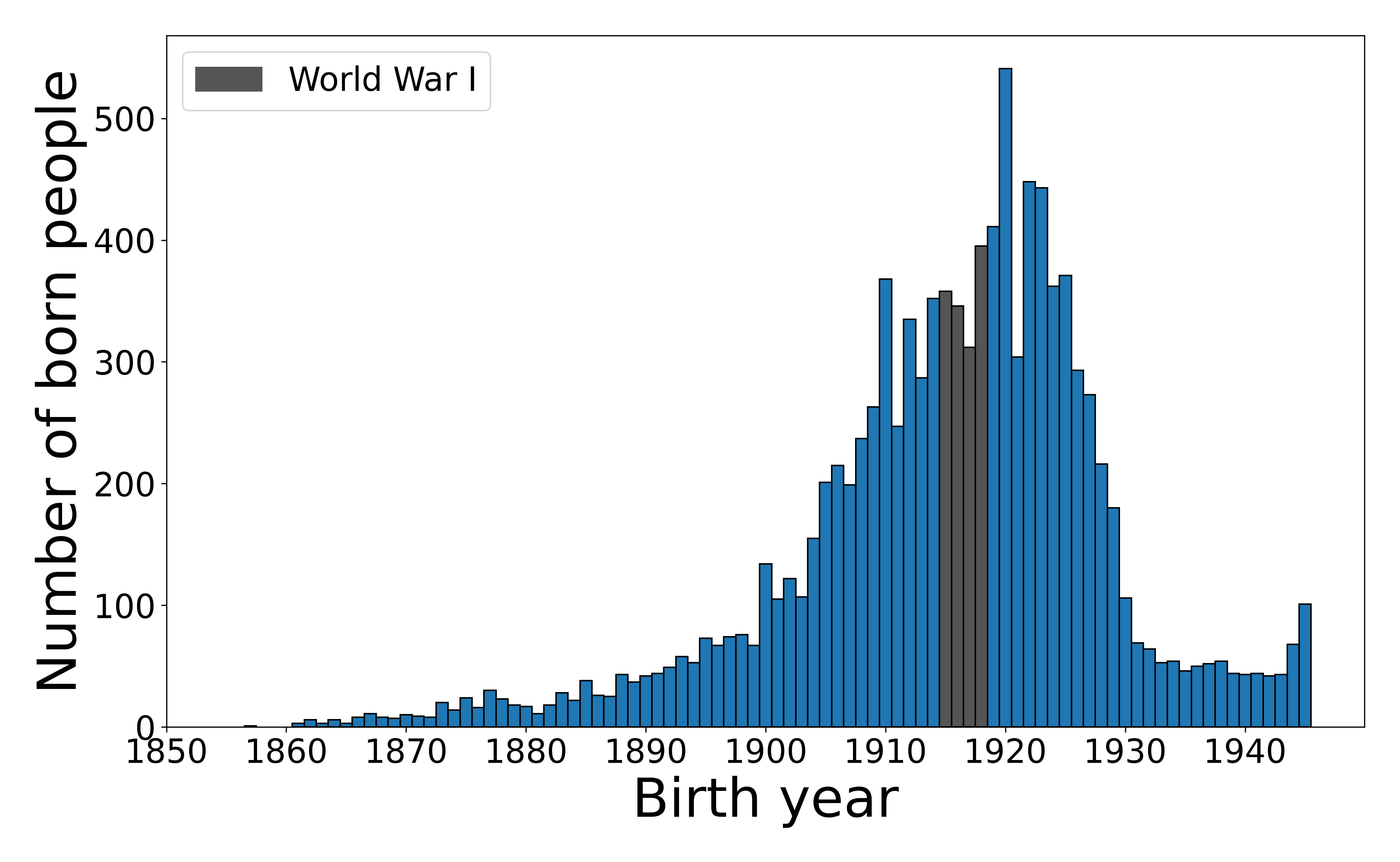}
  \label{fig:20702}
  }%
  \subfloat[]{
  \includegraphics[width=0.48\linewidth]{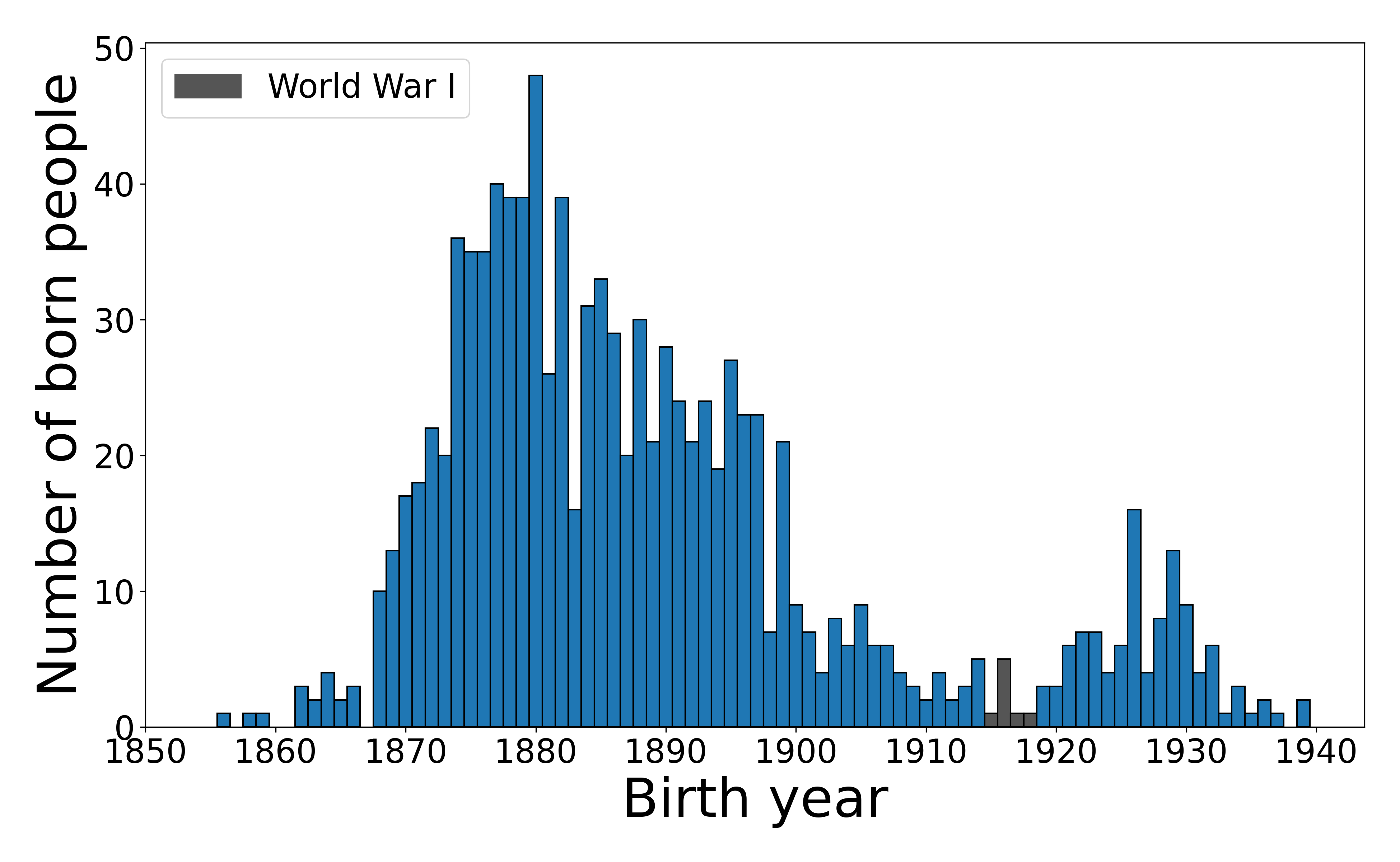}
  \label{fig:35699}
  }
  \\
  \subfloat[]{
  \includegraphics[width=0.48\linewidth]{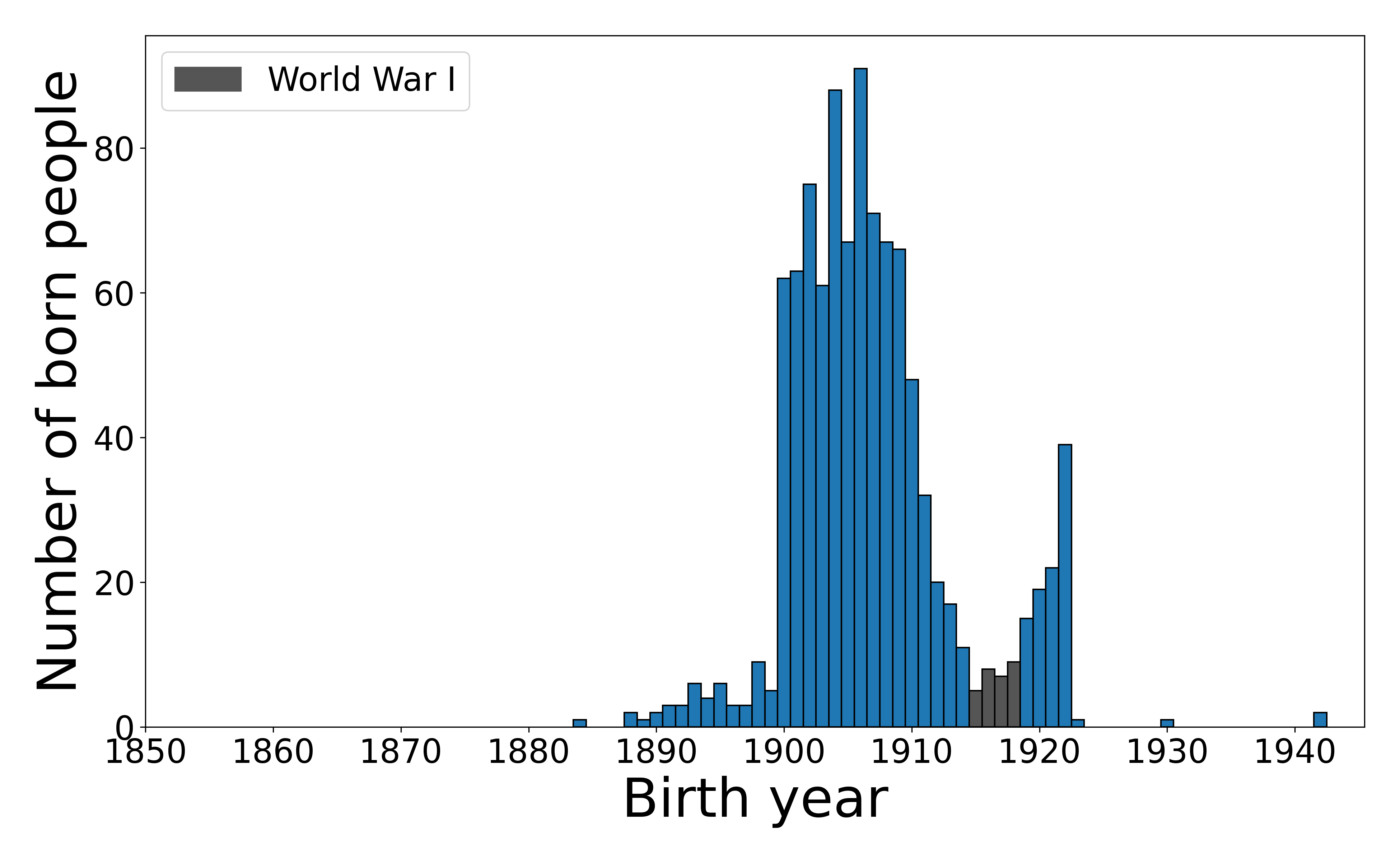}
  \label{fig:20556}
  }%
  \subfloat[]{
  \includegraphics[width=0.48\linewidth]{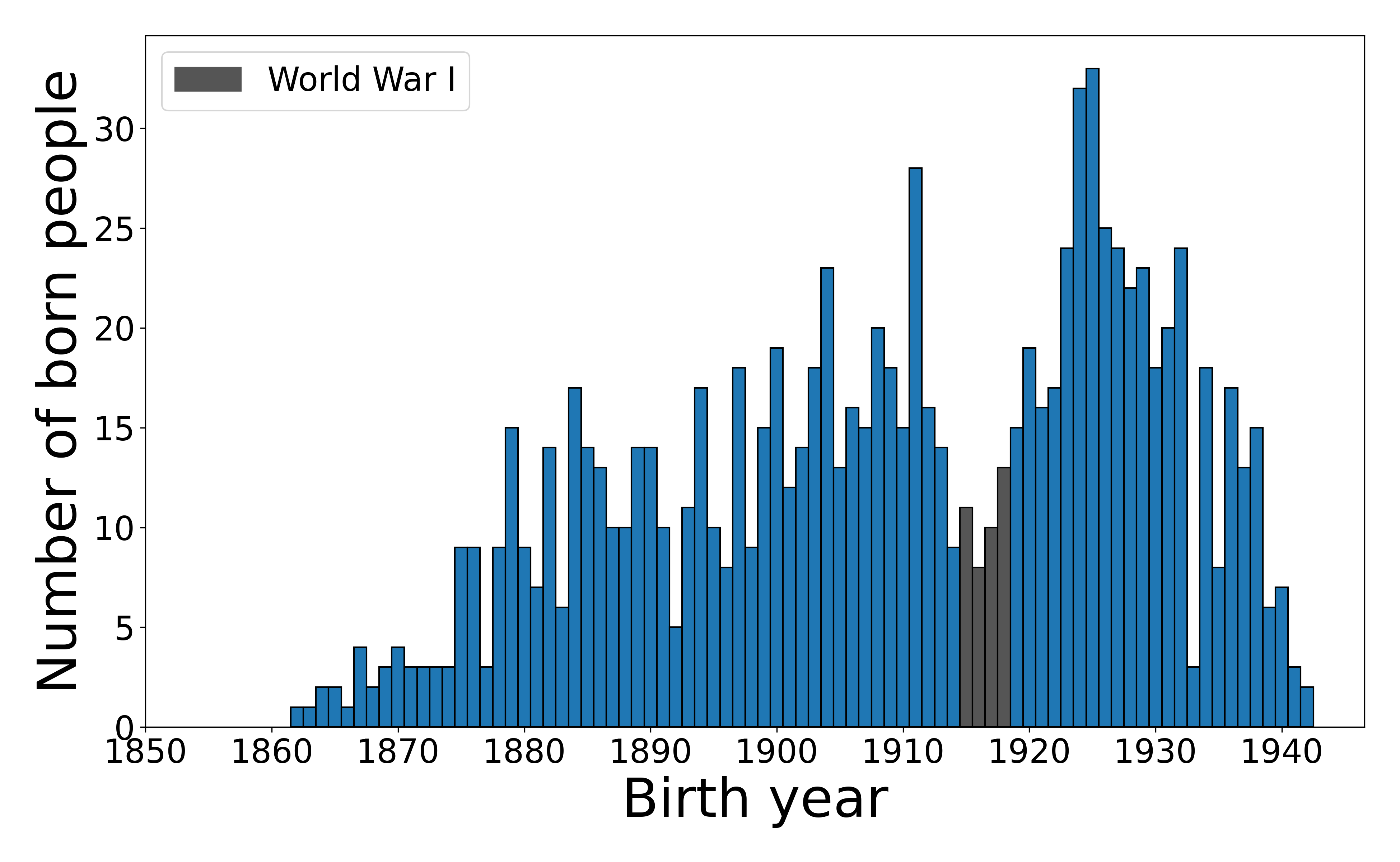}
  \label{fig:38525}
  }

    \caption{An example of histograms of populations with mutually different dominant age groups for which the scores obtained by TVOR are very similar: (a)~[Documents related to repatriation]~\cite{ushmm2021documents}, (b)~CENTRAL-EUROPEAN JEWISH REFUGEES WHO DIED IN SHANGHAI, 1940-1945~\cite{ushmm2021central}, (c)~J{\:{U}}DISCHE GEMEINDE ZU BERLIN Bestand B. 1/9, Nr. 1 DP-Kartei  1945–1949~\cite{ushmm2021juedische}, (d)~Transport Wien 5~\cite{ushmm2021transport}, (e)~[Auschwitz arrival list, June 27, 1942]~\cite{ushmm2021auschwitz}, and (f)~Hohensteiner Str. Nr. 43/45 (Steinmetzg. 12/14)~\cite{ushmm2021hohensteiner}. The scores calculated by TVOR for all these histograms vary from roughly $2.42$ to $2.81$.}
  \label{fig:old_vs_young}
    
\end{figure*}

\section{Reply to ``TVOR preliminaries and USHMM dataset''}
\label{sec:preliminaries}

The critique mentions that ``[t]he technical approach of TVOR relies on the fact that histograms drawn from the same probability distribution will, informally stated, share similar smoothness''. However, this is not the case. Nowhere in our previous paper is there a mention of TVOR's direct reliance on the same probability distribution among all histograms. The only part of our previous paper that may be slightly similar to this is that ``[d]ue to randomness the discrete total variation of $\mathbf{x}_n$, i.e. $\V{\mathbf{x}_n}$ can differ for each sampling, but it should mostly not differ significantly from its expected value $\E{\V{\mathbf{x}_n}}$ for a given $N$ and probabilities $p_i$''. However, this does not imply the critique's claim and it is part of a general discussion before the formal proposal of TVOR and without any imposing limitations. TVOR directly relies only on the behavior of DTV and due to DTV's nature, it can be the same or relatively similar for various probability distributions as explained and demonstrated in more detail in Section~\ref{sec:assumptions}.

The critique also mentions that ``the proportionality constant between the standard error and $\sqrt{N}$ is dependent on the distribution'' and that ``different distributions may lead to different constants''. However, this is not specific enough and it may lead a reader into thinking that the distributions in questions are the histogram bin probability distribution, which is not the case. The distributions in question are the distributions of the DTVs and these can be very similar even for different bin probability distributions as described in Section~\ref{sec:assumptions}. Additionally, in the specific case of distributions of the DTVs, the multiplier for $\sqrt{N}$ can strictly speaking depend on the sample size, but this topic is out of the scope of this paper. The lack of full correctness was already mentioned in our previous paper and it did not have any significant impact. Additional experiments with calculations of the multiplier for $\sqrt{N}$ and carried out by using synthetic data and Monte Carlo simulations had no significant impact on ranking produced by TVOR for the USHMM dataset.

\section{Reply to ``Inapplicability of assumptions''}
\label{sec:assumptions}

\subsection{Probabilities and smoothness}
\label{subsec:probabilities}

The critique raises several objections where the problem is allegedly supposed to be that the probability distributions of samples of some of the histograms from the used United States Holocaust Memorial Museum~(USHMM) dataset differ significantly and that many of the histograms' samples cannot be expected to share the same probability distribution or similar discrete total variation. It is then claimed that this represents a problem for TVOR, but instead of providing firm evidence to support these claims, only a few hand-picked histograms are shown. Such claims can be disproved by taking into account the experimental results already given in our previous paper, by further analyzing some of them, and by providing some new examples to give a clearer explanation. 

\subsubsection{Different probabilities, but same smoothnes}
\label{subsubsec:different}

First of all, having same probability distributions is not a precondition for applying TVOR. This can be shown in several steps of gradually increasing complexity. One of the simplest examples of two different probability distributions that will have the same effect on TVOR due to having the same properties in terms of DTV is shown in Fig.~\ref{fig:same_dtv} with a histogram of an older population in Fig.~\ref{fig:same_dtv_1} and a younger population in Fig.~\ref{fig:same_dtv_2}. The only difference between the histograms in Figs.~\ref{fig:same_dtv_1} and~\ref{fig:same_dtv_2} are the shifted bin probabilities. While metrics such as the Wasserstein distance~\cite{dudley2018real} would be affected by this shift, TVOR will not distinguish between such histograms since their DTVs are exactly the same. Even if the bin probabilities change, but the DTV remains similar or exactly the same, which is the case in Fig.~\ref{fig:same_dtv_3}, TVOR will not be affected. Somewhat similar observations can be made for the histograms of the USHMM dataset as well. For example, TVOR assigns a very similar score to all histograms in Fig.~\ref{fig:old_vs_young} despite the fact that the age distributions there vary significantly. However, since in each histogram the transitions between and among various age groups are relatively similar, the same goes for the properties based on DTV including the score assigned by TVOR to histograms.

\subsubsection{The effect of the sample size}
\label{subsubsec:size}

\begin{figure*}[htbp]
    \centering
    
  \subfloat[]{
  \includegraphics[width=0.48\linewidth]{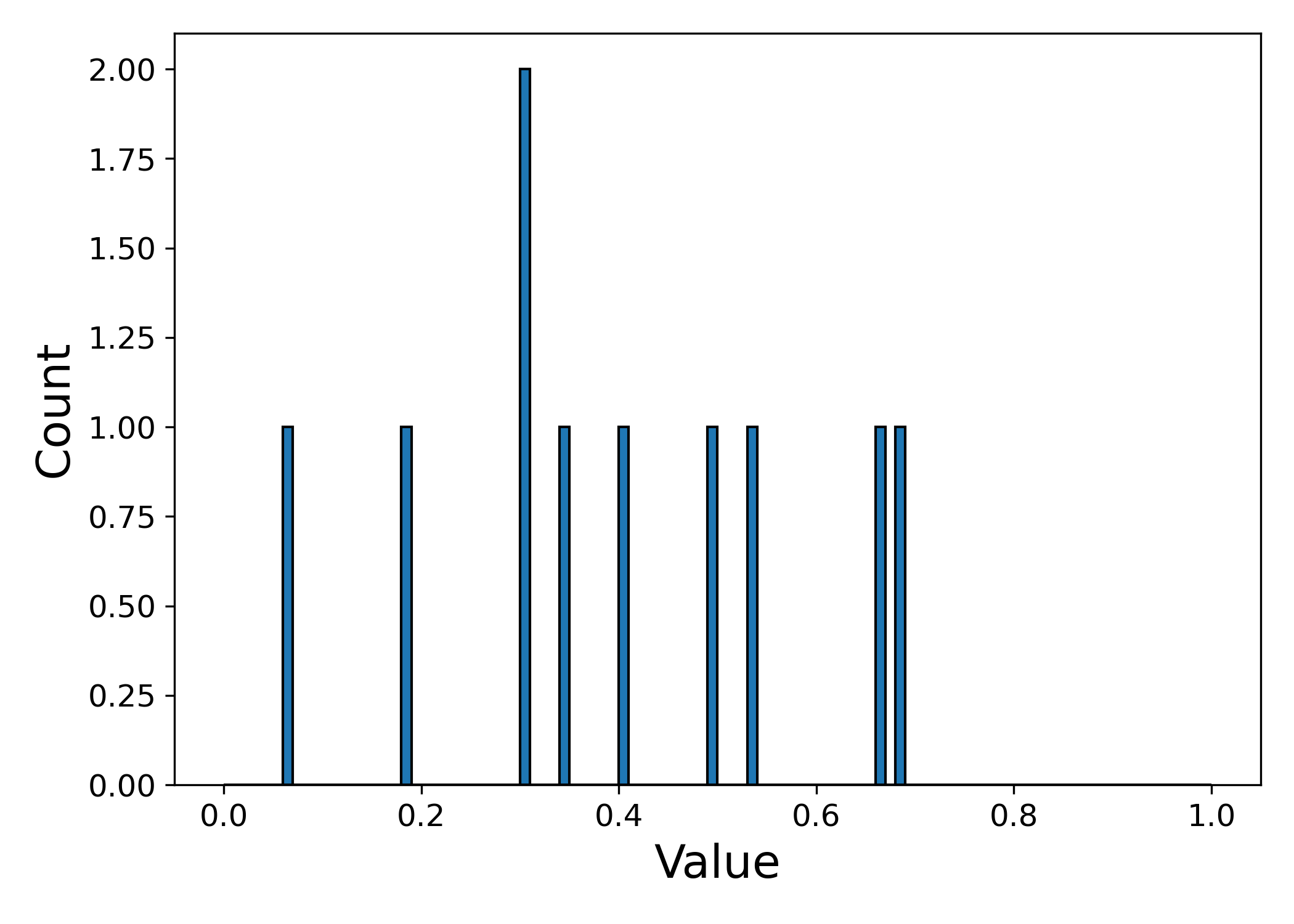}
  \label{fig:beta_sample_10}
  }%
  \subfloat[]{
  \includegraphics[width=0.48\linewidth]{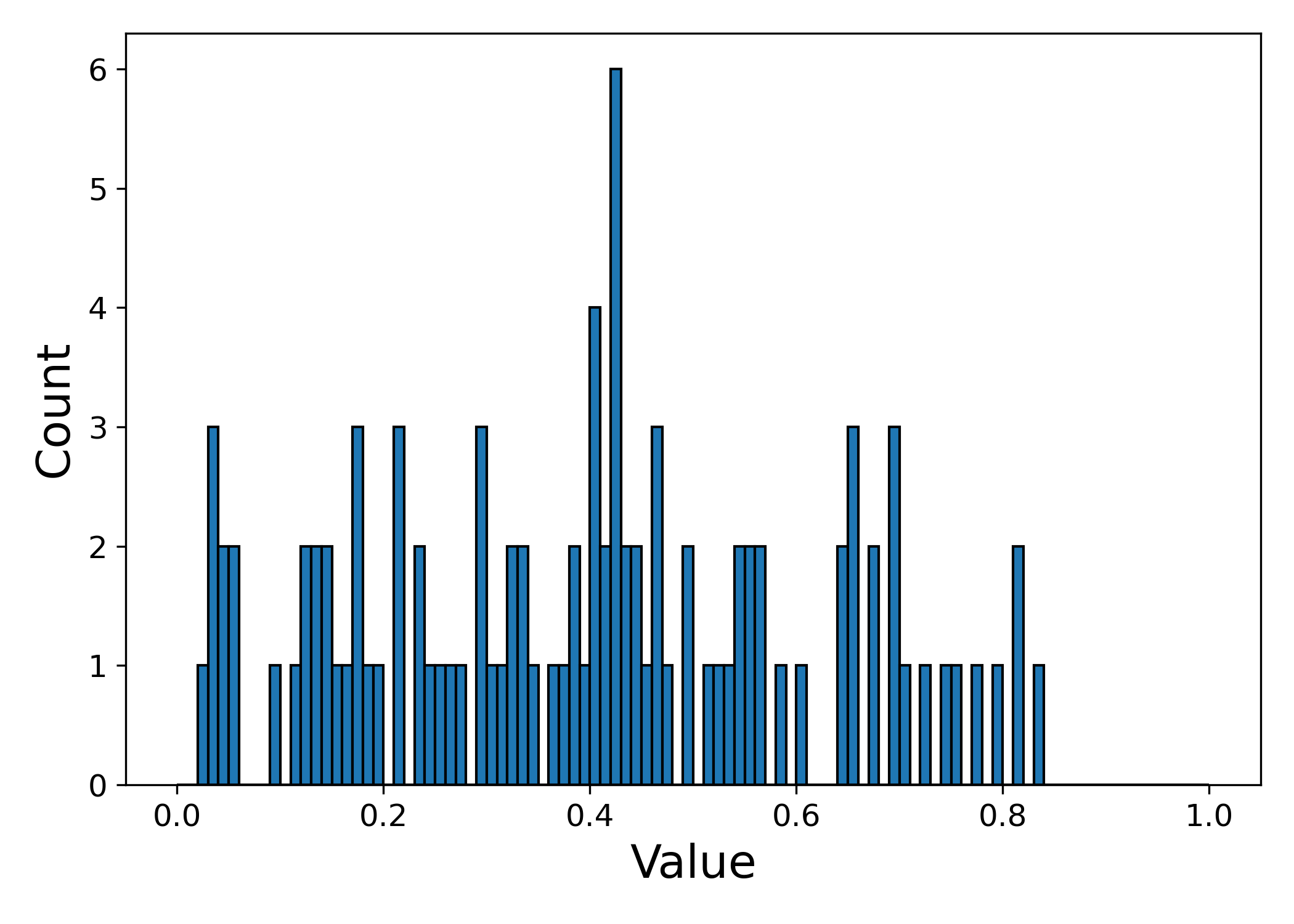}
  \label{fig:beta_sample_100}
  }
  \\
  \subfloat[]{
  \includegraphics[width=0.48\linewidth]{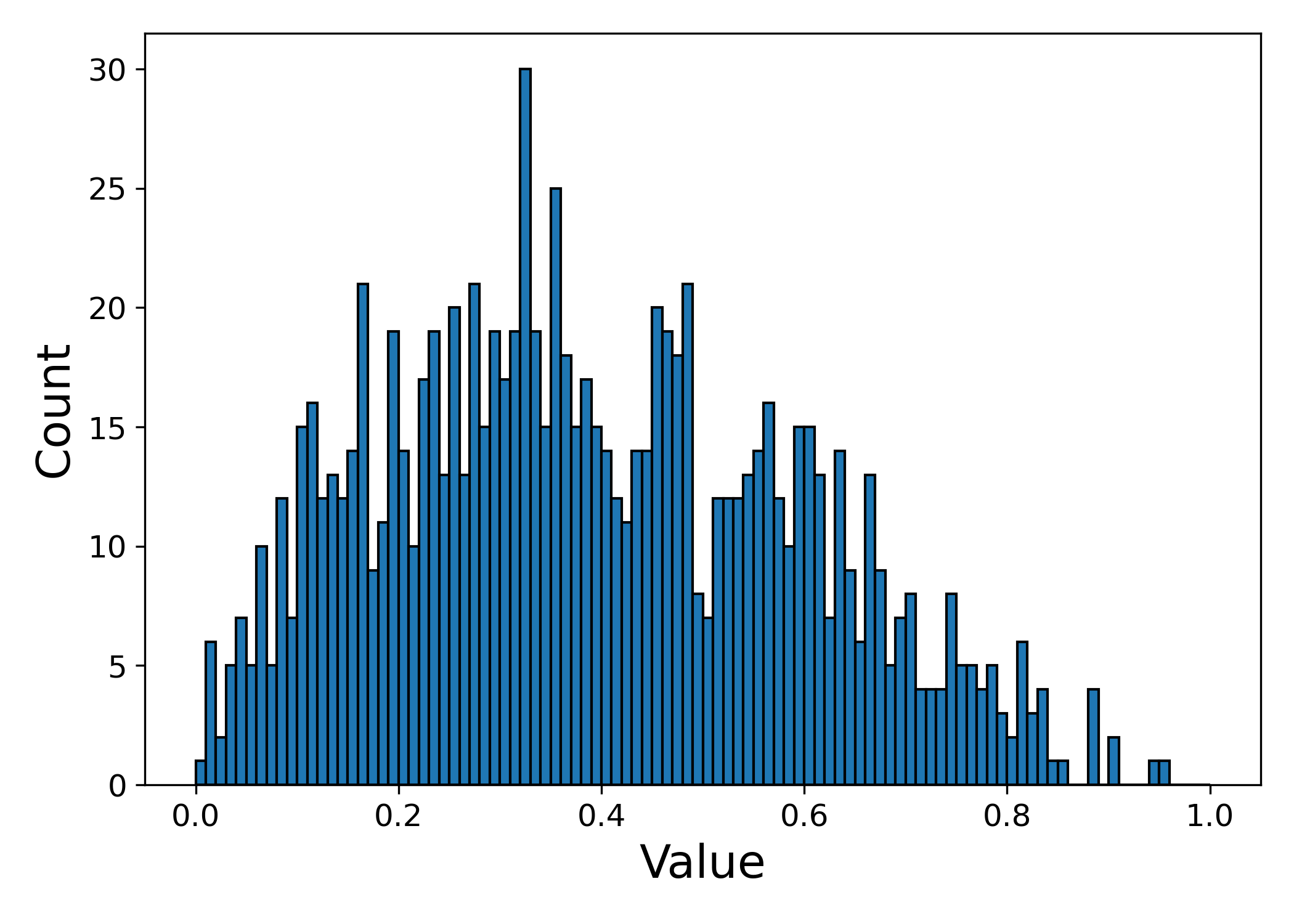}
  \label{fig:beta_sample_1000}
  }%
  \subfloat[]{
  \includegraphics[width=0.48\linewidth]{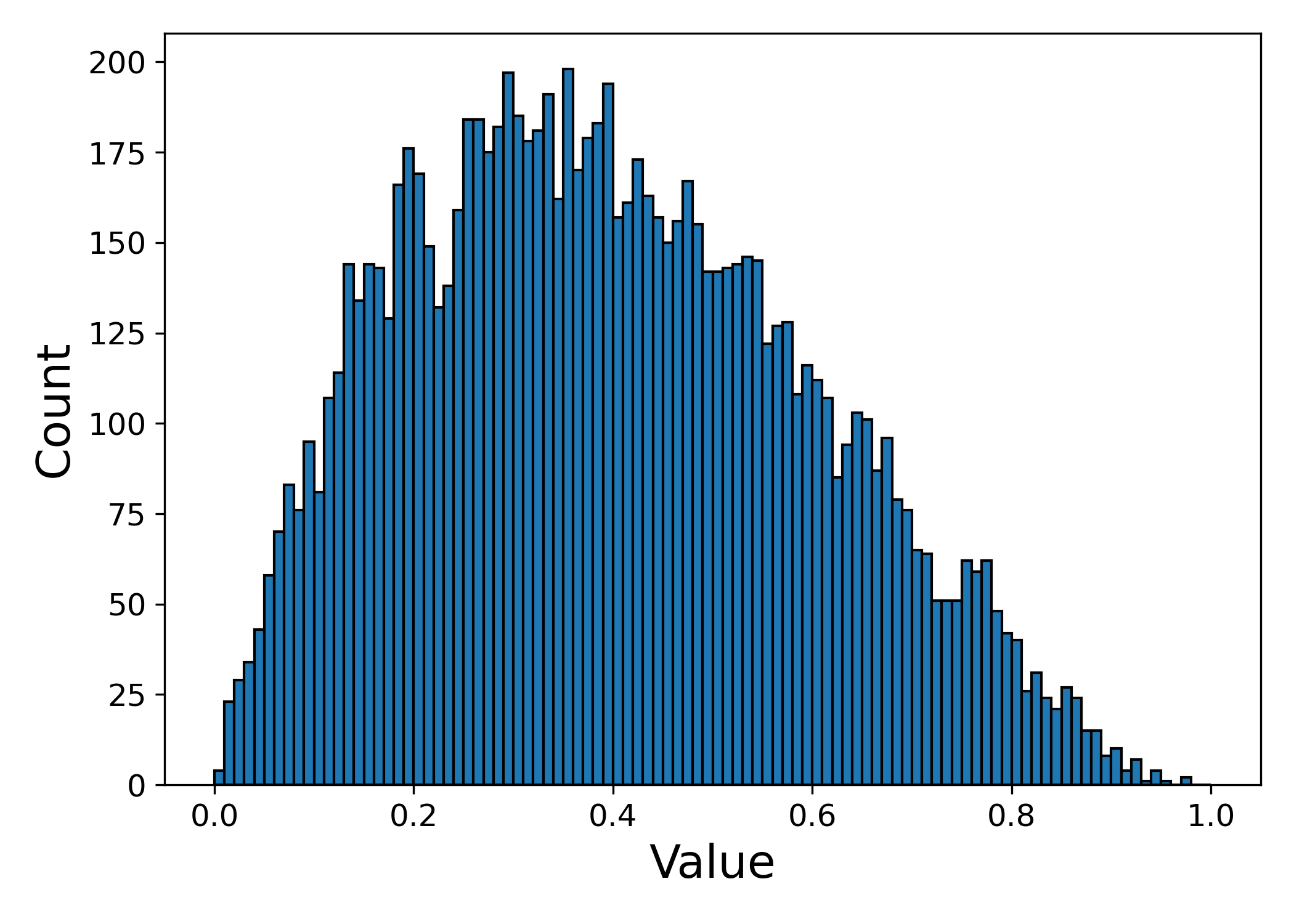}
  \label{fig:beta_sample_10000}
  }
  \\
  \subfloat[]{
  \includegraphics[width=0.48\linewidth]{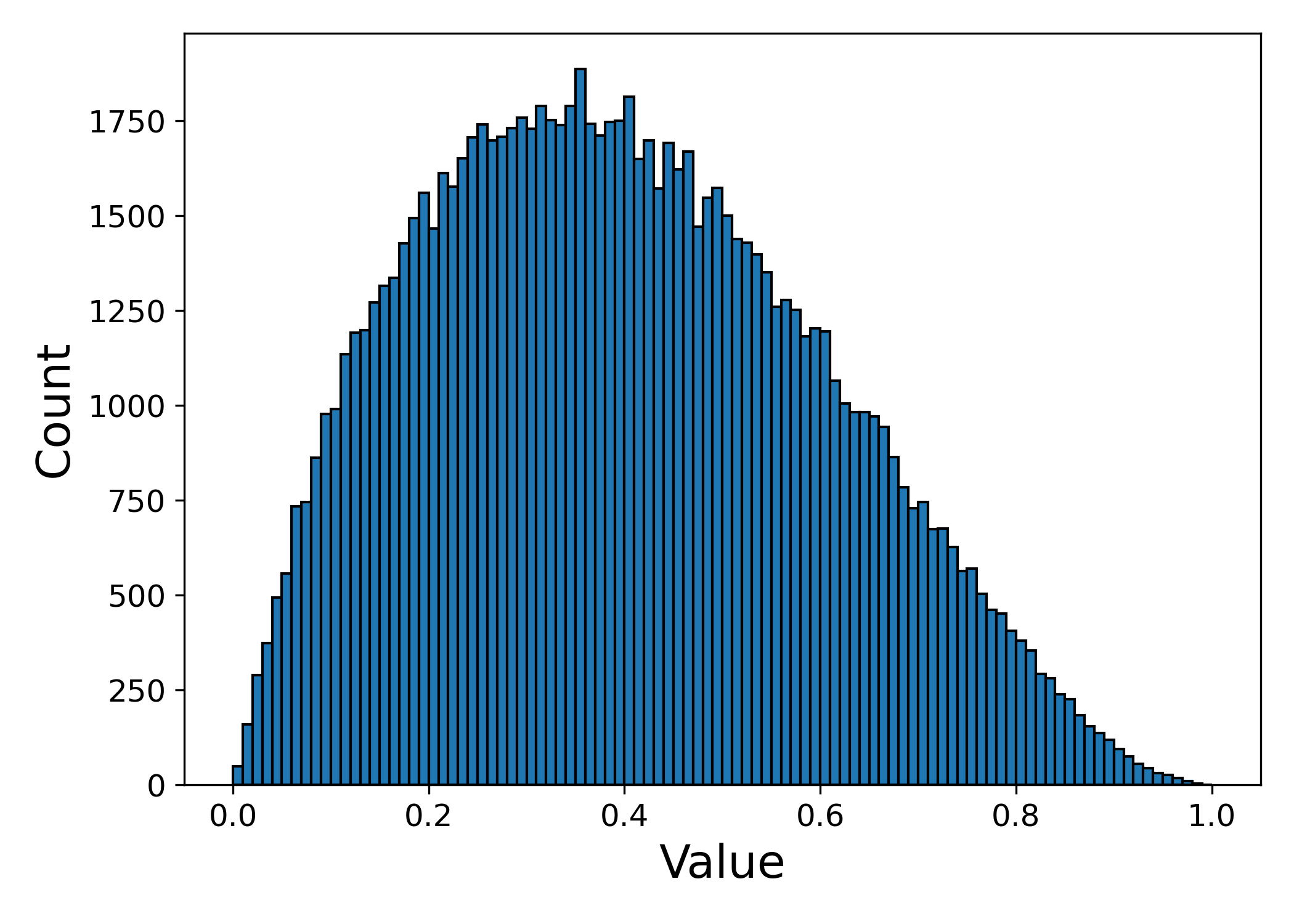}
  \label{fig:beta_sample_100000}
  }%
  \subfloat[]{
  \includegraphics[width=0.48\linewidth]{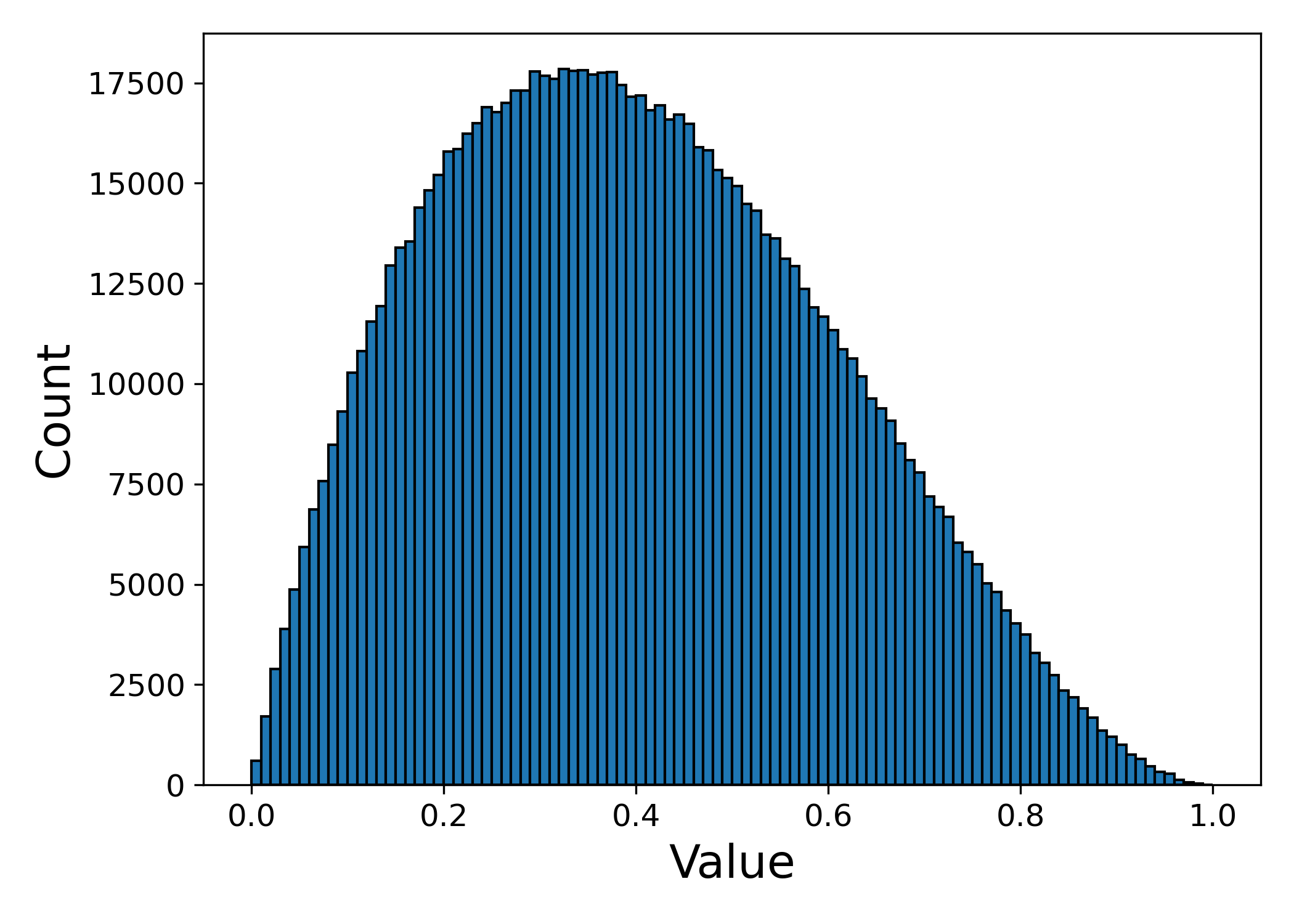}
  \label{fig:beta_sample_1000000}
  }

    \caption{The histogram of a random sample drawn from the beta distribution with $\alpha=2$ and $\beta=3$ when the sample size is: (a)~$10$, (b)~$100$, (c)~$10^3$, (d)~$10^4$, (e)~$10^5$, and (f)~$10^6$. For larger samples the effect of randomness decreases as expected by the Glivenko-Cantelli theorem, but for smaller samples it still exerts a significant influence on the histograms' smoothness.}
  \label{fig:beta_samples}
    
\end{figure*}

When talking about the similarity of smoothness and its impact on TVOR, what may be confusing at first with respect to Fig.~\ref{fig:old_vs_young} is the fact that a relatively smooth histogram as in Fig.~\ref{fig:20702} and a histogram with numerous spikes as in~Fig.\ref{fig:38525} have close TVOR scores. However, this is the expected behavior in the light of Glivenko-Cantelli since randomness is more pronounced on smaller samples and this is also what TVOR takes into account. To visualize this behavior, Fig.~\ref{fig:beta_samples} shows the histograms of random samples of sizes spanning several orders of magnitude and drawn from the same beta distribution with $\alpha=2$ and $\beta=2$, which was also used in our previous paper for demonstrative purposes. While for the largest sample in Fig.~\ref{fig:beta_sample_1000000} the effects of the randomness are relatively small and the histogram is rather smooth, on a smaller sample such as the one in Fig.~\ref{fig:beta_sample_1000}, which is somewhat similar to Fig.~\ref{fig:38525}, the effects of randomness produce clearly visible spikes. However, when the sample size and consequently the effects of randomness are handled properly as in the case of TVOR, it is possible to adequately compare even histograms of various sizes in terms of their smoothness and check for deviations in DTV to detect potential outlier candidates.

\subsubsection{The examples given in the critique}

Having in mind these results, it can easily be seen that the the critique's claim that ``there is also no reason to believe that the total variations in different lists should be similar'' is first of all missed since it does not include the notion of the sample size. However, even if the claim was generalized to replace the mention of DTV with the TVOR score, it would again not hold. Namely, our previous paper's Figs.~15 and especially~17 clearly show how the majority of the scores are relatively close to each other. Since the scores are spanning over a range, as can be expected from a random variable, it is relatively easy to find examples of histograms with different scores that look arbitrarily similar or different. Because of that, it is possible to find histograms of various scores that seem mutually rather different in shape as was done in the critique, but for these very same histograms it is also possible to find histograms of either similar or different shape that will have a similar score. Hence, the critique's example shows the fallacy of incomplete evidence also known as cherry picking.

\subsection{The similarity of populations}

Regarding our paper's claim about the similarity of birth year histograms of populations, in the critique it is said that ``such a claim is easily disproved in general by a wealth of historical evidence, as well as by data analysis.'' The critique did not offer any evidence for such a claim. On the contrary, the demographic and historical data such as~\cite{shryock1980methods,siegel2004methods} can be used to show the opposite. Because of this, it is not fully clear what exactly was the motivation for such an unsupported claim in the critique. Already by comparing sources such as~\cite{jahrbuch1939germany} and~\cite{siegel2004methods}, it can be seen that the populations described there have some quite similar marks such as e.g. the effects of World War I or more general kinds of age transitions, which gives them similar smoothness properties. A similar conclusion can also be reached when comparing other populations.

\section{Reply to ``Histogram size bias''}
\label{sec:bias}

\subsection{The initial claims}
\label{subsec:initial}

	In the section on the alleged histogram size bias, the critique again mentions a supposed normalization's reliance ``on all histograms being drawn from the same probability distribution'', which is not true as described in Sections~\ref{sec:preliminaries} and~\ref{sec:assumptions}.

\subsection{The critique's mathematical mistakes}
\label{subsec:mistakes}

\subsubsection{The theoretical analysis}
\label{subsubsec:theoretical}

	The critique goes on by finding a bias by obtaining a small positive linear correlation between the scores $d'$ given by TVOR and the histograms' sample size for the USHMM dataset. In order to remove the bias, the critique proposes a ``renormalization'', which boils down to dividing the TVOR's score $d'$ by a term dominated by $N$, which would implicate a direct linear dependence of $d'$ on $N$ and this is utterly incorrect and it amounts to a rather bad and flawed 
	statistical practice. The dependence of $d'$ on $N$ cannot be linear since it would directly mean that the discrete total variation~(DTV) has a dependence $N^{\frac{3}{2}}$, which is definitely not the case by the very definition of DTV. The graveness of this mistake should not be underestimated due to its highly detrimental effects on reasoning. Therefore, it should be explained in more detail.

        The discrete total variation has the following behavior:
        \begin{equation}
        \label{behavior1}
	        \V{\mathbf{x}_n}\approx \alpha \cdot N+\beta \sqrt{N}.
        \end{equation}
        Here, $\alpha$ is the variation of the theoretical distribution, denoted in \cite{banic2021tvor} by $\V{\mathcal{D}}$. If all probabilities in adjacent bins are different, then for a sufficiently large $N$, according to Glivenko-Cantelli theorem, we would have 
        $\V{\mathbf{x}_n}= \V{\mathcal{D}} \cdot N$ almost surely. If the probabilities for some adjacent bins are equal, then we have random fluctuation which contribute to the discrete variation with the term proportional to $\sqrt{N}$, as explained in \cite{banic2021tvor}. The same efect of randomness proportional to $\sqrt{N}$ arises for all bins in the case of modest values of $N$.

        The remarks in the critique's Section IV ``Histogram Size Bias'' are crucial to understanding the intent with which~\cite {ornik2021comment} was written. The main intention was to remove the Jasenovac list as the primary outlier. For this purpose, fallacious and unethical methods were used, which can be clearly seen as follows. 
        The critique uses the values of $d'(h_i)$ in a wrong way by first finding the dependence of $d'(h_i)$ on volumes $N_i$ expressed through linear regression in the following form:
        \begin{equation}
        \label{eq:regression1}
        \overline{d}'(N)=4.017\cdot 10^{-5}N+2.128
        \end{equation}
        As the critique says, the coefficient of $4.017\cdot 10^{-5}$ may not seem significant for small $N$, but it may make some difference in cases such as the Jasenovac list, which has $N=78493$ elements. We must cite another excerpt form\cite{ornik2021comment}: ``To show the existence of bias, we use several methods. We first calculate the correlation coefficient between histogram sizes and $d'$ scores. Its value is indeed positive and equals $0.17$. To place this value in comparison, the ``random noise'' correlation coefficient between the dataset histogram sizes
        and 7106 samples chosen from a normal distribution is generally
        smaller by roughly a factor of 10--100.''

       Both remarks are the result of a misunderstanding of TVOR method. The score $d'$ given in Algorithm~1 in~\cite{banic2021tvor} and Eq.~(51) in~\cite{banic2021tvor} is given \textit{with its absolute value}. However, the value of $d'$ which should be taken in calculations of correlation coefficient and linear regression should be taken \textit{with its sign}, since $d'$ describes the deviation from expected value and this deviation could be negative as well. We eventually deliberately choose to take absolute value for better graphical presentation of deviation and easier ranking of outliers.

        If one takes the values $d'_i$ \textit{with their signs}, then the correlation coefficient between $d'_i$ and volume $N_i$ is equal to $-0.0174531$, which is the value comparable to any random simulation of independent quantities. Also, the exact value of linear regression line is not given by \eqref{eq:regression1}, but instead by
        \begin{equation}
        \label{eq:regression2}
        \overline{d}'(N)=-5.67649\cdot 10^{-6} N +0.858288 
        \end{equation}
        Not only is the leading coefficient of the order of magnitude that would be obtained for ``random noise'', but its sign is negative!

\subsubsection{Additive instead of multiplicative bias}
\label{subsubsection:additive}

	This very small bias given in (\ref{eq:regression2})
	can be further reduced through additional adjustment of the values of $a$ and $b$ using an iterative procedure. 
	This means that the accuracy of $d'$ can be improved by subtracting the value of $a_1\sqrt{N}+b_1$ resulting in a negligible change of $a$ and $b$. The order of top outliers after such modification remains the same.

       After establishing a non-existent correlation due to a misinterpretation of the TVOR method, the critique uses an outrageous manipulation by introducing ``renormalization'' by \textit{dividing} the score by $\overline{d}'(N)$.
	For reasons already described, this division is a procedure that cannot be justified by any means, which makes it detrimental for any reasonable discussion. As a matter of fact, it is even unethical.

        Why is this unethical? Before publishing his critique~
        \cite{ornik2021comment}, its author posted it in on arXiv~\cite{ornik2021inapplicability}. One of authors (N.E.) noticed this post and in a private communication informed the critique's author that $d'$ should not be taken with absolute value and that renormalization by dividing is out of any mathematical sense. The critique's author agreed with this, but he nevertheless published his paper in unchanged form.

\subsubsection{Inconsistency on the USHMM data}
\label{subsubsec:inconsistency}

	Even if absolute values are used, which is clearly wrong, the ``renormalization'' proposed in the critique can be shown to obtain results that are in contrast with the critique's assumption of a positive bias for larger lists. Namely, if linear fitting is applied only to the results of the $7$ largest lists in the USHMM dataset, then the obtained slope is $-3.6471\cdot 10^{-5}$, i.e. it is negative. This would indicate a bias towards smaller lists, which is in contrast to the critique's claims. 

\subsection{The alleged sample size issue}
\label{subsubsec:size2}

The critique mentions that our previous paper allegedly alludes the identified bias by ``noting that even the sublist [of the Jasenovac list] with the highest outlier $d'$ score has a lower outlier score than the entire list.'' This mentioning is based on the fact that a smaller sublist had a smaller score $d'$ and on our comment that ``such similar deviations are considered to be less likely on a larger sample and thus the whole Jasenovac list has a slightly larger value of score $d'$.''

\begin{figure*}[htbp]
	\centering

	\subfloat[]{
	\includegraphics[width=0.48\linewidth]{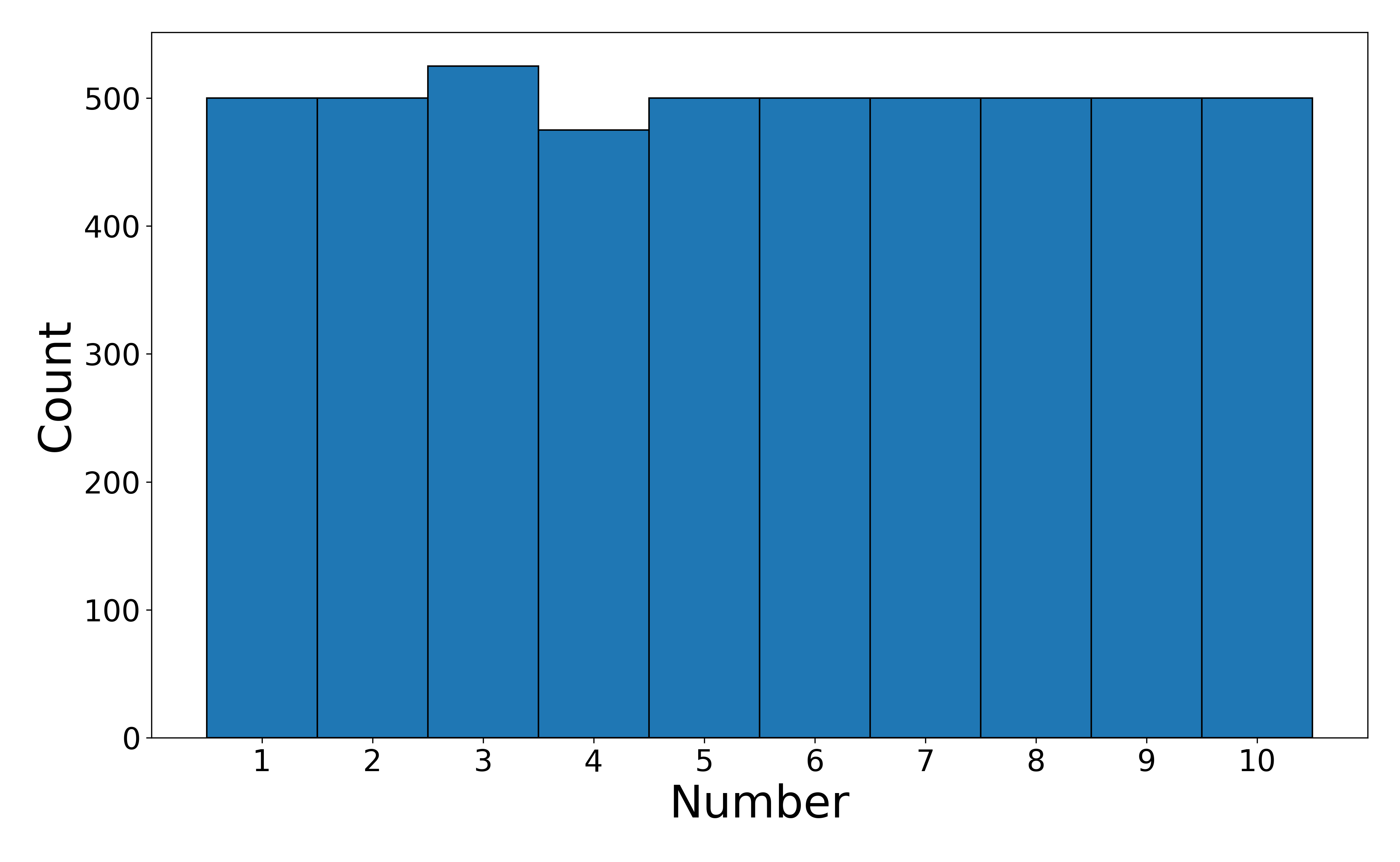}
	\label{fig:almost_uniform_500}
	}%
	\subfloat[]{
	\includegraphics[width=0.48\linewidth]{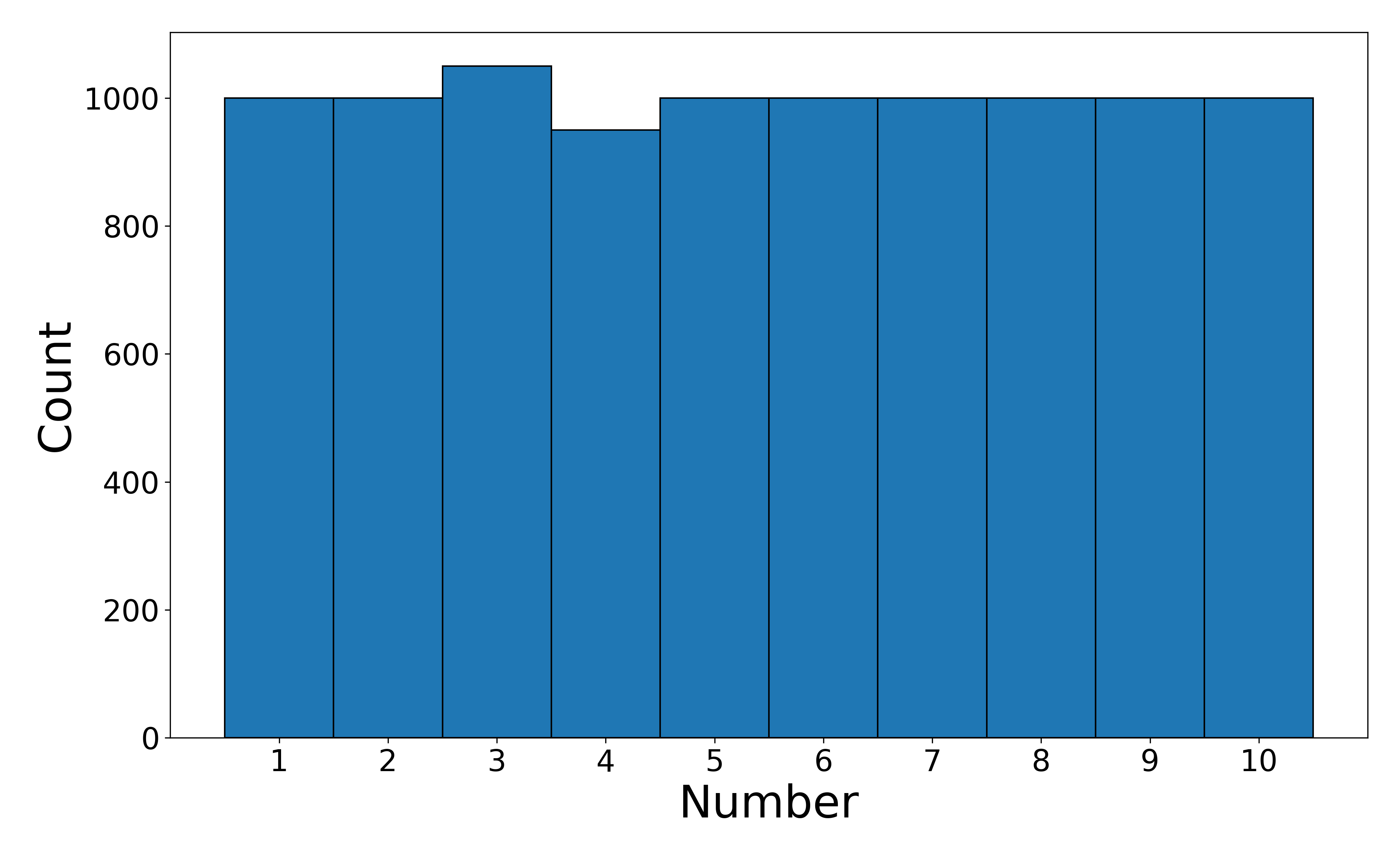}
	\label{fig:almost_uniform_1000}
	}
	\\
	\subfloat[]{
	\includegraphics[width=0.48\linewidth]{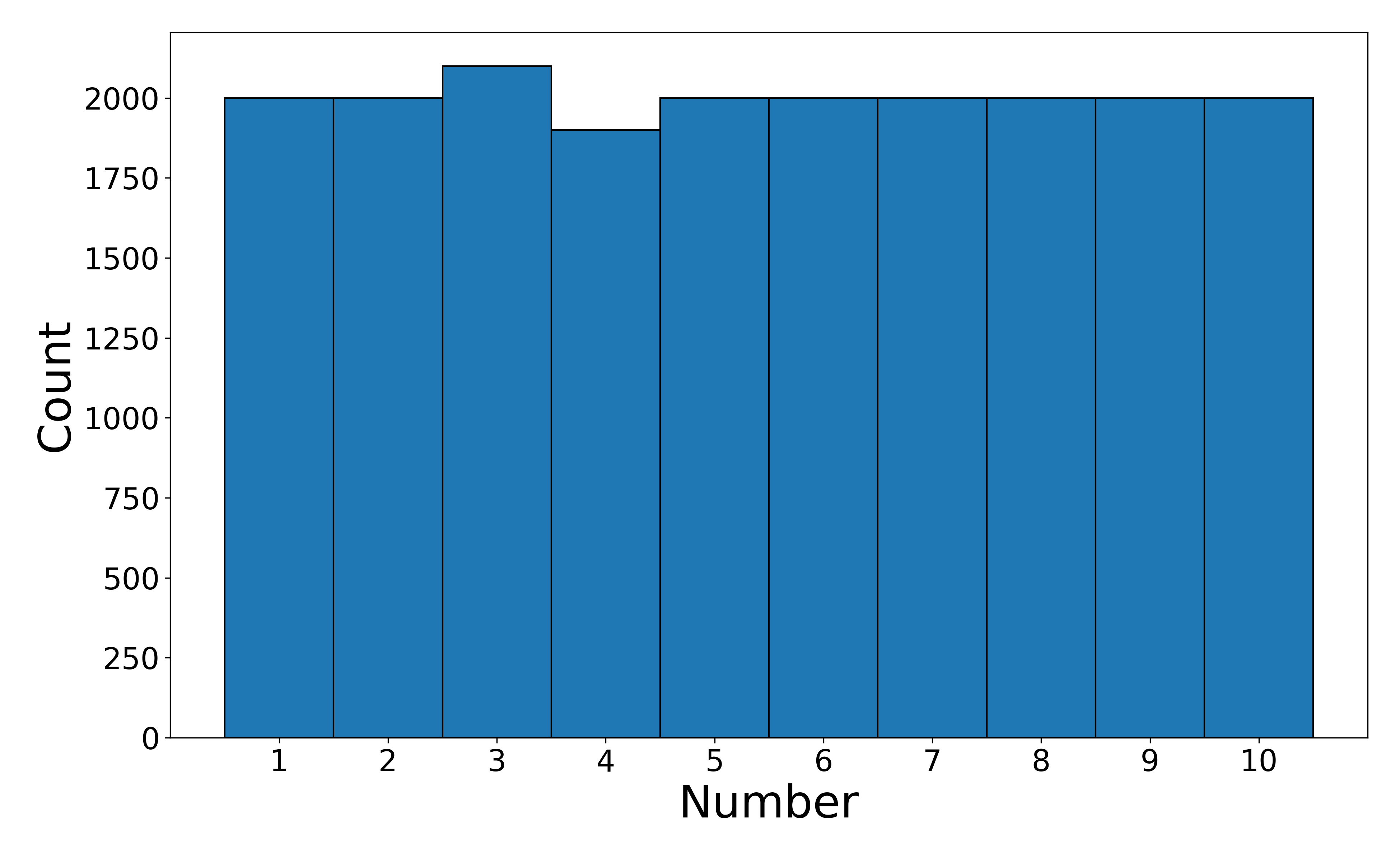}
	\label{fig:almost_uniform_2000}
	}%
	\subfloat[]{
	\includegraphics[width=0.48\linewidth]{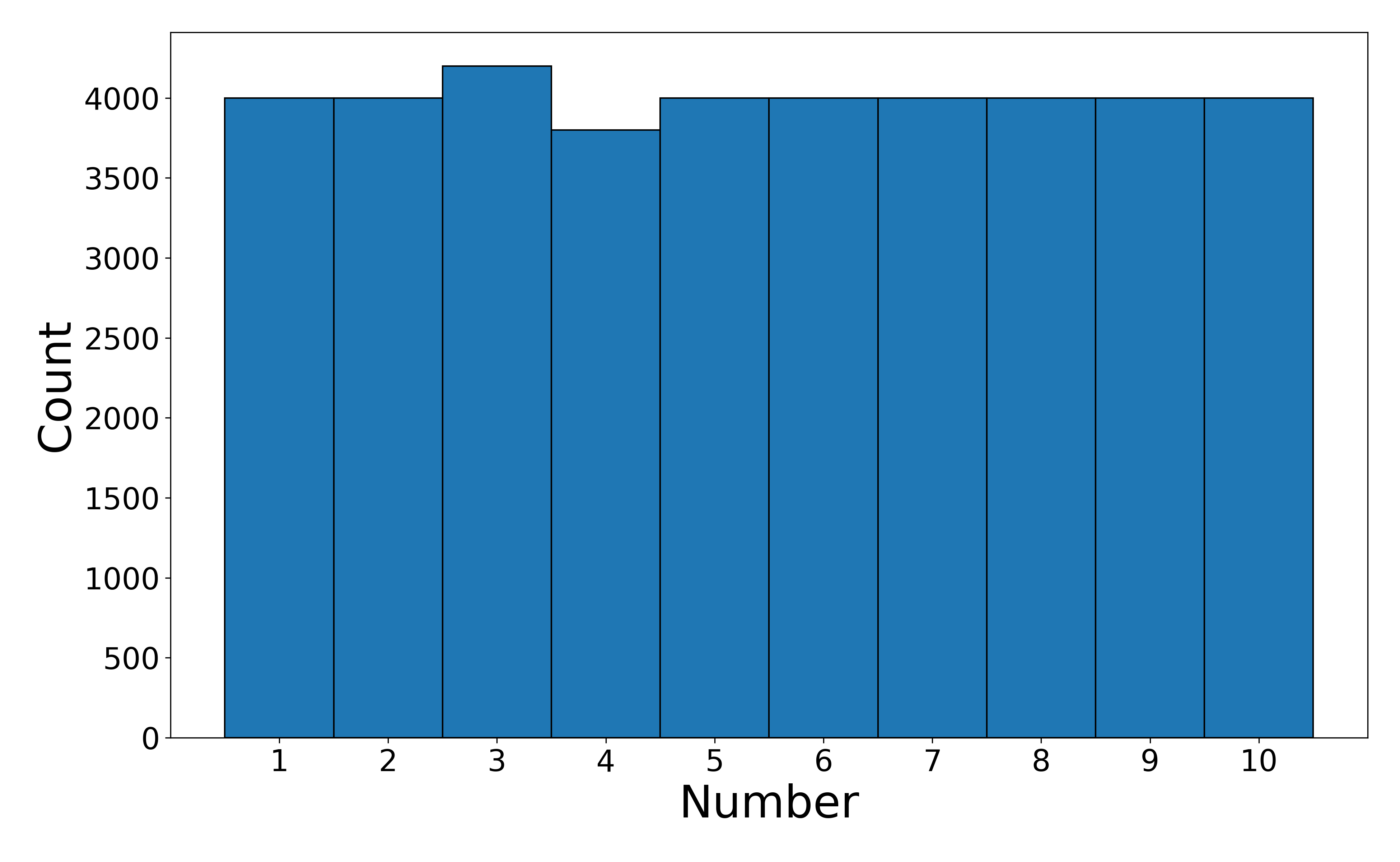}
	\label{fig:almost_uniform_4000}
	}

    \caption{Samples of same shape, but of different sizes and consequently of different likeliness of having been randomly drawn from the uniform distribution. If the Pearson's chi-square for the uniform distribution is applied, then the obtained $p$-values for different sample sizes are a)~$0.9809$ for $5000$, b)~$0.8343$ for $10000$, c)~$0.3505$ for $20000$, and d)~$0.0179$ for $40000$.}
	\label{fig:almost_uniform}
    
\end{figure*}

However, this has nothing to do with an alleged bias, but with a regular behavior that is often associated with various statistical tools. Namely, sample size matters a lot. If the critique's statement were true, than it could mean that tests such as Pearson's chi-square tests are also biased. For example, testing whether a histogram of a small size is in accordance with a prespecified distribution will yield some value $p$ that will not be $1$ if there is some noise present. If another larger histogram with proportionally scaled bin values is also tested in the same way by applying Pearson's chi-square test, then the resulting $p$ value will easily be smaller. This can be shown on an example with uniform distribution shown in Fig.~\ref{fig:almost_uniform}. The samples there are of same shape, but of different size and because of that, the Pearson's chi-square tests gives different $p$-values despite the samples having the exactly same shape. Namely, in accordance with Cantelli-Glivenko theorem, randomness plays an ever smaller role as the sample size increases and so any deviations from the shape expected by a given distribution should be penalized more strictly.



\section{Reply to ``Dearth of relevant histogram data''}
\label{sec:dearth}

The critique mentions that ``the lists of similar size to Jasenovac [...] have significantly different DTVs from each other, again in contravention of the assumptions.'' First of all, as already explained on several places in this paper, as the sample size grows, the effects of randomness become ever smaller as described by the Glivenko-Cantelli theorem, which means that the uncertainty whether a deviation was caused by randomness also diminishes. This in turn means that scoring will become stricter regardless of the similarity or dissimilarity between the histograms, which is in accordance with statistics and similar behavior can be found in tests such as Pearson's chi-square test as shown in Fig.~\ref{fig:almost_uniform}. Because of that, the critique's claim again misses the point.

The critique claims that ``the [USHMM] dataset is sparse around the point of interest'', with the point of interest seemingly being the Jasenovac list. This is then described in a way which hints that this is problematic, but no analysis of the sparseness is given whatsoever in terms of actual results.

To check whether the sparseness effect around the top outlier candidate, i.e., the Jasenovac list and the density at lists of smaller sizes has any significant effect on the final ranking, it is possible to conduct an experiment by ignoring smaller lists and leaving only arbitrarily large lists. In the used USHMM dataset, there are $742$ different unique sizes among $7106$ lists. Hence, it is possible to run TVOR $742$ times, once for each of these sizes. Before conducting each of these TVOR runs, the lists that are smaller than the currently viewed unique size are dropped. It turns out that the Jasenovac lists happens to be the top ranked list in all TVOR runs that do not drop it due to its size, i.e., in $736$ out of $742$ instances the Jasenovac list is the most prominent outlier candidate, while in the remaining $6$ instances it is not included in the dataset subset or largest lists. Therefore, the sparseness of larger lists in the USHMM dataset does not appear to have any particular effect on the final result, and neither does the density of smaller lists.

Although TVOR's scores of larger lists are stricter, i.e., they may have higher values, this does not blur or hide any information about potential outliers among them. For example, if TVOR is applied only to $20$ largest USHMM lists, then the Jasenovac list score clearly fits the standard definition of an outlier based on the interquartile range~\cite{rl1997beta,devore2012modern}. Hence, even if the Jasenovac list is compared only to lists of similar size larger than $32$k, it turns out to be an outlier even in terms of the already established statistical techniques.

The critique goes on to claim that ``[t]he resulting distribution does not seem to have the vast majority of its mass around 0'', but without explaining why this should be the case at all. Our previous paper explicitly avoided probabilistic interpretations and any additional actions that would have to be carried out to adjust the scores and thus there are no explicit expectations about their distribution. The scores are here used only for sorting and as already explained before.

\begin{figure*}[htbp]
	\centering

	\subfloat[]{
	\includegraphics[width=0.48\linewidth]{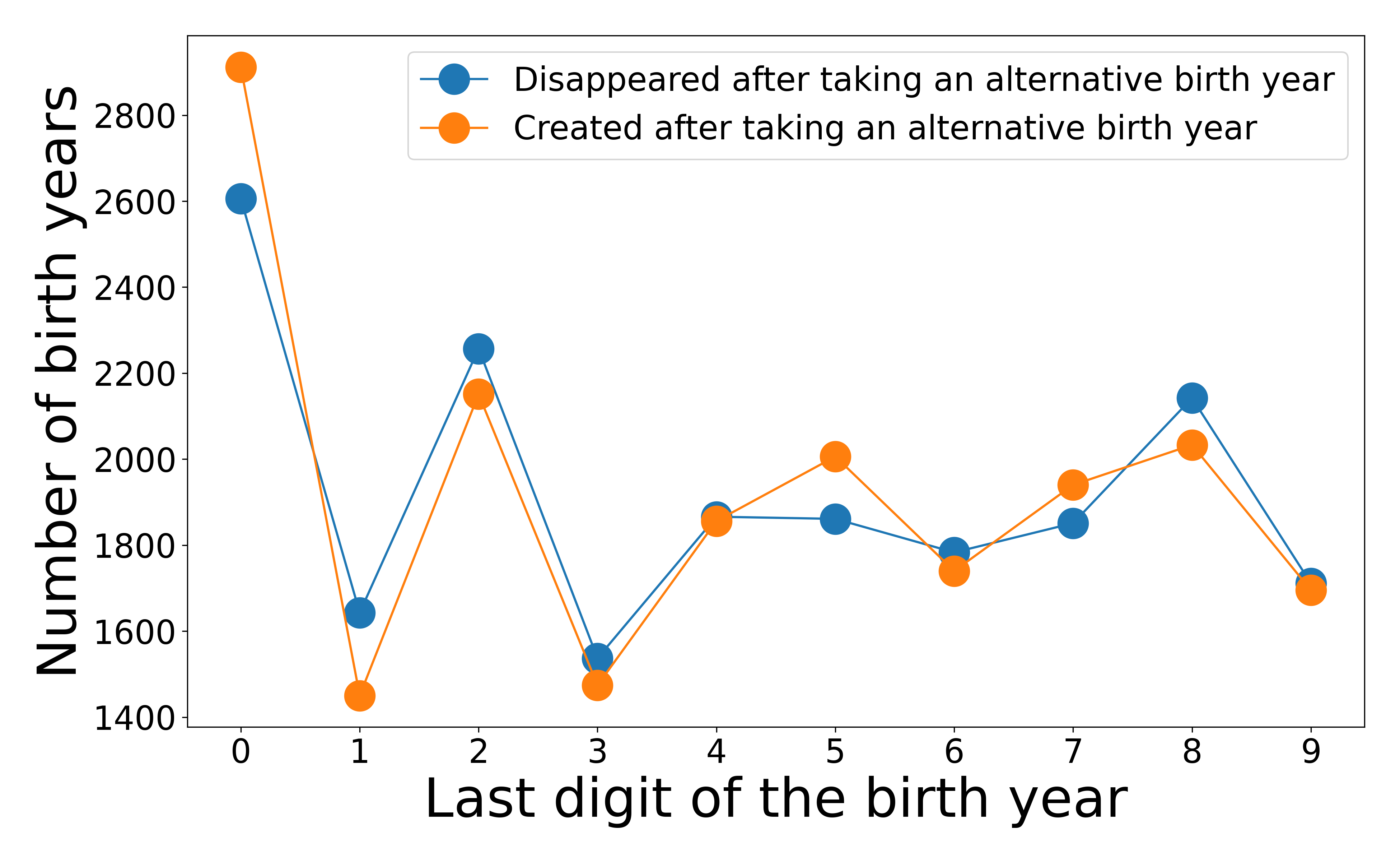}
	\label{fig:digits_before_and_after}
	}%
	\subfloat[]{
	\includegraphics[width=0.48\linewidth]{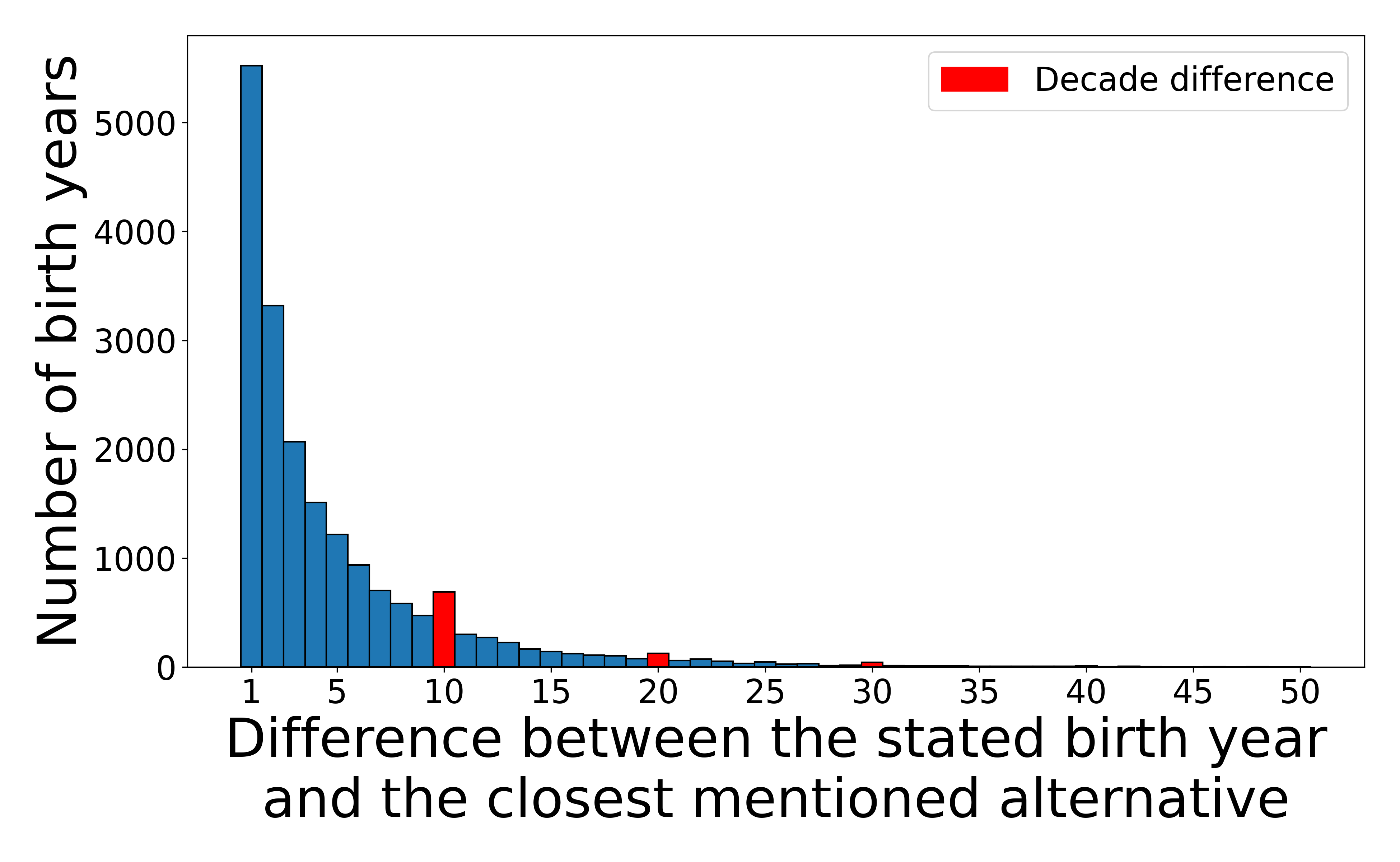}
	\label{fig:year_differences}
	}

    \caption{The effect of substituting the stated birth years of the inmates listed on the Jasenovac list with the closest alternatives mentioned in the source field of each individual inmate: (a)~the number of disappeared and created last digits after the substitution and b)~the distribution of the absolute differences between the stated birth years and their closest alternatives.}
	\label{fig:alternative_birth_years}
    
\end{figure*}

\section{Reply to ``Data processing and analysis issues''}
\label{sec:processing}

The critique mentions that the majority of the USHMM datasets consists of sublist of a single list called \textit{The Elders of the Jews in the {\L}{\'{o}}d{\'{z}} Ghetto}, which is also present in the USHMM dataset. It continues on suggesting that this may be problematic for TVOR, but there is again no evidence at all in terms of numerical results to objectively support such claims. On the contrary, if TVOR is applied to the USHMM dataset without the mentioned sublists, the value of $d'$ for the Jasenovac list turns out to be $44.19$ if all lists with at least one birth year are used and $44.20$ if only lists with at least $100$ birth years are used as was done in the original paper and the given code. These results are even slightly greater that the previously obtained ones, but nothing essentially changed in the case of Jasenovac, which in both cases again turns out to be firmly the top outlier. This in turn all means that the original composition of the USHMM dataset has no significant effect on TVOR's rankings, which confirms TVOR's robustness and further misguidance of the critique.

Skipping several of the claims to deal with them later and going to the last one, the critique claims that our repository ``fails to take into account that some of the years of birth are marked as disputed on the USHMM website''. It is not entirely clear whether this is supposed to be problematic or not. If the information for the Jasenovac list~\cite{ushmm2020jusp} is considered more closely, it can be seen that on the page of every individual inmate, including the ones mentioned in our previous paper~\cite{ushmm2020sternau,ushmm2020razokrak,ushmm2020nickjusp}, there is a note stating that ``[t]he asterisk * indicates there are conflicting reports from different sources''. The asterisk appears next to the birth year of $19431$ inmates, i.e., over $23$\% of them. Rather then a problem, this is possibly an additional confirmation of TVOR's earlier result that the birth years of the Jasenovac list may be problematic.

However, to check whether this ``conflicting reports from different sources'' would have any impact on the highly expressed age heaping of the Jasenovac list if they were used instead, a simple experiment was conducted by substituting the currently stated birth years with their closest alternatives mentioned in the source field. This source field is given on the individual page of every listed inmate and it states, for almost all of them, the data source on which the values of other fields are based, while in the case of the mentioned ``conflicting reports from different sources'' it sometimes also includes the data from this reports. The results of the mentioned experiment are shown in Fig.~\ref{fig:alternative_birth_years}. Based on Fig.~\ref{fig:digits_before_and_after}, it can be concluded that even by adopting the closest alternative birth years, the age heaping properties do not change significantly in the case of the Jasenovac list. As a matter of fact, the age heaping becomes rather slightly more expressed due to an increase of the already numerous birth years ending in $0$ and $2$. While there is a decrease in birth years ending in $5$, it is smaller than the increase in $0$ and $2$. This means that the problematic nature of the Jasenovac list data does not seem to be directly connected to simple ``data processing and analysis issues''. Additionally, Fig.~\ref{fig:year_differences} shows an interesting distribution of the absolute difference between the stated birth year and its closes alternative, which will be used later. The particularly interesting feature is a sudden count rise for the differences of $10$, $20$, and $30$, which may mean that some sources are of lower quality due to potential transcription errors of the decade digit. Nevertheless, these errors have no direct effect on the age heaping problem of the Jasenovac list.

The rest of the critique's claims here are also unfounded. However, since they are interleaved and lead to many new insights, they are covered individually only later in Section~\ref{sec:jasenovac}.

\section{Reply to ``Objections and Answers''}
\label{sec:objections}

The critique tries ``to specifically address some possible objections and concerns'', but most of the answers provided to the questions posed there can be shown to be similarly unfounded and misleading as the rest of the critique for the very same reasons that have already been explained in this paper. Yet, for the sake of better understanding, it is worth to again address each of these answers to possible objections.

The first possible objection mentioned in the critique is the question whether the models for the ``renormalization'' offered by the critique are ``not less valid than TVOR, given that they lack TVOR's theoretical foundation''. The critique starts by mentioning that TVOR allegedly has no theoretical foundation on the USHMM dataset, which is wrong as shown again later in this section. It then goes on by again repeating the false claims about TVOR's connection to probabilities that have already been refuted in Section~\ref{sec:assumptions} in this paper. Next, the critique mentions that the only claim about the ``renormalization'' models is that they have less bias than TVOR and that there is no claim about the models being functional. However, model $0\cdot N$ has even less bias than the critique's proposed models, but despite that it is useless. The only additional feature of the critique's models is that their uselessness is hidden by an apparent mathematical derivation, but as shown in Section~\ref{subsubsec:theoretical} in this paper, this derivation is deeply flawed and consequently utterly wrong. Hence, the critique's claims about this issue are unfounded.

The second mentioned objection is about the validity of the approach of removing all duplicated and incorrect data before applying TVOR. The critique claims that this approach would not be valid because it would allegedly be nearly impossible to access all historical sources and check the validity of all records, and it also claims that TVOR has allegedly no theoretical grounding on the USHMM dataset. First of all, Section~\ref{sec:processing} in this paper clearly shows that removing the duplicate lists has practically no effect whatsoever on the final ranking, which means that the effect of duplicate lists or sublists had no detrimental effect at all. Second, the veracity of the records or any other information related to the birth years is here from a point of pure mathematical calculations irrelevant; what matters are the numbers themselves. DTV takes into account only certain aspects of the histograms such as age heaping, while it simultaneously stays unaffected by numerous other phenomena. If there would be some espetially significant effect caused by any of these, then it would also be reflected on TVOR scores. However, the critique offers no concrete numerical evidence for such claims. Third, the critique mentions again that TVOR has no theoretical grounding on the USHMM dataset, but later in this section it is again explained why this is not the case. Hence, the critique's claims about this issue are unfounded.

The following potential objection is about the same ranking even if only ``lists of similar size and similar locations as the Jasenovac list'' are considered. The critique claims that TVOR's approach would not be valid because of an alleged lack of TVOR's theoretical foundation. The claim about the theoretical foundation is also repeated in the next critique's next potential objection, so there it is again explained to be false in more detail. As for the rankings themselves, Section~\ref{sec:dearth} of this paper clearly shows that no matter how many smaller lists are discarded, the Jasenovac list always turns out to be the top outlier candidate, which may already suggest that is has to do something with its problematic nature rather than with its size. Finally, Section~\ref{sec:dearth} of this paper also checks the scores obtained by TVOR for only $20$ largest lists in the USHMM, which is already a sufficiently big sample size. There it is shown that the score obtained by the Jasenovac list clearly fits the standard definition of an outlier when compared to the scores of other lists. The same is easily repeated for larger samples as well. This means that even standard definitions clearly describe the score of the Jasenovac list as being significantly dissimilar from the scores of lists of similar size. In short, it is a clear outlier. Hence, the critique's claims about this issue are unfounded.

The next potential objection is about the shared population distributions across regions and ethnicities. The critique claims that it is impossible for e.g. lists of children and card holders to have the same distribution and that different population distributions also have different smoothness. The claim about the specific case of the probabilities is true, but for TVOR it is not relevant, while the claim about the smoothness is generally false for such a broad claim. The explanation for why this is so is given in much more detail earlier in this paper. Namely, Sections~\ref{sec:assumptions} covers the probabilities and smoothness issues and Section~\ref{sec:dearth} explains that even when the scores have to be stricter, they are still mutually similar and outliers are easily identified. Because of that, the critique's claims about this issue are unfounded.

The critique goes on to discuss the potential objection of why it disputes the problematic nature of the age heaping on the Jasenovac list. It is claimed that ``the notion of a feature being problematic [...] has never been rigorously defined in'' our previous paper. However, having a highly expressed age heaping problem is very well known to be problematic for the quality of data and this is known common knowledge in the area of demographics. Cases where age heaping is not to be considered problematic are rather exceptions than a rule. The quality of data for the range of Whipple's index that also includes the one for the Jasenovac list is described as ``bad''~\cite{robine2007human} and having bad quality data certainly is problematic. Because of that, the critique's claims about this issue are unfounded.

\begin{figure*}[htbp]
	\centering

	\subfloat[]{
	\includegraphics[width=0.48\linewidth]{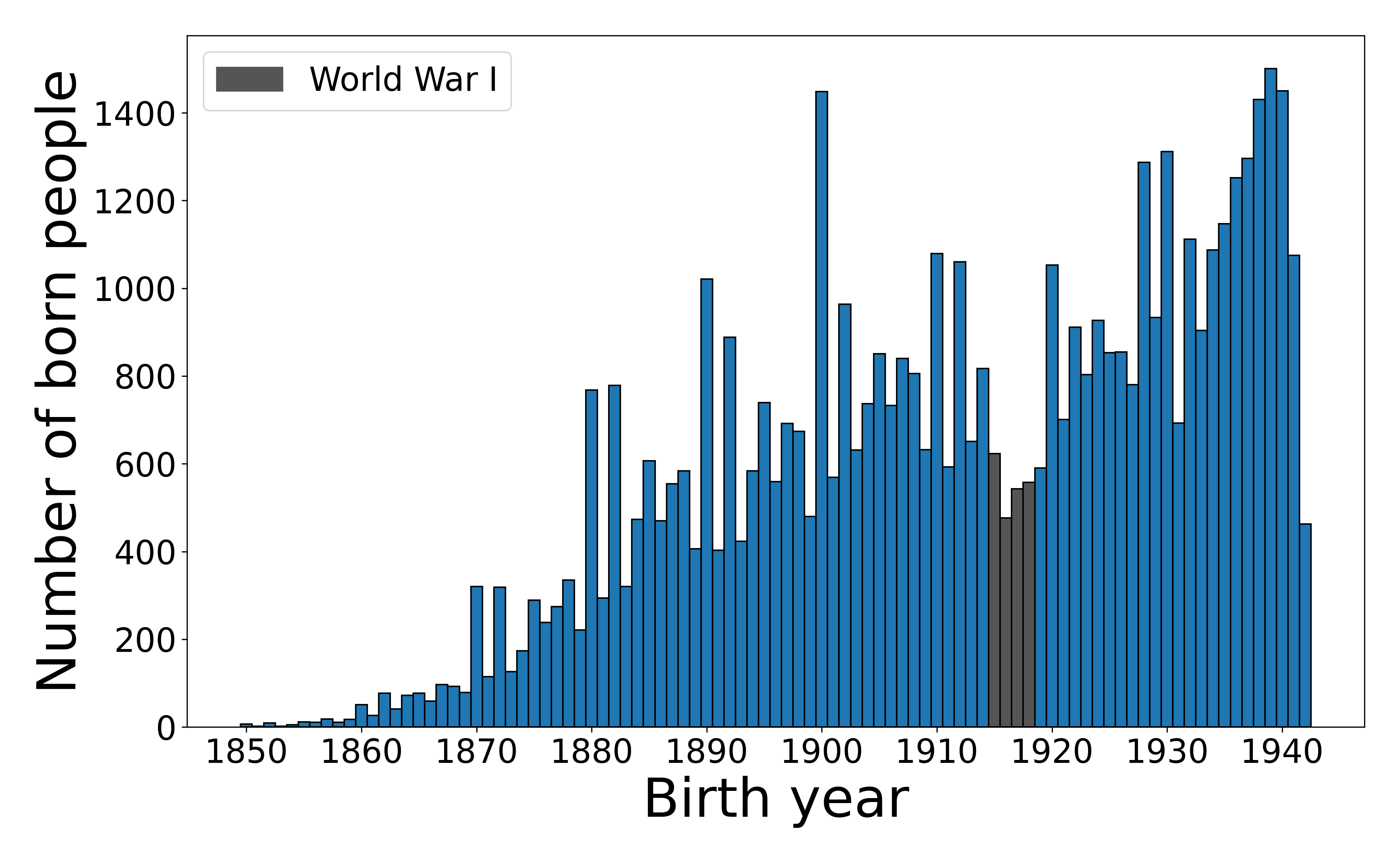}
	\label{fig:jms_1942}
	}%
	\subfloat[]{
	\includegraphics[width=0.48\linewidth]{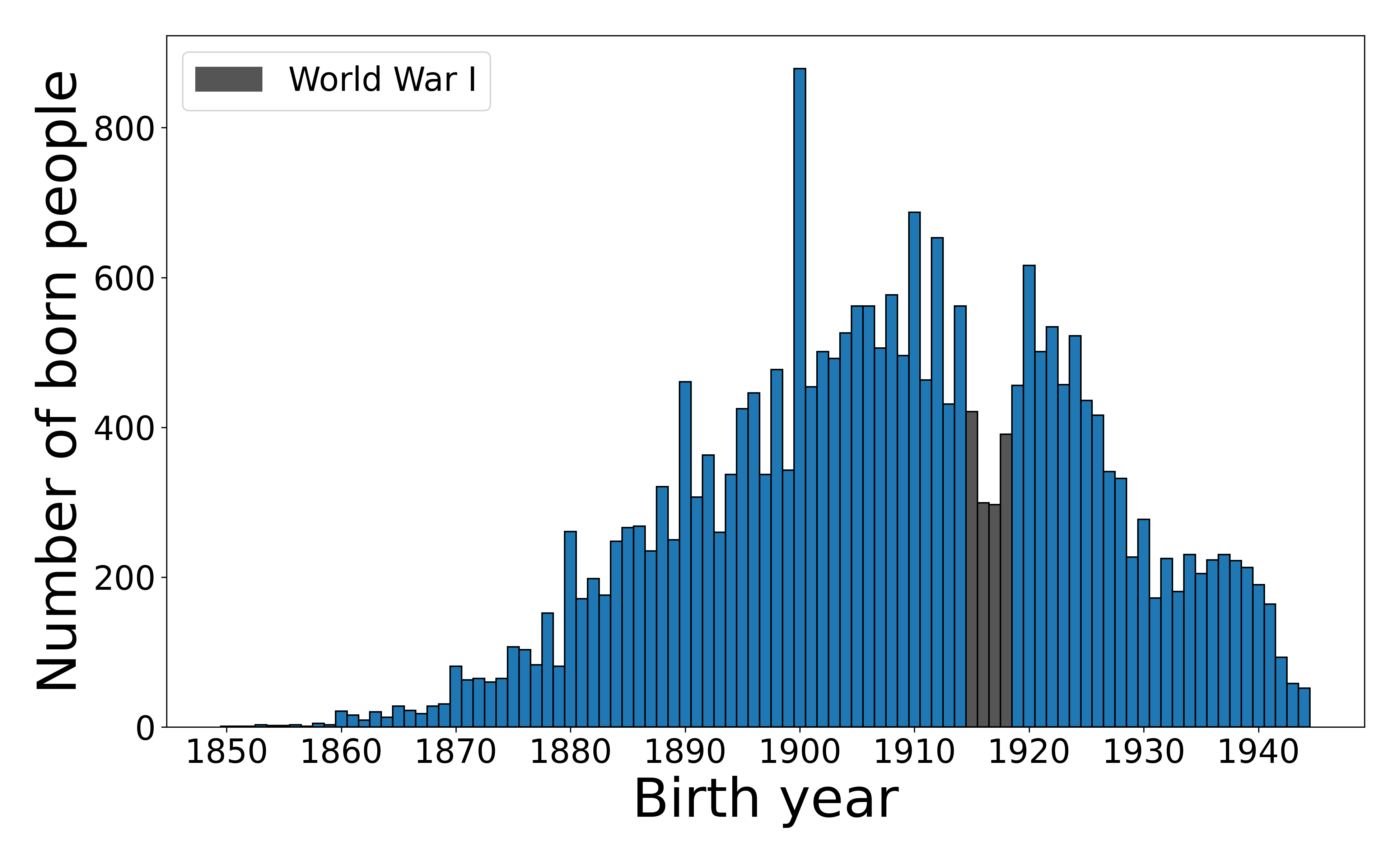}
	\label{fig:jms_no_1942}
	}

    \caption{Birth years of the inmates of the Jasenovac list whose death year is recorded as a)~$1942$ and b)~other than $1942$.}
	\label{fig:jms_comparison}
    
\end{figure*}

The question that is listed as the final potential objection in the critique is whether the critique implies that other statistical tests that ``may also automatically assign a lower score to a smaller sample size [...] are generally worthless''. The critique follows on by claiming that ``if a score produced by a statistical model depends on sample size, this raw score should then not be used across differently sized sample collections to make comparative claims about properties independent of sample size''. Without going in a deeper analysis of this claim, it is possible to dismiss it already by stressing that the properties used in our previous paper are dependent on the sample size. Namely, as already mentioned several times in this paper, larger samples are less affected by randomness in accordance with Cantelli-Glivenko theorem and this is a fact that has to be taken into account when assessing the obtained DTV. Since age heaping also affects DTV, this also applies to the problem of age heaping as well. Because of that, the critique's claims are not applicable here.

\section{The problematic nature of Jasenovac}
\label{sec:jasenovac}

On numerous occasions, the critique finds it problematic that the Jasenovac repeatedly turns out to be the top outlier by TVOR, but in this paper it has been shown that such findings are generally unfounded and unsupported by valid numerical evidence. Nevertheless, since the critique tries to invalidate TVOR's abilities by simultaneously attacking the fact that TVOR ranks the Jasenovac list as the top outlier candidate regardless of the conditions, it is worth to investigate the whole issue in more detail from a data science perspective.

\subsection{Severe age heaping}
\label{subsec:severe}

First of all, by already taking only a glimpse of the histogram of the Jasenovac list birth years, it becomes clear that it represents a case of highly expressed age heaping. Even if a histogram is not at hand, but only the birth years data is available, the value of the Whipple's index calculated for the Jasenovac list clearly falls into the range that describes the data quality as ``bad''~\cite{robine2007human}. It can be expected that having data of good quality is preferable in most disciplines, whereas having data of bad quality is probably less preferable. Now, if the quality of data is bad instead of good, this means that due to that data quality the situation is less preferable instead of being preferable, which can often be a problem. Since this problem occurs due to the data of bad quality, it can be argued that it is correct to label such data at least as potentially problematic and this is also the case with the Jasenovac list.

Since age heaping can by its very definition lead to increased values of DTV, it comes as no surprise that it also influences TVOR's score. While some other lists in the USHMM dataset also receive high TVOR scores, what makes the Jasenovac list particularly outstanding is the fact that its score significantly outweighs others. As already mentioned earlier here in the paper, if only the scores of largest list with, according to the disproved critique's claim, allegedly very differing DTVs are taken, even among them the Jasenovac list clearly fits the standard definition of an outlier due to its outstandingly high DTV caused by its highly expressed age heaping. Therefore, from a demographic point of view, the question is not whether or not the Jasenovac list is problematic, but rather what causes it to be so prominent among the outlier candidates. Already the critique gives possible hints.

Namely, while already the Whipple's index clearly designates the data quality of the Jasenovac lists' birth years as bad, due to its simple definition it takes into account only some of the anomalies on its histogram usually in connection to ages, and in this case also the birth years, ending in $0$ or $5$. However, what also plays a significant role in the case of the Jasenovac list are the birth years ending in $2$, which is not a problem for TVOR, but it is not directly taken into account by the Whipple's index and similar metrics. Therefore, the age heaping of the Jasenovac list relies on significant surpluses of the birth years ending in $0$, $2$, and $5$ with some other features possibly included as well and these are all included in TVOR's score calculation. Since the critique tries to offer an explanation for birth years ending in $2$ and explicitly mentions that in our previous paper we ``fail to identify simple plausible historical and data-collection explanations for them'', the rest of this section is dedicated to this quest.

\subsection{The age heaping of birth years ending in 2}
\label{subsec:1942}

As for the age heaping based on $1942$ as the recorded death year, we agree that the explanation given in the critique about the recorded ages instead of birth years is fully plausible and it only confirms the existence of another unusual age heaping pattern alongside the usual one. This in turn only further confirms the problematic nature of the Jasenovac list, which is thus properly denoted as a top ranking outlier candidate.

If birth year histograms are created separately for those inmates of the Jasenovac list whose death year is recorded as $1942$ and for those whose death is recorded as some other year different from $1942$, then the resulting histograms have visibly different levels of age heaping as shown in Fig.~\ref{fig:jms_comparison}. It can be seen that the histogram in Fig.~\ref{fig:jms_1942}, which can be denoted as the $1942$ histogram for simplicity, shows more prominent signs of age heaping that the histogram in Fig.~\ref{fig:jms_no_1942}, which can be denoted as the non-$1942$ histogram for simplicity. As a matter of fact, if the sample of the $1942$ histogram and the sample of non-$1942$ are separately added to the USHMM dataset and if TVOR is applied to this extended dataset, then the $1942$ histogram slightly outscores the original Jasenovac list histogram to become the most likely outlier candidate, while the non-$1942$ histogram ends up being ranked thirtieth. Since with its less than $54$k inmates it is nevertheless a more likely outlier candidate according to TVOR's scores, it can be concluded that the majority of the potentially problematic data of the Jasenovac list lies in the sample for the $1942$ histogram as it is also suggested by the critique. Interestingly, however, the $1942$ histogram sample is not only the main source of the surplus of birth years ending in $2$, but also the main source of all other birth year surpluses.

While this may simply explain features such a significant surplus of birth years ending in $2$ as a manifestation of rounding the ages, it also opens new questions with one of them being why is there such a feature discrepancy between the $1942$ histogram sample and the non-$1942$ histogram sample.

From a data science perspective, the least that can be said here is that having $1942$ as the recorded death year increases the likeliness of the birth year to be potentially problematic.

\subsection{Factual data veracity}

\subsubsection{Motivation}
\label{subsubsec:motivation}

In our previous paper~\cite{banic2021tvor}, after the Jasenovac list was identified as the top outlier candidate, its data was additionally analyzed and after some cross-checking with the other lists in the dataset as well as another publicly available source about Jasenovac, three examples of persons falsely described by the Jasenovac list were given as a simple example of how TVOR may be able to find outliers that contain data that, alongside age heaping, may also have other problems such as factual data veracity. This was also required to justify the potential requirement of compliance with the General Data Protection Regulation after satisfying one of the reviewers' requests. No further analysis was conducted because in terms of data veracity, the paper was primarily focused on the age heaping problem with only some hints at other potential problems.

However, as explicitly stated in the critique, one of its goals is ''to indicate the lack of importance of the claim in~\cite{banic2021tvor} about the three individuals on the Jasenovac list''. Since we did not find similar problems with other lists in the USHMM dataset, we think of such a brief and swift disregard of these potentially useful listed findings as somewhat unfounded and reminiscent of the critique's previous erroneous claims. Therefore, to properly address this critique's claim as well, it was decided to investigate the issue in somewhat more detail.

\subsubsection{Comparison with data from other sources}
\label{subsubsec:comparison}

To check whether the listed errors were only some isolated examples that may occur on regular basis on lists as large as the Jasenovac list or whether they are a potential sign of some other underlying problem, additional effort was put into investigating this issue. In order to remain in the scope of the techniques used in the TVOR paper, the investigation was limited only to simple personal data comparison. The compared data included the Jasenovac list data and the publicly available data of related historical institutions or related sites.

In the scope of the website of the Yad Vashem Vashem, the world holocaust remembrance center, there is The Central Database of Shoah Victims' Names~\cite{yv2021yvng} with numerous publicly available data, which also includes scanned and digitally processed forms, i.e., pages of testimony about the fate of various victims of the holocaust. By checking whether there are persons with the same name, surname, father's name, birth year, and birth place for which the Jasenovac list in the USHMM and pages of testimony in the Yad Vashem's Central Database claim different fates, it was concluded that there are at least $84$ such cases. Taking into account the highly expressed age heaping present in the Jasenovac list, which was already shown in our previous paper and is again illustrated here in Fig.~\ref{fig:alternative_birth_years}, and by allowing for the absolute differences in the birth years being $1$, additional $40$ matches can be found. If the absolute difference of $2$ is allowed, additional $29$ matches can be found. An absolute differences of $3$ gives $24$ matches and, as intuitively expected in accordance with Fig.~\ref{fig:year_differences}, the number of matches decreases as the allowed difference increases. The exact results with all these matches and the information for the relevant links appropriate for an independent verification are in Appendix.

The next interesting source that was checked was the list of Yugoslav victims of 1941-1945, which is publicly available in the PDF form on the website of the Museum of genocide victims in Belgrade, Serbia~\cite{mvg2021szsj}. This list has been chosen because it is mentioned as one of the sources for most of the inmates listed on the Jasenovac list in the USHMM dataset. By performing the same comparison as with the Yad Vashem's Central Database, it is possible to find over two thousand exact matches, which are given in Appendix. By allowing for small birth year differences, further matches can be found. However, they were not listed here because already the presented ones are sufficient to demonstrate the larger scale of the potential veracity issues of the Jasenovac list.

It turns out that there is a large number of other sources that also contain claims about fates that are contrary to the ones mentioned in the Jasenovac list for individual inmates. Listing all the matches in all of them is not necessary because for the purpose of proving the critique's claim unfounded, the numerous matches from two previous sources already suffices. Nevertheless, to demonstrate the disagreement of these other sources with the Jasenovac list, a single example of contrary claim is given from each of them in Appendix.

\subsubsection{Historians' interpretation}
\label{subsubsec:interpretation}

Since the critique mentions the lack of historical explanation for some of the findings from the area of data science that happened to deal with historical data, this time we asked several historians with expertise in the area in question for a comment. We strictly stress, however, that this is out of the original scope of this paper and that is has been added here only to answer one of the critique's many unfounded claims.

The first historian we asked for a comment was Dr. Esther Gitman, an American historian and expert on the holocaust in Yugoslavia who also happens to have survived it. When asked about the contradictory and somewhat confusing reports made by various sources for the same persons, in the resulting correspondence she repeated what was confirmed to be one of her earlier public claims that she has seen similar discrepancies while working on the archive material of the so called Projekt Dotr{\v{s}}{\'{c}}ina~\cite{grakalic1988projekt} where she found that numerous documents directly contradicted the Jasenovac list claims. Additionally, one of her responses included the following quote of another expert: ''In no country of Axis occupied Europe is the fate of Jewry more difficult to trace than in Jugoslavia. In no country are the final figures of losses and survivals more difficult to asses.''~\cite{reitlinger1987final}. Finally, there was also the suggestion to refer to her book~\cite{gitman2011courage} to find more detail.

Having a confirmation of findings of similar discrepancies by a top expert may be encouraging and being provided the explanation for why these discrepancies occur may be useful, but it may also not be enough for a proper data veracity interpretation. Because of that, we contacted Dr. Vladimir Geiger, an expert in the area of human war losses as well as a reviewer on topics related to the Jasenovac list~\cite{geiger2020issue,geiger2020ivo}. When asked about the conflicting reports about the fate of same persons, one among many explanations that he gave was that if there are several sources claiming different fates of a certain person, then the source claiming that a person lived longer should be trusted more if it can be confirmed that it is same person. Additionally. Dr. Geiger also referred us to one of his papers~\cite{geiger2017lazo} in which he also deals with a specific case of what he already knew before was the problematic nature of the Jasenovac list birth years and inconsistent fate claims.

For whatever reason, the Jasenovac list seems to contain numerous data of highly questionable veracity to say the least. Since looking for these reasons is out of the scope of this paper, it is sufficient to say that this problem of data veracity is maybe connected to the problem of age heaping.

\subsubsection{Other lists}
\label{subsubsec:other}

The critique notes ''that the authors of \cite{banic2021tvor} make no effort to identify or prune similar inconsistencies in other lists in the dataset, even when seemingly more egregious and more obvious''. We strongly disagree with this claim, which is totally unfounded. Namely, we were not able to find similar problems with other lists in the USHMM datasets and therefore there was nothing to report. On the other hand, the critique also did not offer an example of similar scale and this kind of faulty data veracity on any other of the USHMM lists.

\section{Conclusions}
\label{sec:conclusions}

After summarizing the TVOR method presented in our previous paper~\cite{banic2021tvor}, we have demonstrated that none of the claims made in the critique~\cite{ornik2021comment} against TVOR or its application to the USHMM dataset are correct. The critique begins by incorrectly overlooking the connection between sample probabilities, DTV, and their use by TVOR. It continues by proposing mathematically incorrect bias models, claiming an alleged data dearth problem, and insisting on alleged repercussions of USHMM data processing issues, but without giving numerical evidence. All these claims are easily disproved even by using standard statistical definitions. What is concerning is that the whole critique seems to be purposefully ignoring the Glivenko-Cantelly theorem.
In this paper we have addressed the claims in the critique~\cite{ornik2021comment} point by point and shown that they are neither proven nor justified.

\section*{Appendix}
\label{sec:appendix}

\subsection{The Yad Vashem's Central Database}
\label{subsec:yv}

For the sake of simplicity, the matches are given in the following format:\\ (USHMM\_ID, YVNG\_ID). Each such match means that in the USHMM dataset on the webpage\\ https://www.ushmm.org/online/hsv/person\_view.php?PersonId=USHMM\_ID\\ there is a claim about a certain Jasenovac inmate's fate that is directly contested by the webpage\\ https://yvng.yadvashem.org/nameDetails.html?language=en\&itemId=YVNG\_ID. An example of such a match is (\href{https://www.ushmm.org/online/hsv/person_view.php?PersonId=7517979}{7517979}, \href{https://yvng.yadvashem.org/nameDetails.html?language=en&itemId=5228098}{5228098}). Namely, the relevant USHMM dataset webpage claims that a certain Bencion Kaveson born to Jakob in Sarajevo in 1909 allegedly died in Jasenovac in 1942, while the relevant Yad Vashem webpage contains the page of testimony of his spouse Sara Kaveson claiming that he died in southern Italy on May 10, 1945. A similar example of the match for the same person is (\href{https://www.ushmm.org/online/hsv/person_view.php?PersonId=7517979}{7517979}, \href{https://yvng.yadvashem.org/nameDetails.html?language=en&itemId=10507552}{10507552}) with the page of testimony given by his niece claiming the death time and place being May 12, 1945 in an English hospital on Malta. In such cases, only one match is listed here. Due to the problematic nature of the Jasenovac list data specified even on the list itself, altered name versions as well as alternative values for specific fields given in the source field were also taken into account during the data matching.

\subsubsection{Full matches}
\label{subsubsec:yvng_full}

The full matches with the exception of the person's fate given in the previously described format are as follows: (7509901, 10108575), (7513200, 3630352), (7513326, 1091164), (7515941, 1267494), (7515976, 1262061), (7516078, 2033397), (7516149, 1267477), (7516160, 1267434), (7516164, 1267476), (7516190, 726250), (7516264, 1269292), (7516385, 1269503), (7516508, 1828659), (7516762, 374423), (7516804, 1398484), (7517221, 5650215), (7517294, 1740307), (7517298, 10106950), (7517367, 1035591), (7517581, 1247188), (7517778, 435246), (7517780, 801158), (7517939, 1191743), (7517979, 5228098), (7518413, 10108595), (7518497, 6880609), (7518712, 966512), (7518816, 1331685), (7518823, 1331689), (7518934, 815932), (7519083, 732782), (7519139, 5228103), (7519156, 704047), (7519275, 3586962), (7519374, 1543907), (7519545, 826050), (7519629, 895793), (7519643, 1969809), (7524314, 703459), (7525330, 909419), (7525366, 1840269), (7526480, 4428914), (7531919, 537073), (7534868, 792425), (7534886, 1342315), (7535146, 729111), (7540111, 736797), (7540527, 1662408), (7540534, 1718130), (7540580, 792611), (7540855, 1169997), (7540856, 2010620), (7540861, 2010615), (7551607, 1438976), (7553245, 529460), (7553246, 1653082), (7553350, 744681), (7553767, 3891851), (7553810, 5858833), (7554174, 1228000), (7559132, 626878), (7559307, 334699), (7559894, 988166), (7560714, 724490), (7561513, 1783349), (7562214, 6338619), (7562288, 1976527), (7562361, 3913858), (7563304, 4332260), (7563371, 870836), (7563408, 1486623), (7563518, 497255), (7563918, 684804), (7563965, 1636127), (7565194, 403926), (7565442, 1266987), (7567956, 1676420), (7568116, 2001745), (7568136, 440904), (7568148, 840880), (7568472, 1493653), (7569068, 1460546), (7574706, 6972966), and (7576225, 5228101). 

\subsubsection{Birth year difference of $1$}
\label{subsubsec:yvng_1}

If the birth year in the records for the supposedly same person in the USHMM dataset and the Yad Vashem's digital archive is allowed to be $1$, then the matches with the exception of the person's fate given in the previously described format are as follows:
(7496549, 1264619), (7511658, 1463315), (7514188, 1459104), (7514204, 401591), (7515939, 1256510), (7516044, 11417693), (7516095, 1267411), (7516438, 1269504), (7516552, 1268921), (7516750, 3629761), (7516772, 755668), (7517015, 629600), (7517271, 1036415), (7517297, 10106949), (7517351, 1017005), (7517429, 5642105), (7517682, 416352), (7517922, 1638175), (7518112, 1965922), (7518460, 829653), (7518509, 840758), (7519637, 1859121), (7519642, 722858), (7524372, 10239570), (7535051, 1773979), (7540493, 2008551), (7540882, 11853030), (7541588, 791858), (7553809, 1980525), (7553892, 1995948), (7559895, 980309), (7560609, 621134), (7560673, 683243), (7562309, 683065), (7568132, 394679), (7568419, 548263), (7568505, 5252150), (7568507, 5252142), (7573408, 3625019), and (7574083, 1256502).

\subsubsection{Birth year difference of $2$}
\label{subsubsec:yvng_2}

If the birth year in the records for the supposedly same person in the USHMM dataset and the Yad Vashem's digital archive is allowed to be $2$, then the matches with the exception of the person's fate given in the previously described format are as follows: (7493074, 5866278), (7513313, 5252142), (7516089, 1377115), (7516274, 912783), (7516303, 10507539), (7516440, 3962054), (7516464, 3962054), (7516745, 1266982), (7517366, 1035591), (7517430, 5642105), (7517902, 416754), (7518130, 442255), (7518183, 551837), (7518202, 3904601), (7518467, 3617705), (7521842, 919204), (7535068, 1504614), (7535085, 1404901), (7560612, 830579), (7560688, 869486), (7560778, 1909905), (7560800, 3934530), (7561355, 1882572), (7568278, 887884), (7574125, 1269326), (7574143, 1965830), (7574279, 824295), (7574693, 6880610), and (7574978, 1269479).

\subsubsection{Birth year difference of $3$}
\label{subsubsec:yvng_3}

If the birth year in the records for the supposedly same person in the USHMM dataset and the Yad Vashem's digital archive is allowed to be $3$, then the matches with the exception of the person's fate given in the previously described format are as follows: (7510960, 772754), (7511636, 1957095), (7511977, 1093837), (7515948, 3572735), (7516025, 1859708), (7516163, 1267446), (7516355, 1269602), (7516457, 1269504), (7516588, 1268910), (7517929, 1608589), (7518306, 3912521), (7518612, 958397), (7519276, 3586962), (7519289, 609332), (7519589, 2026617), (7519843, 1631816), (7526549, 1231064), (7535092, 1778896), (7540514, 1730946), (7540994, 4011323), (7553559, 869467), (7562320, 748363), (7563055, 5328596), and (7568111, 755583).

\subsection{The list of Yugoslav victims}
\label{subsec:l64}

The publicly available list of Yugoslav victims does not have an individual webpage for every listed person. Instead, it offers several PDF files with the data of all victims. Because of that, the matches cannot be given in the simple form as in the case of the digital archive of the Yad Vashem. Nevertheless, they can still be given in a slightly extended version. For the sake of simplicity, the matches are given in the following format: name surname (father's name, birth year, birth place). They are additionally grouped by the place of death mentioned by the Yugoslav list, which directly counters the claims made by the Jasenovac list in the USHMM dataset. Like in the case of Yad Vashem, altered name versions as well as alternative values for specific fields given in the source field were also taken into account during the data matching.

The matches are organized by groups marked in underlined and bold text that represent the death place claimed by the Yugoslav list and followed by the persons for which this claim is made and for which the Jasenovac list simultaneously claims they died in Jasenovac. Each person is described by in the following form: name surname (father's name, birth place, birth year, USHMM\_ID) where the USHMM\_ID can be used fetch the webpage\\ https://www.ushmm.org/online/hsv/person\_view.php?PersonId=USHMM\_ID\\ with the original record in the Jasenovac list. For the sake of an easier verification, the groups are given here in their original form used in the Yugoslav list. This means that e.g. the term Au{\v{s}}vic is used instead of Auschwitz.

\subsubsection{Full matches}
\label{subsubsec:l64_full}

\textbf{\underline{2.KRAJI{\v{S}}KA BRIGADA}}: Veljko Radeti{\'{c}} (Stojan, Jablan, 1919, 7575940). 

\textbf{\underline{ADA{\v{S}}EVCI}}: {\v{Z}}ivan Pogrmi{\'{c}} (Lazar, Ada{\v{s}}evci, 1882, 7495492). 

\textbf{\underline{AU{\v{S}}VIC}}: Irena Adam (-, Vukovar, 1892, 7560658), Ivan Adam (-, Vukovar, 1922, 7560659), Elza Angelus (-, Vukovar, 1887, 7560667), Julius Angelus (-, Vukovar, 1872, 7560668), Marko Angelus (-, Vukovar, 1882, 7560669), Ljudevit Berger (-, Osijek, 1900, 7560897), Malvina Fischer (-, Vukovar, 1862, 7560698), Ru{\v{z}}a Fischer (-, Vukovar, 1872, 7560697), Leopold Gross (-, Vukovar, 1882, 7560711), Hinko Hermann (-, Osijek, 1869, 7561676), Liza Herzog (-, Vukovar, 1892, 7560715), Olga Irvas (-, Vukovar, 1892, 7560720), Jakob Jakobovi{\'{c}} (-, Vukovar, 1892, 7560723), Dragutin Jurkovi{\'{c}} (-, Osijek, 1888, 7561656), Irma Jurkovi{\'{c}} (-, Osijek, 1895, 7561663), Ivana Kaiser (-, Vukovar, 1867, 7560734), {\v{Z}}iga Katz (-, Osijek, 1916, 7561584), Berta Kohn (-, Vukovar, 1877, 7560726), Ljudevit Kohn (-, Vukovar, 1882, 7560725), Samuel Kohn (-, Vukovar, 1872, 7560730), Irena Kraus (-, Osijek, 1870, 7535044), Roza Kraus (-, Osijek, 1894, 7561643), Vlado Kraus (Hugo, Osijek, 1924, 7561648), Irma Reinitz (-, Osijek, 1921, 7561505), Mi{\v{s}}o Roth (-, Vukovar, 1872, 7541679), Mirjana Schwarzenberg (Albert, Zagreb, 1922, 7560580), Leo Schön (-, Vukovar, 1897, 7560770), Olga Schön (-, Vukovar, 1907, 7560771), Marija Singer (-, Vukovar, 1912, 7560768), Samuel Singer (-, Vukovar, 1872, 7560769), Marija Spitzer (-, Vukovar, 1872, 7560808), Jakob Stein (-, Osijek, 1877, 7561501), Josip Stein (-, Vukovar, 1872, 7560803), Mari{\v{s}}ka Stein (-, Vukovar, 1892, 7560802), Matilda Veli{\'{c}} (-, Vukovar, 1872, 7560779), Arnold Weiner (-, Vukovar, 1882, 7560782), Riza Weiner (-, Vukovar, 1887, 7560785), Ela Winter (-, Vukovar, 1872, 7560777), Hermina Winter (-, Vukovar, 1892, 7560776), Anka Abinun (-, Zagreb, -, 7553140), Petar Matija{\v{s}}{\v{c}}i{\'{c}} (-, Otru{\v{s}}evec, -, 7568370), Vid Matija{\v{s}}{\v{c}}i{\'{c}} (-, Otru{\v{s}}evec, -, 7568371), Izidor Winter (-, Bijeljina, 1912, 7563304), Isidor Fuchs (-, Vukovar, 1877, 7560691), Blanka-Branka Hochstädter (Josip, Zapre{\v{s}}i{\'{c}}, 1919, 7563844), Ljerka Hochstädter (Josip, Zapre{\v{s}}i{\'{c}}, 1922, 7563846), Roza Katz (-, Osijek, 1896, 7561582), Nada Weinberger (Mavro, Sunja, 1924, 7571579), {\DJ}urica-{\DJ}ura Weiss (Maksim, Ruma, 1935, 7494309), Robert Deutsch-Maceljski (Vilim, Zagreb, 1884, 7553287), Anka Abinun (-, Zagreb, -, 7553141), Helga Adler ({\v{S}}andor, Slavonski Brod, 1923, 7559254), Matilda Albahari (-, Vukovar, 1891, 7560664), Salamon Albahari (-, Vukovar, 1861, 7560665), Lea Danon (Bencion, Sarajevo, 1885, 7517015), Viktor Diamant (-, Vukovar, 1896, 7560689), Nada Feigenbaum (-, Vukovar, 1877, 7560707), Johana Mandl (-, Vukovar, 1906, 7560748), Emil Mozes (-, Vukovar, 1881, 7560751), Sarika Mozes (-, Vukovar, 1891, 7560752), Marta Rosenfeld (-, Vukovar, 1893, 7560764), Norbert Weinberger (-, Vukovar, 1881, 7560787), Josip Zeiger (-, Vukovar, 1881, 7560685), Erna Atijas (Jako, Biha{\'{c}}, -, 7568086), Frida Bauer (Leopold, Zagreb, 1889, 7553196), Ferdo Gleisinger (Lavoslav, Zagreb, 1933, 7568229), Johana Bienenstock (-, Osijek, 1870, 7560901), Otto Kastl (-, Osijek, 1892, 7561589), Julio Adut (-, Vukovar, 1867, 7560661), Marijana Bichler (Dragutin, Osijek, 1912, 7560898), Hinko Schönwald (-, Krapina, 1881, 7552294), Hinko-Vinko Fischbein (Salamon, Zagreb, 1893, 7569146), Ilona-Jelena Fischbein (Izidor, Zagreb, 1906, 7569147), Vera Fischbein (Radivoje, Zagreb, 1927, 7569145), Nela Fischer (-, Vukovar, 1892, 7560702), Ivka-Hana Gleisinger (Salamon, Daruvar, 1902, 7563927), Jakob Kajon (Elias, Sarajevo, 1884, 7517737), Nn Kollin (-, Osijek, -, 7561617), Gizela Pollak (-, Vukovar, 1897, 7560754), Mira Pollak (Ljudevit, Vukovar, 1922, 7560756), Zlata Schrenger (-, Pakrac, 1903, 7526529), Moric Vogel (Herman, Banja Luka, 1910, 7496876), {\v{Z}}aneta Bresslauer (-, Vukovar, 1867, 7560681), Zlatica Epstein (-, Daruvar, 1910, 7565190), Rozalija Fischhoff (-, Vukovar, 1872, 7560696), Ivan Guttmann (-, Vukovar, 1886, 7541595), Bela Haberfeld (-, Vukovar, 1920, 7560712), Ignac Kellert (-, Vukovar, 1882, 7560738), Gizela Neumann (-, Osijek, 1894, 7561554), Filip Rosendorn (-, Vukovar, 1902, 7560761), Hinko Rosendorn (-, Vukovar, 1892, 7560762), Kika Rosendorn (-, Vukovar, 1914, 7560763), Hedviga Schlossberger (-, Vukovar, 1902, 7560774), Jakob Schnitzler (-, Vukovar, 1862, 7560809), Nada Schnitzler (-, Vukovar, 1872, 7560810), Josip Stern (Marko, Ore{\v{s}}ac, 1917, 7527187). 

\textbf{\underline{B.GRADI{\v{S}}KA}}: Savo Trivali{\'{c}} (Stojan, Gornja Ilova, 1913, 7575174). 

\textbf{\underline{BABINAC}}: Jovo Kerkez (Stanko, Babinac, 1880, 7555516). 

\textbf{\underline{BAJINCI}}: Ignjatije Popovi{\'{c}} (Mihajlo, Bajinci, 1888, 7575126). 

\textbf{\underline{BANJA KOVILJA{\v{C}}A}}: Stevan Gaji{\'{c}} (Jovo, Ku{\v{c}}i{\'{c}} Kula, 1909, 7576074). 

\textbf{\underline{BANJA LUKA}}: Pero Babi{\'{c}} (Ostoja, Bistrica, 1903, 7566281), Janko Jankovi{\'{c}} (Danilo, Dera, 1903, 7566239), Savan Jankovi{\'{c}} (Simo, Dera, 1912, 7566240), Bogdan Lazi{\'{c}} (Du{\v{s}}an, Malo Bla{\v{s}}ko, 1923, 7565226), Jura Majstorovi{\'{c}} (Marko, Banja Luka, 1921, 7575282), Jovo Marjanovi{\'{c}} (Ostoja, Laminci Dubrave, 1902, 7569415), Stanoje Stijakovi{\'{c}} (Todor, Motike, 1905, 7575281), Niko Bijak (Vaso, Rebrovac, 1900, 7575273), Ostoja Dubravac (Nedo, Kokori, 1917, 7575068), {\DJ}ura{\dj} Mileti{\'{c}} (Jovan, Rekavice, 1902, 7575308), Mirko Radi{\'{c}} (Risto, Ostru{\v{z}}nja Donja, 1909, 7575185), Vaskrsije Radi{\'{c}} (Risto, Ostru{\v{z}}nja Donja, 1908, 7575184), Vaso Todorovi{\'{c}} (Stevan, Ljuba{\v{c}}evo, 1910, 7575305), Marko {\'{C}}osi{\'{c}} (Pejo, Rebrovac, 1907, 7575274), Mirko Bo{\v{z}}i{\'{c}} (Jevto, Drago{\v{c}}aj, 1915, 7496899), Rafael Poljokan (Leon, Banja Luka, 1881, 7496835), Nikola Miji{\'{c}} (Stevan, Ahmetovci, 1898, 7570779), Petar Bursa{\'{c}} ({\DJ}ura{\dj}, Peto{\v{s}}evci, 1912, 7565341), Mira Popovi{\'{c}} (-, Banja Luka, -, 7568061), Pavle {\v{C}}eko ({\DJ}uran, Ljuba{\v{c}}evo, 1917, 7575302), {\DJ}uran {\v{C}}eko (Stevan, Ljuba{\v{c}}evo, 1875, 7575303), Jovo {\v{C}}oli{\'{c}} (Lazar, Ponir, 1889, 7496930), Dmitar Vranje{\v{s}} (Pejo, Ljuba{\v{c}}evo, 1899, 7575304). 

\textbf{\underline{BANJEVI{\'{C}}I}}: Luka Simi{\'{c}} (Drago, Lije{\v{s}}anj, 1938, 7568911), Marica Simi{\'{c}} (Ljubo, Lije{\v{s}}anj, 1932, 7568902), Milan Simi{\'{c}} (Drago, Lije{\v{s}}anj, 1936, 7568910). 

\textbf{\underline{BANOVA JARUGA}}: Milan Anti{\v{c}}ek (Ivan, Zagreb, 1911, 7568254). 

\textbf{\underline{BATO{\v{C}}INA}}: Rade Panteli{\'{c}} (-, Hrti{\'{c}}, 1923, 7557717). 

\textbf{\underline{BA{\'{C}}IN}}: Nikola Bosio{\v{c}}i{\'{c}} (Stojan, Veliko Dvori{\v{s}}te, 1938, 7500337), Pava Bosio{\v{c}}i{\'{c}} (Stojan, Veliko Dvori{\v{s}}te, 1940, 7500338). 

\textbf{\underline{BA{\v{C}}EVI{\'{C}}I}}: Vidoje Sudar (Pero, Ba{\v{c}}evi{\'{c}}i, 1915, 7562995), Stevan {\v{S}}koro (Jovo, Ba{\v{c}}evi{\'{c}}i, 1890, 7562984). 

\textbf{\underline{BEKTE{\v{Z}}}}: Milan Radovanovi{\'{c}} (-, Jurkovac, -, 7562454). 

\textbf{\underline{BEOGRAD}}: Jovanka Luki{\'{c}} (-, Milo{\v{s}}evo Brdo, -, 7566269), Aron Finci (Jahiel, Sarajevo, 1884, 7517294), Ljubomir Majki{\'{c}} (-, Murati, 1927, 7499582), Jozef Abinun (Jakov, Sarajevo, 1914, 7515964). 

\textbf{\underline{BEOGRAD BANJICA}}: Avram Altarac (Mo{\v{s}}o, Sarajevo, 1906, 7516537), Boro Markovi{\'{c}} (Jovo, Tr{\v{s}}i{\'{c}}, -, 7569022), Relica Filker (Jo{\v{s}}ko, Mostar, 1927, 7555147), Tilda Finci (-, Split, 1915, 7565125), Jakov Gaon (Santo, Sarajevo, 1910, 7517445). 

\textbf{\underline{BE{\v{S}}ENOVO}}: Marko Petrovi{\'{c}} (Panto, Be{\v{s}}enovo, 1912, 7494343). 

\textbf{\underline{BIHA{\'{C}}}}: Ale Rami{\'{c}} (Suljo, Biha{\'{c}}, 1907, 7574907). 

\textbf{\underline{BIJELJINA}}: Luka Mijatovi{\'{c}} (Sekula, Zagoni, 1886, 7576030), Marko Blagojevi{\'{c}} (Blagoje, Zvornik, 1923, 7575894), Milan Blagojevi{\'{c}} (Jovan, Zvornik, 1920, 7575895), Mitar Blagojevi{\'{c}} (Jovan, Zvornik, 1925, 7575896), Kojo Peji{\'{c}} (Jovo, Zvornik, 1918, 7575893), Bo{\v{s}}ko Stojanovi{\'{c}} (Pero, Zvornik, 1913, 7575892). 

\textbf{\underline{BILE{\'{C}}A}}: Mi{\v{s}}o Popovi{\'{c}} (-, Sarajevo, -, 7569348). 

\textbf{\underline{BILOGORA}}: Vladimir Kos (Bla{\v{z}}, Zagreb, 1923, 7568433). 

\textbf{\underline{BILOGORSKI ODRED}}: Gli{\v{s}}o Trbojevi{\'{c}} (-, Velika Peratovica, -, 7576551). 

\textbf{\underline{BISTRICA}}: Milja Laji{\'{c}} (Dragoje, Bistrica, 1905, 7569256), Branko Vu{\v{c}}kovac ({\DJ}uro, Bistrica, 1890, 7566310), Ljubinko Laji{\'{c}} (Stanoje, Bistrica, 1937, 7569244). 

\textbf{\underline{BJELANOVAC}}: Ankica Romani{\'{c}} (-, Bjelanovac, 1935, 7574320), Ljuba Romani{\'{c}} (-, Bjelanovac, -, 7574322), Mileva Romani{\'{c}} (-, Bjelanovac, 1932, 7574323), Ilija Vlajkovi{\'{c}} (-, Bjelanovac, 1942, 7574338), Ljubica Vlajkovi{\'{c}} (-, Bjelanovac, 1937, 7574334), Du{\v{s}}an Vuji{\'{c}} (-, Bjelanovac, 1937, 7574342), Milka {\v{S}}eatovi{\'{c}} (-, Bjelanovac, 1891, 7564588), {\DJ}uja Vuji{\'{c}} (-, Bjelanovac, 1897, 7574348), Mara {\v{S}}eatovi{\'{c}} (-, Bjelanovac, 1910, 7564591), {\DJ}uka {\v{S}}eatovi{\'{c}} (-, Bjelanovac, 1937, 7564593), {\DJ}eko Vuji{\'{c}} (-, Bjelanovac, 1875, 7574347), Dragi{\'{c}} Vlajkovi{\'{c}} (-, Bjelanovac, 1938, 7574335). 

\textbf{\underline{BJELJENA}}: Petar Mi{\'{c}}anovi{\'{c}} (Luka, Bogutovo Selo, 1922, 7515483). 

\textbf{\underline{BJELJINA}}: Kostadin Bogdanovi{\'{c}} (Nikola, Priboj, 1920, 7560981). 

\textbf{\underline{BJELOVAC}}: Jovan Ma{\v{s}}i{\'{c}} (-, Drinja{\v{c}}a, 1926, 7568718). 

\textbf{\underline{BJELOVAR}}: Ljubica Jezdi{\'{c}} (Stevan, Mala Peratovica, 1895, 7576455), Ljuban Ratkovi{\'{c}} (Jandrija, Mala Peratovica, 1907, 7576447), {\DJ}uro Ba{\v{s}}i{\'{c}} (-, {\v{Z}}ivaja, 1905, 7572891), Cvijo Rabat (Ostoja, Pobr{\dj}ani, 1940, 7507938), Kata {\v{S}}tekovi{\'{c}} (Petar, Cremu{\v{s}}ina, 1919, 7576347), Mikailo Blagojevi{\'{c}} (Stevo, Bjelajci, 1929, 7497249). 

\textbf{\underline{BLINJSI KUT}}: Simo Juzba{\v{s}}i{\'{c}} (-, Jo{\v{s}}avica, 1900, 7567107). 

\textbf{\underline{BOLJANI{\'{C}}}}: Petar Ani{\'{c}} (Luka, Boljani{\'{c}}, 1924, 7575813), Lazo Pani{\'{c}} (Stanko, Boljani{\'{c}}, 1905, 7575796). 

\textbf{\underline{BOROJEVI{\'{C}} BANIJ}}: Adam Kljaji{\'{c}} (Dmitar, Strmen, 1905, 7555303). 

\textbf{\underline{BOROJEVI{\'{C}}I}}: {\DJ}uro Borojevi{\'{c}} ({\'{C}}iro, Borojevi{\'{c}}i, 1898, 7576660). 

\textbf{\underline{BOROVIK}}: {\DJ}uro Pinter (-, Borovik, -, 7565461), Bosiljka Stani{\'{c}} (-, Borovik, -, 7565464), Zorka Stani{\'{c}} (-, Borovik, -, 7565465). 

\textbf{\underline{BOS GRADI{\v{S}}KA}}: Branko Novkovi{\'{c}} (Rajko, Lakta{\v{s}}i, 1914, 7575947). 

\textbf{\underline{BOS-DUBICA}}: Mirko Samard{\v{z}}ija (Stojan, Pucari, 1909, 7558310). 

\textbf{\underline{BOS. {\v{S}}AMAC}}: Moric Altarac (Elijas, Travnik, 1920, 7575514). 

\textbf{\underline{BOSANSKA DUBICA}}: Voja Breki{\'{c}} (Jovo, Drakseni{\'{c}}, 1885, 7556630), Danica Bundalo (Simo, Novoselci, 1930, 7499655), Draginja Nikoli{\'{c}} (Stanko, Pobr{\dj}ani, 1901, 7558258), Milja Premasunac (Trivun, Komlenac, 1924, 7573211), {\DJ}uro Radi{\v{s}}i{\'{c}} (Nikola, {\v{Z}}ivaja, 1927, 7545902), Stana Vukas (-, Klekovci, 1903, 7557067), Mile Zori{\'{c}} (Pero, Gornji Jelovac, 1884, 7569372), Mladen Vuki{\'{c}} (Mile, Donja Gradina, 1927, 7498462), Slavko Lazi{\'{c}} (Jovan, Jasenje, 1920, 7498622), Sava Petkovi{\'{c}} (-, Klekovci, 1907, 7557165), Ljubica Baki{\'{c}} (Simo, Vrioci, 1924, 7573201), Fahrudin Berberovi{\'{c}} ({\v{S}}aban, Biha{\'{c}}, 1913, 7562660), Dobrila Stojni{\'{c}} (Stevan, Dobro Selo, 1913, 7558987), Ana Vlajni{\'{c}} (-, Drakseni{\'{c}}, 1926, 7556648). 

\textbf{\underline{BOSANSKA GRADI{\v{S}}KA}}: Milan E{\'{c}}imovi{\'{c}} (Vaso, Vrbovljani, 1921, 7554613), Pava Jandri{\'{c}} (Bo{\v{s}}ko, Strigova, 1926, 7500159), Mileva Kele{\v{c}}evi{\'{c}} (Nikola, Seferovci, 1923, 7573597), Jefto Preli{\'{c}} (Mile, Milo{\v{s}}evo Brdo, 1920, 7572246), Draga Stanojevi{\'{c}} (Stevo, Donji Bogi{\'{c}}evci, 1918, 7533528), Ile Vu{\v{c}}i{\'{c}} (Kosta, {\v{C}}ipulji{\'{c}}, 1904, 7574966), Sava Jandri{\'{c}} (Bo{\v{s}}ko, Strigova, 1938, 7500161), An{\dj}elko Vu{\v{c}}i{\'{c}} (Antonije, {\v{C}}ipulji{\'{c}}, 1892, 7574965), {\v{Z}}ivko {\v{S}}u{\v{s}}njara (Aleksa, Kukulje, 1908, 7575127), Mirko Milinovi{\'{c}} (Pane, Prijedor, 1901, 7509603), Niko Bori{\'{c}} (Kosta, {\v{C}}ipulji{\'{c}}, 1903, 7574970), Ana Pavlovi{\'{c}} (-, Donji Bogi{\'{c}}evci, 1912, 7562909), Vojislav Cviji{\'{c}} (Sretko, Gornji Jelovac, 1928, 7509148), Dragica Radi{\'{c}} (Nikola, Ba{\v{c}}ka Topola, 1926, 7496536), Rudolf {\v{Z}}agrovec (Josip, Sarajevo, 1920, 7575222). 

\textbf{\underline{BOSANSKA KOSTAJ}}: Jovo Korosi{\'{c}} (Ostoja, Prevr{\v{s}}ac, 1919, 7576700), Pero Korosi{\'{c}} (Branko, Prevr{\v{s}}ac, 1922, 7576701). 

\textbf{\underline{BOSANSKA KOSTANJICA}}: Dragan Muharem (Jovo, Mrakodol, 1939, 7560432), Ostoja Muharem (Jovo, Mrakodol, 1934, 7507876), Zorka Muharem (Jovo, Mrakodol, 1941, 7560435), Simo Baljak (Mile, Prevr{\v{s}}ac, 1880, 7576702), Pero {\DJ}uri{\'{c}} (Nikola, Gornje Vodi{\v{c}}evo, 1925, 7565763). 

\textbf{\underline{BOSANSKI JANKOVAC}}: Ilija Zec (Mile, Donja Dolina, 1902, 7573023). 

\textbf{\underline{BOSANSKI KOBA{\v{S}}}}: Stanko Gu{\v{z}}vi{\'{c}} (Rade, Vlaknica, 1880, 7575130), Jovan Kne{\v{z}}evi{\'{c}} (Stevo, Vlaknica, 1904, 7575129), Slavko Kne{\v{z}}evi{\'{c}} (Simeun, Vlaknica, 1916, 7575145), Milan Miki{\'{c}} (Savo, Vlaknica, 1903, 7575133), Du{\v{s}}an Drini{\'{c}} (Stevo, Vlaknica, 1900, 7575132), Savo Ljubojevi{\'{c}} (Miko, Vlaknica, 1875, 7575134), Bo{\v{z}}o Smiljani{\'{c}} (Andrija, Brusnik, 1904, 7575139), Petar {\v{C}}oli{\'{c}} (Stevan, Vlaknica, 1889, 7575135), Ilija Grumi{\'{c}} (Stevo, Brusnik, 1903, 7575141), Vladimir Grumi{\'{c}} (Stevo, Brusnik, 1899, 7575140). 

\textbf{\underline{BOSANSKI NOVI}}: Svetozar Gvozden (Lazo, Maslovare, 1908, 7576622), Simo Balti{\'{c}} (Tomo, Maslovare, 1900, 7507842), Pero Stankovi{\'{c}} (Stojan, Sokoli{\v{s}}te, 1912, 7508031), Janko Vujasin (Marko, Sokoli{\v{s}}te, 1904, 7508037), Petar Gradi{\v{c}}ak (Mijo, Mokrice, 1922, 7559543), Ilija Balti{\'{c}} (Pero, Maslovare, 1901, 7507841), {\DJ}or{\dj}o Stojanovi{\'{c}} (Milo{\v{s}}, Kr{\v{s}}lje, 1920, 7570873). 

\textbf{\underline{BOSANSKI PETROVAC}}: {\DJ}or{\dj}e Pe{\'{c}}anac ({\DJ}uran, Mileti{\'{c}}evo, 1923, 7572875). 

\textbf{\underline{BOSNA}}: Ljubica Dejanovi{\'{c}} (Marko, {\v{S}}iroka Rijeka, 1908, 7560354), Anka Dragi{\'{c}} ({\DJ}uro, Mracelj, 1938, 7560258), Mile Dragi{\'{c}} ({\DJ}uro, Mracelj, 1936, 7560262), Marta Dudukovi{\'{c}} (Milo{\v{s}}, Mracelj, 1887, 7560265), Milka Dudukovi{\'{c}} (Janko, Mracelj, 1939, 7560267), Nevenka Dudukovi{\'{c}} (Janko, Mracelj, 1935, 7560268), Sofija Dudukovi{\'{c}} (Janko, Mracelj, 1935, 7560269), Bosiljka Paji{\'{c}} (Du{\v{s}}an, {\v{S}}iroka Rijeka, 1935, 7560371), Ilija Paji{\'{c}} ({\DJ}uro, {\v{S}}iroka Rijeka, 1935, 7560373), Milivoj {\DJ}akovi{\'{c}} (Stevan, Mracelj, 1924, 7560274), Kata Ma{\'{c}}e{\v{s}}i{\'{c}} (Dragi{\'{c}}, Mracelj, 1904, 7560278), Ana Mrk{\v{s}}i{\'{c}} (Nikola, {\v{S}}iroka Rijeka, 1937, 7560365), Mirko Eckstein (-, Zagreb, 1930, 7564652), Milica Kotaranin (Mili{\'{c}}, Mracelj, 1899, 7560277). 

\textbf{\underline{BOVK KOD KIRINA}}: Milan Cico (Teodor, Ostro{\v{z}}in, 1900, 7558664). 

\textbf{\underline{BO{\'{C}}IN}}: Ljuba Gaji{\'{c}} (Stevan, Komlenac, 1914, 7573215), Stana Gaji{\'{c}} (-, Komlenac, -, 7573209). 

\textbf{\underline{BRDO}}: Mirko Risti{\'{c}} (Risto, Tr{\v{s}}i{\'{c}}, -, 7569016). 

\textbf{\underline{BRESTA{\v{C}}}}: Bo{\v{z}}a Jovanovi{\'{c}} (-, Bresta{\v{c}}, -, 7568517). 

\textbf{\underline{BREZIK}}: Cvijetin Goganovi{\'{c}} (Simo, Si{\v{z}}je, 1923, 7575169). 

\textbf{\underline{BREZOVE DANE}}: Simo Spasojevi{\'{c}} (Mitar, Brezove Dane, 1890, 7575594). 

\textbf{\underline{BRGULE}}: Petra Petrovi{\'{c}} (Du{\v{s}}an, Brgule, 1934, 7575372). 

\textbf{\underline{BRO{\v{C}}ICE}}: Dane Vlajsevi{\'{c}} (Nikola, Bo{\v{z}}i{\'{c}}i, 1895, 7497537). 

\textbf{\underline{BR{\v{C}}KO}}: Sava Jovi{\v{c}}i{\'{c}} (Krsto, Buzekara, 1897, 7575868), Saveta Stojkovi{\'{c}} (Uro{\v{s}}, Bosut, 1929, 7494538). 

\textbf{\underline{BUCHENWALD}}: Marija Bala{\'{c}} (Nikola, Mlaka, 1929, 7547088), Sofija Bjelivuk (Ilija, Jagrovac, 1927, 7530174), Grozda Ma{\v{c}}ki{\'{c}} (Stevo, Jasenovac, 1922, 7546722), Milivoj Radulovi{\'{c}} (Nikola, U{\v{s}}tica, 1939, 7555454), Stoja Romani{\'{c}} (-, Bjelanovac, 1914, 7564559), {\v{Z}}ivka Vu{\v{c}}enovi{\'{c}} (-, {\v{C}}enkovo, -, 7565492), Mara Bobi{\'{c}} (Iso, Li{\v{c}}ko Petrovo Selo, 1900, 7557529). 

\textbf{\underline{BUGONJO}}: Savo Livopoljac (Tomo, Ko{\v{s}}{\'{c}}ani, 1923, 7566128), Jovo Livopoljac (Luka, Ko{\v{s}}{\'{c}}ani, 1900, 7508184), Svatko Kisin (Isak, Novo Selo, 1890, 7574964), Drago Livopoljac (Ile, Ko{\v{s}}{\'{c}}ani, 1898, 7566127). 

\textbf{\underline{BUJAVICA}}: Mirko Ili{\'{c}} (Teodor, Bujavica, 1885, 7565374). 

\textbf{\underline{BUKVIK}}: Vaso Pjani{\'{c}} (Nikola, Cerovljani, 1928, 7501422). 

\textbf{\underline{BULO{\v{Z}}ANI}}: Branko Radi{\'{c}} (Simo, Bulo{\v{z}}ani, 1926, 7575877). 

\textbf{\underline{BUNA}}: Du{\v{s}}an Golo (Risto, Ba{\v{c}}evi{\'{c}}i, 1907, 7563102), Sava Golo (Milisav, Ba{\v{c}}evi{\'{c}}i, 1920, 7563101). 

\textbf{\underline{BU{\v{C}}I{\'{C}}I}}: Stojan Simeti{\'{c}} (Ostoja, Bri{\v{s}}i{\'{c}}i, 1907, 7575000). 

\textbf{\underline{CEROVLJANI}}: Mile {\'{C}}uri{\'{c}} (Nikola, Novoselci, 1874, 7558183), Milka Gligi{\'{c}} (Lazo, Pucari, 1886, 7558284). 

\textbf{\underline{CREMU{\v{S}}INA}}: {\DJ}uro Vidi{\'{c}} (-, Cremu{\v{s}}ina, -, 7576356). 

\textbf{\underline{CREMU{\v{S}}NICA}}: {\DJ}uro Radanovi{\'{c}} (Simo, {\v{C}}remu{\v{s}}nica, 1923, 7558523). 

\textbf{\underline{CRKVENI BOK}}: Veno Toma{\v{s}}ek (Franjo, Grbavac, 1927, 7576415), Milica Be{\v{c}} (-, Crkveni Bok, 1941, 7549405), Stojan {\DJ}uri{\v{c}}i{\'{c}} (-, Crkveni Bok, 1942, 7549431), Janja Turajli{\'{c}} (-, Crkveni Bok, -, 7549524). 

\textbf{\underline{CRNA GORA}}: Meho Hamuli{\'{c}} (Ibro, {\v{C}}arakovo, 1918, 7566360), Milan Svilar (Budo, Pe{\'{c}}ane, 1924, 7559610). 

\textbf{\underline{DACHAU}}: {\DJ}uja Kordi{\'{c}} (Jakov, Donji Cerovljani, 1914, 7576751), David Perera (Mojsije, Sarajevo, 1876, 7519454), Bogdan Boksi{\'{c}} (-, Bestrma, -, 7570178), Smiljka Komlenac (-, Dereza, 1915, 7564484), Todor Stjepanovi{\'{c}} (Simo, Strije{\v{z}}evica, 1890, 7575595). 

\textbf{\underline{DALMACIJA}}: Erdonja Levi (Sadik, Zenica, 1922, 7513240). 

\textbf{\underline{DAP{\v{C}}EVICA}}: Mile Tesla (-, Mala Peratovica, -, 7576445). 

\textbf{\underline{DARUVAR}}: Rosa Jankovi{\'{c}} (Rade, Bijakovac, 1941, 7555558), Slobodanka Milisavljevi{\'{c}} (Jovo, Pucari, 1937, 7558294), Nikola Slijep{\v{c}}evi{\'{c}} (-, Gakovo, -, 7576384). 

\textbf{\underline{DEMIROVAC}}: Milo{\v{s}} Danilovi{\'{c}} (-, Demirovac, 1908, 7556125). 

\textbf{\underline{DEREZA}}: Marko D{\v{z}}akula (Stanko, Dereza, 1872, 7556826), Pajo D{\v{z}}akula (Pavle, Dereza, 1938, 7556819), Saveta D{\v{z}}akula (Dmitar, Dereza, 1920, 7556822), Bo{\v{s}}ko Ga{\'{c}}e{\v{s}}a (Mile, Dereza, 1939, 7564163), Jula Ga{\'{c}}e{\v{s}}a (Matija, Dereza, 1899, 7556761), Mara Ga{\'{c}}e{\v{s}}a (Mile, Dereza, 1941, 7564166), Stojan Ga{\'{c}}e{\v{s}}a (Pavle, Dereza, 1939, 7564173), Luka Jovanovi{\'{c}} ({\DJ}uro, Dereza, 1937, 7564464), Senija Jovanovi{\'{c}} (Jakov, Dereza, 1888, 7556777), Stana Jovanovi{\'{c}} (Jovo, Dereza, 1902, 7525682), Ljubica Markovi{\'{c}} (Marko, Dereza, 1904, 7556798), Pela Markovi{\'{c}} (Marko, Dereza, 1915, 7556799), Stana Vukmanovi{\'{c}} (Stanko, Dereza, 1923, 7556738), Stana An{\dj}eli{\'{c}} (-, Dereza, 1898, 7556734), Avram Ga{\'{c}}e{\v{s}}a (Stanko, Dereza, 1880, 7556757), Bosiljka Maleti{\'{c}} (Pavle, Dereza, 1941, 7564123), Mile {\DJ}ur{\dj}evi{\'{c}} (Nikola, Dereza, 1895, 7556771), Stojan D{\v{z}}akula (Petar, Dereza, 1941, 7525675), Teja Jovanovi{\'{c}} (Pane, Dereza, 1878, 7525684), Jovanka Jovanovi{\'{c}} (-, Dereza, 1888, 7556778), Smilja Maleti{\'{c}} (Pavle, Dereza, 1937, 7564511), Stojan D{\v{z}}akula (Lazo, Dereza, 1907, 7525672). 

\textbf{\underline{DIVI{\'{C}}}}: Janja Ili{\'{c}} (-, Novo Selo, -, 7568824). 

\textbf{\underline{DJAKOVO}}: Mira Finci (-, Zemun, 1907, 7559119). 

\textbf{\underline{DOBOJ}}: Ankica Dasovi{\'{c}} (Nikola, {\v{S}}vica, 1925, 7528314). 

\textbf{\underline{DOBO{\v{S}}NICA}}: Jovo Vukovi{\'{c}} (Ilija, Pura{\v{c}}i{\'{c}}, 1895, 7575173). 

\textbf{\underline{DOBRLJIN}}: Vid Daki{\'{c}} ({\DJ}or{\dj}ija, Donja Jurkovica, 1927, 7564638). 

\textbf{\underline{DONJA GRADI{\v{S}}KA}}: Milka Bijeli{\'{c}} (Stole, Rakovica, 1880, 7558323). 

\textbf{\underline{DONJA SLABINJA}}: Simo Popovi{\'{c}} (-, Donja Slabinja, -, 7567864). 

\textbf{\underline{DONJE VODI{\v{C}}EVO}}: Mileva {\v{Z}}uni{\'{c}} (Jandro, Donje Vodi{\v{c}}evo, 1912, 7507663). 

\textbf{\underline{DONJI JELOVAC}}: Dragutin {\'{C}}urguz (Milan, Donji Jelovac, 1938, 7556549). 

\textbf{\underline{DONJI KLAKAR}}: Sava Gro{\v{s}}evi{\'{c}} (Vaskrsije, Klakar Donji, 1880, 7565984), Pejo Me{\dj}edovi{\'{c}} (Jovan, Klakar Donji, 1925, 7566089). 

\textbf{\underline{DONJI RAKANI}}: Dragomir Vajagi{\'{c}} (Vlado, Donji Rakani, 1927, 7508014). 

\textbf{\underline{DORA}}: Stana Vukomanovi{\'{c}} (-, {\v{C}}enkovo, -, 7559728). 

\textbf{\underline{DRINJA{\v{C}}A}}: Miladin Mirkovi{\'{c}} (Mi{\'{c}}an, Tr{\v{s}}i{\'{c}}, -, 7569019). 

\textbf{\underline{DRLJA{\v{C}}A}}: Vaso Vu{\v{c}}enovi{\'{c}} (-, {\v{C}}etvrtkovac, 1938, 7549612). 

\textbf{\underline{DRNEK}}: Imbro Ivi{\v{c}}ar (Petar, Drnek, 1894, 7568456). 

\textbf{\underline{DRVAR}}: {\DJ}uro Dragi{\'{c}} (Mile, Mracelj, 1910, 7560260), Milan Golubovi{\'{c}} ({\DJ}or{\dj}e, Kosjerovo, 1902, 7565311). 

\textbf{\underline{DUBICA}}: Milica Stan{\v{c}}evi{\'{c}} (Nikola, Donji Cerovljani, 1886, 7576754), Milja Baki{\'{c}} (-, Vrioci, -, 7573202), Jagoda Dikli{\'{c}} (-, Donji Hrastovac, 1906, 7573156), Ljeposava Dikli{\'{c}} (-, Donji Hrastovac, 1926, 7564867), {\v{Z}}arko Dikli{\'{c}} (-, Donji Hrastovac, 1928, 7573153), {\DJ}uro Papi{\'{c}} (-, Papi{\'{c}}i, 1933, 7573199), Ivan Abaza (Pavao, Ba{\'{c}}in, 1924, 7576762), Todor Gaji{\'{c}} (Petar, Velika {\v{Z}}uljevica, 1885, 7508080), Matija Zaklanac ({\v{Z}}ivko, Donji Cerovljani, 1904, 7576745), Mi{\'{c}}o Svilokos (-, Papi{\'{c}}i, -, 7573197), Leposava Bijeli{\'{c}} (Vlado, Rakovica, 1930, 7499890), Milja Zaklanac ({\v{Z}}ivko, Donji Cerovljani, 1904, 7576746), Marija-Maca Ko{\v{s}}uti{\'{c}} (Milo{\v{s}}, Slovinci, 1934, 7573194), Nada Srbljanin (-, Donji Hrastovac, 1941, 7573133), Petar Srbljanin (-, Donji Hrastovac, 1937, 7573124), Ljuban Vrlini{\'{c}} (-, Donji Hrastovac, 1896, 7573143), Ankica Srbljanin (-, Donji Hrastovac, 1932, 7573137), Branka Srbljanin (-, Donji Hrastovac, 1944, 7573126), Du{\v{s}}anka Srbljanin (-, Donji Hrastovac, 1934, 7573131), Jula Srbljanin (-, Donji Hrastovac, 1906, 7573138), Milja Srbljanin (-, Donji Hrastovac, 1937, 7573117), Vladimir Srbljanin (-, Donji Hrastovac, 1889, 7573108). 

\textbf{\underline{DUBICAHRV.}}: Ljuba Milkovi{\'{c}} (-, {\v{Z}}ivaja, -, 7572931). 

\textbf{\underline{DUBNICA}}: Ilija Peri{\'{c}} (Ilija, Dubnica, 1885, 7576065), Lazar Ivanovi{\'{c}} (-, Dubnica, -, 7569225), Vladan Mali{\v{s}}evi{\'{c}} (-, Dubnica, 1918, 7561130). 

\textbf{\underline{DUBOKA}}: Aleksa Brane{\v{z}}ac (-, Duboka, -, 7570182). 

\textbf{\underline{DUBRAVA NOSKOVA{\v{C}}KA}}: Radojka Bezbradica (Stevo, Rakovica, 1942, 7499886), Milan Jankovi{\'{c}} (Mirko, Demirovac, 1942, 7497747), Dragica Grgi{\'{c}} (-, Rakovica, 1941, 7558328), Mile Miljevi{\'{c}} (-, Rakovica, 1940, 7558333). 

\textbf{\underline{DVOR}}: Milo{\v{s}} Dikli{\'{c}} (-, Donji Hrastovac, 1904, 7573157), Rade Drageljevi{\'{c}} (Petar, Matijevi{\'{c}}i, 1888, 7543657). 

\textbf{\underline{ERDEVIK}}: Milka Mila{\v{s}}inovi{\'{c}} (-, Ratkovac, -, 7554508). 

\textbf{\underline{FO{\v{C}}A}}: Devla Aganovi{\'{c}} (Zulfo, Fo{\v{c}}a, 1910, 7562275). 

\textbf{\underline{G. STUB.}}: Stjepan Jak{\v{s}}i{\'{c}} (Josip, Jak{\v{s}}inec, 1915, 7559545). 

\textbf{\underline{GACKO MOSTAR}}: Marija Mileti{\'{c}} (Mi{\'{c}}o, Gacko, 1898, 7555136). 

\textbf{\underline{GAKOVO}}: Stojan Novakovi{\'{c}} (Mitar, Gakovo, 1922, 7576392). 

\textbf{\underline{GARA{\v{S}}NICA}}: Manojlo Milnovi{\'{c}} (Simo, Verija, 1885, 7500459), Pane Milnovi{\'{c}} (Manojlo, Verija, 1918, 7500457). 

\textbf{\underline{GARE{\v{S}}NICA}}: Stevo Vurin (Nikola, Verija, 1933, 7558792). 

\textbf{\underline{GA{\v{S}}NICA}}: Leposava Danilovi{\'{c}} (Mi{\'{c}}o, Ga{\v{s}}nica, 1932, 7502251), Anka Vidovi{\'{c}} (Du{\v{s}}an, Ga{\v{s}}nica, 1918, 7574055). 

\textbf{\underline{GLAMO{\v{C}}}}: Mile Orli{\'{c}} (Rade, Had{\v{z}}in Potok, 1900, 7576621). 

\textbf{\underline{GLINEN}}: Emil Domany (-, Tuzla, -, 7574424). 

\textbf{\underline{GLINSKO NOVO SE}}: Du{\v{s}}an Dabi{\'{c}} (Milo{\v{s}}, Komogovina, 1898, 7545262). 

\textbf{\underline{GLOGOVICA}}: Aleksa Grabovac (Andrija, Glogovica, 1860, 7575181). 

\textbf{\underline{GORAD{\v{Z}}E}}: Panto Vukovi{\'{c}} (Ne{\dj}o, Nekopi, 1910, 7512329). 

\textbf{\underline{GORNJA JUTROGO{\v{S}}TA}}: Anka Petri{\'{c}} (Du{\v{s}}an, Gornji Jelovac, 1941, 7569388). 

\textbf{\underline{GORNJA RIJEKA}}: Radosava Zrni{\'{c}} (Milan, Turjak, 1935, 7573050), Slavko Zrni{\'{c}} (Du{\v{s}}an, Turjak, 1940, 7573051). 

\textbf{\underline{GORNJA SLABINJA}}: Stanko Palija ({\DJ}uran, Gornja Slabinja, 1910, 7571036). 

\textbf{\underline{GORNJE VODI{\v{C}}EVO}}: Marko Kolund{\v{z}}ija (Stojan, Gornje Vodi{\v{c}}evo, 1890, 7565769), Stojan {\DJ}uri{\'{c}} (Mi{\'{c}}o, Gornje Vodi{\v{c}}evo, 1925, 7565766), Ljuba {\v{S}}urlan (Ilija, Gornje Vodi{\v{c}}evo, 1926, 7507717). 

\textbf{\underline{GORNJI JAVORANJ}}: Miladin Cvetojevi{\'{c}} (-, Javnica, 1900, 7557724). 

\textbf{\underline{GORNJI JELOVAC}}: Mirko Dobrijevi{\'{c}} (Ilija, Gornji Jelovac, 1898, 7569362), Marko Luki{\'{c}} (Simo, Gornji Jelovac, 1915, 7569366), Stevan Macura (Simo, Gornji Jelovac, 1887, 7569373), Milan Maleti{\'{c}} (Nikola, Gornji Jelovac, 1928, 7509171), Kosa Marin (Jovan, Gornji Jelovac, 1914, 7569386), Du{\v{s}}an Vuki{\'{c}} (Vaso, Gornji Jelovac, 1907, 7569360), Milan Vuki{\'{c}} (Vaso, Gornji Jelovac, 1910, 7569361), Stanka Vuleti{\'{c}} (Rade, Gornji Jelovac, 1940, 7569359), Mile Zori{\'{c}} (Janko, Gornji Jelovac, 1892, 7569364), Mara {\v{S}}ormaz (Petar, Gornji Jelovac, 1898, 7569402), Smilja {\v{S}}ormaz (Rade, Gornji Jelovac, 1941, 7569404), Rade Mudrini{\'{c}} (Du{\v{s}}an, Gornji Jelovac, 1938, 7566190), Rade Vuleti{\'{c}} (Petko, Gornji Jelovac, 1902, 7566386), Dragoja Cviji{\'{c}} (Antonije, Gornji Jelovac, 1890, 7566381), Ljubko Cviji{\'{c}} (Jovan, Gornji Jelovac, 1888, 7566382). 

\textbf{\underline{GORNJI LOKANJ}}: Ljubo Savi{\'{c}} (-, Gornji Lokanj, -, 7569002). 

\textbf{\underline{GORNJI PODGRADCI}}: Mirko Milanovi{\'{c}} ({\v{Z}}ivko, Bakinci, 1940, 7565273), Milutin Vuji{\'{c}} (Simeun, Gornji Podgradci, 1933, 7503490). 

\textbf{\underline{GORNJI SJENI{\v{C}}AK}}: Marko Manojlovi{\'{c}} (Lazo, Gornji Sjeni{\v{c}}ak, 1886, 7528659), Mili{\'{c}} Manojlovi{\'{c}} ({\DJ}ura{\dj}, Gornji Sjeni{\v{c}}ak, 1893, 7528658). 

\textbf{\underline{GOSPI{\'{C}}}}: Ljubo {\v{S}}e{\v{s}}um (Ilija, Rasti{\v{c}}evo, 1902, 7566136), Branko Ga{\v{c}}i{\'{c}} (Spaso, Jo{\v{s}}ava, 1905, 7507756), Simo Ga{\v{c}}i{\'{c}} (Petar, Jo{\v{s}}ava, 1888, 7507755), {\DJ}or{\dj}e Kapetanovi{\'{c}} ({\DJ}ura{\dj}, Jo{\v{s}}ava, 1893, 7507759), Veljko Medan (Luka, Mostar, 1907, 7555135), Petar Vujasin (Luka, Radomirovac, 1887, 7507985), Petar Vasiljevi{\'{c}} (-, Bosanska Krupa, 1914, 7568509), Damjan {\DJ}eri{\'{c}} (Spasoje, Nevesinje, 1901, 7555108), Proko Balaban (Jovo, Jo{\v{s}}ava, 1882, 7507750), {\DJ}or{\dj}e Berak (Ilija, Ra{\v{s}}ka Gora, 1905, 7555100), Maksim Ivica (Todo, Donji Vakuf, 1883, 7508171), Mihajlo Kisi{\'{c}} ({\DJ}or{\dj}e, Mostar, 1896, 7563093), Stevo Balaban (Proko, Jo{\v{s}}ava, 1918, 7507751). 

\textbf{\underline{GOSTOVI{\'{C}}}}: Stojan Kenji{\'{c}} (Simo, Podvolujak, 1910, 7575768). 

\textbf{\underline{GRABOVAC}}: Branko Poropati{\'{c}} (Janko, Grabovac Banski, 1926, 7566972), Janko Poropati{\'{c}} (Ilija, Grabovac Banski, 1890, 7566971). 

\textbf{\underline{GRAC}}: Mirko Pilipovi{\'{c}} (Mirko, Donje Vodi{\v{c}}evo, 1904, 7507641). 

\textbf{\underline{GRADAC}}: Mara {\DJ}uri{\v{c}}i{\'{c}} (-, Crkveni Bok, 1943, 7549428). 

\textbf{\underline{GRE{\DJ}ANI OKU{\v{C}}ANSKI}}: Armin Sandberg (Martin, Slavonski Brod, 1930, 7537344). 

\textbf{\underline{GRME{\v{C}}}}: Uro{\v{s}} Maksimovi{\'{c}} (Nikola, Krivaja, 1900, 7566276). 

\textbf{\underline{GROMILE-{\v{S}}IPOVO}}: Nikola Kaurin (Niko, {\v{C}}ifluk, 1870, 7575022). 

\textbf{\underline{GRUBI{\v{S}}NO POLJE}}: Filip Domitrovi{\'{c}} (Joco, Mala Barna, 1897, 7576437), Milan Dra{\v{z}}i{\'{c}} (-, Grubi{\v{s}}no Polje, -, 7576577). 

\textbf{\underline{HAN PESAK}}: {\v{Z}}ivko Santra{\v{c}} (Stevo, Risovac, 1920, 7574879). 

\textbf{\underline{HA{\DJ}ER}}: Du{\v{s}}an {\v{C}}upi{\'{c}} (-, Majske Poljane, 1929, 7544239). 

\textbf{\underline{HORVATI}}: Nikola Mavrovi{\'{c}} (Josip, Hudi Bitek, 1926, 7568362), Nikola Mavrovi{\'{c}} (Josip, Oto{\v{c}}ec Zaprudski, 1924, 7552470). 

\textbf{\underline{HRASNICA}}: Salko Had{\v{z}}i{\'{c}} (Mehmed, Hrasnica, 1926, 7562346). 

\textbf{\underline{HRVATSKA}}: Stjepan Pavli{\v{s}}ko (Imbro, Zagreb, 1899, 7571691). 

\textbf{\underline{HRVATSKA DUBICA}}: An{\dj}a Premasunac (-, Komlenac, -, 7573208), Du{\v{s}}an Vrebac (-, Komlenac, -, 7567856), Soka Sopi{\'{c}} (Gligo, Donji Cerovljani, 1894, 7576760), Bo{\v{z}}o {\v{S}}kondri{\'{c}} (Stevo, Bo{\v{z}}i{\'{c}}i, 1925, 7555781). 

\textbf{\underline{HRVATSKA KOSTANJICA}}: Mile Kukrika (Milan, Mrakodol, 1920, 7507872). 

\textbf{\underline{IRIG}}: Mileva Stojni{\'{c}} (Ilija, Nova Gradi{\v{s}}ka, 1922, 7533911). 

\textbf{\underline{IVANJSKI BOK}}: Pavle Ba{\v{s}}i{\'{c}} (-, Ivanjski Bok, 1865, 7555336). 

\textbf{\underline{IVANKOVO}}: Marko Ro{\v{z}}dijevac (Ilija, Ivankovo, 1917, 7537905). 

\textbf{\underline{IVANOVO SELO}}: Vladimir Vaji{\'{c}} (Vaso, Stara Krivaja, 1927, 7561792). 

\textbf{\underline{IZ ZATVORA}}: Salih Jamakosmanovi{\'{c}} (Mustafa, Sarajevo, -, 7562680). 

\textbf{\underline{JABLANICA}}: Mirko Savi{\'{c}} ({\v{S}}{\'{c}}epo, Dobri{\v{c}}, 1909, 7574847), Dragoljub Popovi{\'{c}} (Ljuban, Jablanica, 1934, 7572234), Marija Popovi{\'{c}} (Ljuban, Jablanica, 1937, 7572236), Rodoljub Popovi{\'{c}} (Ljuban, Jablanica, 1932, 7572233). 

\textbf{\underline{JADOVNO}}: Miljenko Aleksander ({\DJ}uro, Zagreb, 1921, 7553150), Milan D{\v{z}}abi{\'{c}} (Luka, Vinska, 1878, 7565902), Ilija Jovanovi{\'{c}} (Joco, Trije{\v{s}}nica, 1898, 7576028), Ivo Kell (Rudolf, Zagreb, 1924, 7553583), Sigismund Klugmann (Salamon, Zagreb, 1887, 7560503), Du{\v{s}}an Miloj{\v{c}}i{\'{c}} (Pantelija, Slavonski Koba{\v{s}}, 1897, 7537385), Jakob Montiljo (Juda, Travnik, 1906, 7512825), Dimitrije Panteli{\'{c}} (Luka, Brodac Gornji, 1911, 7576035), Milivoj Suboti{\'{c}} (Nikola, Su{\v{s}}ine, 1908, 7557518), Stevo Umljenovi{\'{c}} (Svetozar, Slavonski Koba{\v{s}}, 1902, 7562778), Ljubomir {\v{C}}avi{\'{c}} (Simo, Travnik, 1890, 7575520), Luka Borota (-, Gornja Ra{\v{s}}enica, -, 7576417), Nikola Bosanac (Pavle, Katinac, -, 7569651), Simo Brkovi{\'{c}} (-, Gornja Ra{\v{s}}enica, -, 7576418), Nikola Jonica (-, Grada{\v{c}}ac, 1897, 7561280), {\DJ}uro Kova{\v{c}}evi{\'{c}} (-, Velika Barna, -, 7576588), Milenko Markovi{\'{c}} (-, Pakrac, 1878, 7574368), Lazo Mudri{\'{c}} (-, Pakrac, -, 7574367), Zaharije Petrovi{\'{c}} (-, Slavonski Koba{\v{s}}, 1892, 7562766), Ostoja Trivanovi{\'{c}} (-, Slavonski Koba{\v{s}}, -, 7565722), Vladimir Weiss (Maks, Pribi{\'{c}}, 1891, 7552163), Leon Weisser (Me{\v{s}}ulam, Travnik, 1910, 7512895), Milan D{\v{z}}abi{\'{c}} (Luka, Vinska, 1877, 7510691), Rade Ivkovi{\'{c}} (Luka, Od{\v{z}}ak, 1896, 7555121), Savo Jankovi{\'{c}} (Simo, Klju{\v{c}}, 1901, 7574953), Gojko Stanoj{\v{c}}i{\'{c}} (Jovan, Donji Vakuf, 1900, 7508177), {\v{S}}piro Vujinovi{\'{c}} ({\DJ}uro, Ra{\v{s}}tani, 1898, 7576008), Jefto Pani{\'{c}} (Joco, Slavonski Koba{\v{s}}, 1897, 7562765), Maksim Baki{\'{c}} (Pavle, Su{\v{s}}ine, 1904, 7557510), Jozef Danon (Jehuda-Haim, Sarajevo, 1906, 7517040), Stojan Janji{\'{c}} ({\v{S}}{\'{c}}epan, Goranci, 1913, 7511940), Branko Kerpner (Jakob, Zagreb, 1921, 7568685), Damjan Milinovi{\'{c}} (Ignjatija, Su{\v{s}}ine, 1913, 7557514), Spasoje {\v{S}}ipovac (Risto, Rast, 1900, 7512090), Petar Zlokas (Stanko, Gre{\dj}ani, 1900, 7558620), Nedeljko Kova{\v{c}}evi{\'{c}} (Nikola, Mostar, 1902, 7563092), Savo Kova{\v{c}}evi{\'{c}} (Mile, Gornja Ra{\v{s}}enica, 1912, 7523407), Velemir Kova{\v{c}}evi{\'{c}} (Petar, Mostar, 1919, 7512003), Manojlo Adamovi{\'{c}} (Savo, Donji Volar, 1900, 7509024), Mihailo Laki{\'{c}} (-, Donja Trnova, -, 7561080), Mihajlo Stanimirovi{\'{c}} (Stako, Brodac Donji, 1891, 7576036), Milivoj Vasiljevi{\'{c}} (Pero, Mostar, 1890, 7562977), Stevo {\'{C}}ur{\v{c}}i{\'{c}} (Mile, Podum, 1896, 7529631), Branko {\v{C}}upi{\'{c}} (Gavro, Travnik, 1921, 7575522), Gligorije {\v{S}}krpan (Danilo, Slavonski Koba{\v{s}}, 1902, 7562769), Jakob Hirschl (Mavro, Kri{\v{z}}evci, 1868, 7525322), Oskar Friedmann (Jakob, Tuzla, 1917, 7569155), {\DJ}uro Klein-Klaji{\'{c}} (Ljudevit, Grubi{\v{s}}no Polje, 1911, 7563795), Josip Neumann (Jakob, {\DJ}akovo, 1900, 7554723), Leopold Erbsenhaut (Ferbus, Turbe, 1923, 7512902), Haim Rosenblatt (Samuel, Sarajevo, 1921, 7519670). 

\textbf{\underline{JAJCE}}: Gojko Bu{\v{z}}anin (Ile, Gornji Mujd{\v{z}}i{\'{c}}i, 1914, 7575083), Simo Lon{\v{c}}ar ({\v{S}}piro, Gorica, 1920, 7575910), Lazar Vojinovi{\'{c}} (Mile, Gorica, 1900, 7575923), Vojislav Vojinovi{\'{c}} (Sava, Gorica, 1922, 7575909), Cvijan Bu{\v{z}}anin (Ile, Gornji Mujd{\v{z}}i{\'{c}}i, 1906, 7575082), Simo Dobrati{\'{c}} (Stojan, Volari, 1885, 7575054), {\DJ}uro Dobrati{\'{c}} (Simo, Volari, 1906, 7575053), Ilija Grahovac (Cvijan, Volari, 1909, 7575081), Jovan Jandri{\'{c}} (Vaso, Mo{\v{c}}ioci, 1894, 7575913), Jovo Lon{\v{c}}ar (Jovo, Gorica, 1911, 7575911), Milan Lon{\v{c}}ar (Lazo, Gorica, 1919, 7575912), {\v{Z}}ivko Radoja (Pero, {\'{C}}usine, 1908, 7574995), Mitar Vojinovi{\'{c}} (Simo, Gorica, 1893, 7575924), Petar Vojinovi{\'{c}} (Mile, Gorica, 1910, 7575922), Sava Vojinovi{\'{c}} (Simo, Gorica, 1878, 7575908), Ljupko Jandri{\'{c}} (Manojlo, Gorica, 1920, 7575920), Milo{\v{s}} Lovri{\'{c}} (Spasoja, Klekovci, -, 7567873), Krstan {\v{S}}olaja (Blagoje, Volari, 1900, 7575057). 

\textbf{\underline{JASENICE}}: Mehmed Grahovi{\'{c}} (Red{\v{z}}o, Miljkovi{\'{c}}i, 1900, 7574889), Bo{\v{z}}o Tomi{\'{c}} (Savo, Gornja Paklenica, 1901, 7575586). 

\textbf{\underline{JASENJE}}: {\v{Z}}ivko Alavuk (Mile, Jasenje, 1939, 7556974), Stevo Lazi{\'{c}} (Marko, Jasenje, 1892, 7557005). 

\textbf{\underline{JASTREBARSKO}}: {\DJ}uro Brdar (Janko, Lu{\v{s}}{\v{c}}ani, 1938, 7548638), Stana Brki{\'{c}} (Rade, Lu{\v{s}}{\v{c}}ani, 1930, 7548657), Danica Cviji{\'{c}} (Stanko, Lu{\v{s}}{\v{c}}ani, 1933, 7548662), Pantelija Daki{\'{c}} (Kosta, Donja Jurkovica, 1935, 7501625), Rade Gaji{\'{c}} (Stanko, Bukovac, 1939, 7501391), Evica Janjanin (Luka, Gornja Ba{\v{c}}uga, 1936, 7548417), Nada Janjanin (Stevo, Gornja Ba{\v{c}}uga, 1937, 7548435), Zorka Janjanin (Nikola, Gornja Ba{\v{c}}uga, 1938, 7548426), Du{\v{s}}an Jeki{\'{c}} (Mirko, Grabovac Banski, 1931, 7548529), Ljuban Kovarba{\v{s}}i{\'{c}} (Vujo, Petrinja, 1935, 7548997), Ljubica Kozi{\'{c}} (Pavle, Lu{\v{s}}{\v{c}}ani, 1933, 7548712), {\v{Z}}ivka Ku{\v{c}}uk (Petar, Kijevci, 1933, 7504437), {\v{Z}}ivko Luji{\'{c}} (Petar, Dobrogo{\v{s}}{\'{c}}e, 1936, 7536388), Ljubica Marki{\v{s}} (Bo{\v{z}}o, Lu{\v{s}}{\v{c}}ani, 1930, 7548746), Marija Mirjani{\'{c}} (Trivun, Turjak, 1941, 7506769), Vladimir Poluga (Ljuboje, Pakrani, 1934, 7522434), Branislava Spasojevi{\'{c}} (Du{\v{s}}an, Gornja Jurkovica, 1937, 7502831), Du{\v{s}}anka Spasojevi{\'{c}} (Luka, Gornja Jurkovica, 1937, 7502836), Dragutin Stameni{\'{c}} ({\DJ}uro, Lu{\v{s}}{\v{c}}ani, 1936, 7548827), Cvijeta Tulum (Lazo, Velika Peratovica, 1940, 7576547), Rade Turopoljac (Nikola, Petrinja, 1936, 7549024), Zdravko Ugrenovi{\'{c}} (Bo{\v{s}}ko, Dragelji, 1933, 7572785), Branislava Vukmirovi{\'{c}} (Nikola, Jasenovac, 1935, 7546789), Mladen Vuksan ({\DJ}uro, U{\v{s}}tica, 1935, 7555501), Latinka Vuleti{\'{c}} (Ilija, Jasenovac, 1933, 7546793), Kata Vu{\v{c}}i{\v{c}}evi{\'{c}} (Petar, U{\v{s}}tica, 1937, 7555492), Mira Zvonar (Cvijo, Turjak, 1939, 7506948), Branislava {\DJ}ur{\dj}evi{\'{c}} (Rade, Kijevci, 1938, 7504392), Cvijo {\v{S}}ok{\v{c}}evi{\'{c}} (Milan, Turjak, 1937, 7506909), Dragica {\v{S}}ok{\v{c}}evi{\'{c}} (Veljko, Kijevci, 1941, 7504532), Lazar {\v{S}}ok{\v{c}}evi{\'{c}} (Milan, Turjak, 1937, 7506910), Stanko Vujinovi{\'{c}} (Milo{\v{s}}, Dobrljin, 1934, 7507611), Stana Bo{\v{z}}i{\'{c}} (Adam, Petrinja, 1930, 7548984), Kosta Duki{\'{c}} (Mijat, Omarska, 1934, 7509356), Zora Grube{\v{s}}i{\'{c}} (Uro{\v{s}}, Turjak, 1935, 7506677), Mile Kljukovnica (Dragan, Voji{\v{s}}nica, 1933, 7530729), Milan Opa{\v{c}}i{\'{c}} ({\DJ}uro, Lu{\v{s}}{\v{c}}ani, 1937, 7548790), Mile Stanojevi{\'{c}} (Milan, Grbavci, 1935, 7503692), Milorad Zvijerac (Uro{\v{s}}, Bistrica, 1931, 7501039), Mara {\DJ}uri{\'{c}} ({\DJ}uro, Pakrani, 1937, 7561817), Ilija Duki{\'{c}} (Teodor, Omarska, 1938, 7509362), Marija Janjetovi{\'{c}} (Miajlo, Donji Podgradci, 1936, 7501781), Zorka Vodogaz (Milutin, Jablanica, 1940, 7504293), Danica Metla{\v{s}} (Mihailo, Jasenovac, 1936, 7546728), Du{\v{s}}anka Metla{\v{s}} (Mihailo, Jasenovac, 1935, 7557174), Jelena Pani{\'{c}} (Mihajlo, Trebovljani, 1935, 7506272), Dragoljub Savi{\'{c}} (Manojlo, Ga{\v{s}}nica, 1931, 7572330), Mikajlo Ugrenovi{\'{c}} (Bo{\v{s}}ko, Dragelji, 1934, 7572786), Miodrag Let{\'{c}} (-, Glina, 1938, 7564305), Stojan {\DJ}uri{\v{c}}i{\'{c}} (Milan, Romanovci, 1934, 7566137), Ru{\v{z}}a Dragorajac (Stevan, Ga{\v{s}}nica, 1937, 7502260), Mile Grubje{\v{s}}i{\'{c}} (Jovan, Lu{\v{s}}{\v{c}}ani, 1936, 7548690), David Kanitz-Kani{\'{c}} (Ignac, {\DJ}akovo, 1877, 7563811), An{\dj}elija Radovi{\'{c}} (Nikola, Voji{\v{s}}nica, 1934, 7560491). 

\textbf{\underline{JEDINICA NEPOZNATA}}: Mihael Lon{\v{c}}arevi{\'{c}} (Josip, Plesmo, 1919, 7547954). 

\textbf{\underline{JESENICE}}: Rudo Mohari{\'{c}} (Josip, Zagreb, 1925, 7568305). 

\textbf{\underline{JESENJE}}: Vilim Plevnik (-, Toma{\v{s}}evec, 1922, 7559555). 

\textbf{\underline{JOSIPOVAC}}: Stojan Pavlovi{\'{c}} (-, Branjina, 1915, 7568159). 

\textbf{\underline{JO{\v{S}}AVICA}}: Sava Paunovi{\'{c}} (-, Jo{\v{s}}avica, 1900, 7564851). 

\textbf{\underline{KAKMU{\v{Z}}}}: Cvijetin Stjepanovi{\'{c}} (Spasoje, Kakmu{\v{z}}, 1880, 7575821). 

\textbf{\underline{KALASIJA MEMI{\'{C}}I}}: Bo{\v{z}}o Lazi{\'{c}} (David, Baljkovica, 1900, 7575885). 

\textbf{\underline{KALENDERI}}: Mikan Borojevi{\'{c}} (Milan, Kalenderi, 1925, 7571048). 

\textbf{\underline{KALESIJA}}: Simo Koji{\'{c}} (Obren, Kulina, 1920, 7576063), Petar Vukovi{\'{c}} ({\DJ}uro, Zelina, 1885, 7576064), Jovan {\DJ}uri{\v{c}}i{\'{c}} ({\DJ}or{\dj}o, Zelina, 1896, 7568916), Timotija Mihajlovi{\'{c}} (Todor, Zolje, 1900, 7567630). 

\textbf{\underline{KARLOBAG}}: Mavro Kollmann (Rihard, Koprivnica, 1904, 7563766). 

\textbf{\underline{KARLOVAC}}: In{\dj}ija Keser (Simo, Keserov Potok, 1899, 7560055), Vladimir Ma{\'{c}}e{\v{s}}i{\'{c}} (Dragi{\'{c}}, Kupljensko, 1925, 7560475), Jovan Napijalo (Milo{\v{s}}, Kupljensko, 1899, 7560477), Nikola Paji{\'{c}} (Stevan, Vojni{\'{c}}, 1897, 7560434), Rade Rebi{\'{c}} (Petar, Gornja Trebinja, 1913, 7560028), Branko Vuji{\'{c}} (Stanko, Brezova Glava, 1914, 7560388), {\DJ}uro Vukobratovi{\'{c}} (Mili{\'{c}}, Kupljensko, 1897, 7560482), Du{\v{s}}an Petrovi{\'{c}} (To{\v{s}}o, Brezova Glava, 1894, 7560386), Stanko Markovi{\'{c}} (Teodor, Velika Crkvina, 1873, 7560061), Du{\v{s}}an Trkulja (Mihajlo, Brezova Glava, 1908, 7560387), Du{\v{s}}an Karapand{\v{z}}a (Cvijan, Gornja Trebinja, 1904, 7560410), Cvijan Karapand{\v{z}}a (Petar, Gornja Trebinja, 1888, 7560409), Stjepan Mihelinec (Mijo, Samobor, 1920, 7568318), Du{\v{s}}an Mihajlovi{\'{c}} (Josip, Vojni{\'{c}}, 1923, 7560422). 

\textbf{\underline{KERESTINEC}}: Matija Dumbovi{\'{c}} (-, Gu{\v{s}}{\'{c}}e, 1904, 7575952), Vladimir Korsky-Kestenbaum (Robert, Osijek, 1906, 7535031). 

\textbf{\underline{KESEROV POTOK}}: {\DJ}uro Keser (Cvijan, Keserov Potok, 1892, 7560054). 

\textbf{\underline{KLADU{\v{S}}A}}: Rade Ostoji{\'{c}} (Jovan, Ostoji{\'{c}}i, 1912, 7543916). 

\textbf{\underline{KLAKAR DONJI}}: Savo Puri{\'{c}} (Pejo, Klakar Donji, 1887, 7565948), Stevo Me{\dj}edovi{\'{c}} (Jovan, Klakar Donji, 1938, 7566091). 

\textbf{\underline{KLA{\v{S}}NICE}}: Petar Smiljani{\'{c}} (-, Jakupovci, 1900, 7565257). 

\textbf{\underline{KNIN}}: Albert Vladislovi{\'{c}} (Viktor, Bakar, 1924, 7543528). 

\textbf{\underline{KOBATOVCI, LAKT}}: Bosa Laki{\'{c}} (-, Kobatovci, 1920, 7565330). 

\textbf{\underline{KOD GAGLOVA}}: Jula Torbica (-, Mala Peratovica, -, 7576448). 

\textbf{\underline{KOD GORJEVCA}}: {\DJ}uro Grbi{\'{c}} (Mile, Gorjevac, 1896, 7574900). 

\textbf{\underline{KOD GRADI{\v{S}}KE}}: Risto Pani{\'{c}} (Gajo, Velika Obarska, 1926, 7567573). 

\textbf{\underline{KOD KU{\'{C}}E}}: Jovo Ga{\v{c}}i{\'{c}} (Simo, {\v{Z}}itomisli{\'{c}}i, 1908, 7563113), Stanko Ga{\v{c}}i{\'{c}} (Simo, {\v{Z}}itomisli{\'{c}}i, 1912, 7563106), {\DJ}oko Ga{\v{c}}i{\'{c}} (Pavle, {\v{Z}}itomisli{\'{c}}i, 1876, 7563114), Branko Puhalo (Mitar, {\v{Z}}itomisli{\'{c}}i, 1919, 7563034), Bo{\v{z}}o Simi{\'{c}} (Bogdan, {\v{Z}}itomisli{\'{c}}i, 1896, 7563014), Risto Simi{\'{c}} (Jovo, {\v{Z}}itomisli{\'{c}}i, 1904, 7563007), Savo Simi{\'{c}} (Spasoje, {\v{Z}}itomisli{\'{c}}i, 1885, 7563006), Du{\v{s}}an Sjeran (Nenad, {\v{Z}}itomisli{\'{c}}i, 1907, 7563004), Vasan Torbica (Ilija, Ostro{\v{z}}in, 1899, 7558671), An{\dj}elko Simi{\'{c}} (Dragutin, {\v{Z}}itomisli{\'{c}}i, 1921, 7563015), Branko Svrdlin (Jovo, {\v{Z}}itomisli{\'{c}}i, 1905, 7562999). 

\textbf{\underline{KOD SPLITA}}: Mento Papo (-, Sarajevo, -, 7518934). 

\textbf{\underline{KOMLENAC}}: Stevo Gaji{\'{c}} (Stevan, Komlenac, 1908, 7557205). 

\textbf{\underline{KOMOGOVINA}}: Jovan Dabi{\'{c}} (-, Komogovina, -, 7562890), Milo{\v{s}} Dmitrovi{\'{c}} (Jovan, Komogovina, 1900, 7576665). 

\textbf{\underline{KONJIC}}: Dobrila Toma{\v{s}}evi{\'{c}} (Milan, Konjic, 1928, 7574842). 

\textbf{\underline{KOPRIVNICA}}: {\DJ}uro Blanu{\v{s}}a (Gavro, Pakrac, 1896, 7526408), Mirko E{\v{s}}pek (Luka, Osijek, 1896, 7574229), Du{\v{s}}an Laketi{\'{c}} (Lazo, Srijemske Laze, 1914, 7539859), Stana Preli{\'{c}} (Mirko, Milo{\v{s}}evo Brdo, 1941, 7572244), Milo{\v{s}} Preradovi{\'{c}} (Aleksa, Mi{\v{s}}kovci, 1904, 7561560), Milo{\v{s}} Tepi{\'{c}} (Mile, Mi{\v{s}}kovci, 1901, 7561562), Ivan Valpoti{\'{c}} (Antun, Zagreb, 1917, 7571580), Milan Ba{\v{c}}anovi{\'{c}} (To{\v{s}}o, {\v{C}}akovci, 1886, 7540827), {\DJ}or{\dj}e Kiti{\'{c}} (Todor, Ljeskovica, 1884, 7575749), Nikola Kosijer (Lazo, Brusnik, 1892, 7525638), Mile Cvijanovi{\'{c}} (Gavro, Mi{\v{s}}kovci, 1901, 7561553), Nikola Vujnovi{\'{c}} (Simo, Vujnovi{\'{c}}i, 1898, 7530865), Sailo Pa{\v{s}}ali{\'{c}} (Stanko, Vozu{\'{c}}a, 1888, 7575778), Pavle-Pajo Jovanovi{\'{c}} (-, {\v{S}}abac, 1911, 7568479), Aleksa Manojlovi{\'{c}} ({\DJ}uro, Kula Banjer, 1907, 7575387). 

\textbf{\underline{KORENITA}}: Maksim Torlakovi{\'{c}} (Jovan, Korenita, 1897, 7561023). 

\textbf{\underline{KOSOVSKA MITROVICA}}: Kosta Kljaji{\'{c}} (-, Zvornik, -, 7568952). 

\textbf{\underline{KOSTANJICA}}: Dragutin Babi{\'{c}} (Spasoje, Gornji Podgradci, 1938, 7502898), Vukosava Babi{\'{c}} (Spasoje, Gornji Podgradci, 1937, 7502899), {\v{Z}}arko Dabi{\'{c}} (Matija, Komogovina, 1923, 7554438), Milan Kukrika ({\DJ}ura{\dj}, Mrakodol, 1882, 7507868), Mile Stanoj{\v{c}}i{\'{c}} (-, Pobr{\dj}ani, -, 7570196), Ljuban Balaban (Ostoja, Rakovac, 1917, 7507986), Dragutin Kukrika (Milan, Mrakodol, 1903, 7507870), Marko Kukrika (Milan, Mrakodol, 1910, 7507871), Milovan {\v{S}}urlan (Jovo, Gornje Vodi{\v{c}}evo, 1898, 7507718). 

\textbf{\underline{KOVIN}}: Lazo Toma{\v{s}}i{\'{c}} (Jovan, Zrinska, 1901, 7574547). 

\textbf{\underline{KOZARA}}: Sreto A{\'{c}}imovi{\'{c}} (Mile, Me{\dj}uvo{\dj}e, 1929, 7574250), Zorka Babi{\'{c}} ({\v{Z}}ivko, Rakovica, 1930, 7558320), Milan Bera (Lazo, Mazi{\'{c}}, 1923, 7570892), Mirko Bera (Lazo, Mazi{\'{c}}, 1900, 7570898), Pero Bera (Stojan, Mazi{\'{c}}, 1903, 7570895), Stojan Bera (Stojan, Mazi{\'{c}}, 1882, 7570893), Vladimir Bera (Jovan, Mazi{\'{c}}, 1909, 7570899), Rade Beri{\'{c}} (Teodor, Lamovita, 1912, 7566808), Petar Bosio{\v{c}}i{\'{c}} (Dragutin, Pobr{\dj}ani, 1930, 7558247), Stevo Bundalo (Mile, Pobr{\dj}ani, 1907, 7499720), {\DJ}uro Jak{\v{s}}i{\'{c}} (Ostoja, {\v{S}}evarlije, 1885, 7558643), Dragutin Jan{\v{c}}i{\'{c}} ({\DJ}uro, Donje Vodi{\v{c}}evo, 1915, 7507631), Milan Kne{\v{z}}evi{\'{c}} (Dmitar, Kriva Rijeka, 1911, 7568241), Ostoja Kos (Stevo, Ba{\v{c}}vani, 1925, 7497073), Mile Mari{\'{c}} (Stanko, Omarska, 1908, 7566248), Obrad Mari{\'{c}} (Stanko, Omarska, 1908, 7566245), Draginja Mudrini{\'{c}} (Dmitar, Gornji Jelovac, 1941, 7569376), Mirko Mudrini{\'{c}} (Dmitar, Gornji Jelovac, 1939, 7569377), Branko Paspalj (Ostoja, Bistrica, 1926, 7566325), Gojko Paukovi{\'{c}} (Stevan, Cerovica, 1920, 7570803), Stevan Paukovi{\'{c}} (Ilija, Cerovica, 1895, 7507528), Novak Popovi{\'{c}} (Bo{\v{z}}o, Vojskova, 1938, 7558966), Ostoja Popovi{\'{c}} (Milan, Trebovljani, 1903, 7507223), Dragomir Sad{\v{z}}ak (Pavle, Gornja Gradina, 1936, 7556766), Stanko Sekuli{\'{c}} (Pavle, Mrakodol, 1896, 7507878), Milo{\v{s}} Slijepac (Milan, Donja Slabinja, 1920, 7556608), Piljo Stani{\v{s}}ljevi{\'{c}} ({\DJ}uro, Turjak, 1919, 7568055), Stojan Stojakovi{\'{c}} (Janko, Tavija, 1919, 7508073), {\v{Z}}ivko Stojakovi{\'{c}} (Stojan, Omarska, 1925, 7566241), Jela Sumrak (Marjan, Me{\dj}e{\dj}a, 1939, 7557971), Milan Ugrenovi{\'{c}} (Pero, Petrinja, 1892, 7565822), Branko Vuleti{\'{c}} (Vasilije, Gornji Jelovac, 1929, 7569355), Jovo {\'{C}}ur{\v{c}}ija (Pajo, Prusci, 1892, 7507965), {\v{Z}}arko {\DJ}uri{\'{c}} (Ranko, Prusci, 1923, 7507966), Adem Bajagilovi{\'{c}} (-, Banja Luka, -, 7568062), Milan Radi{\'{c}} (-, Suvaja, -, 7573210), Ilija Slijepac (-, Donja Slabinja, -, 7567897), Margita Vujatovi{\'{c}} (-, Pakrac, 1919, 7571513), Stanko {\v{S}}tekovi{\'{c}} (Mile, Podo{\v{s}}ka, -, 7571094), Stanko Gligi{\'{c}} (Nikola, Petrinja, 1898, 7507901), Vlado Graoni{\'{c}} (Petar, Velika {\v{Z}}uljevica, 1926, 7508086), Mla{\dj}an Ka{\v{c}}avenda (Mile, Jaruge, 1892, 7566226), Stole Kne{\v{z}}evi{\'{c}} (Gligo, Prusci, 1885, 7507968), Branko Koji{\'{c}} (Bo{\v{z}}o, Lamovita, 1925, 7566351), Stojan Kosti{\'{c}} ({\DJ}uro, Jaruge, 1890, 7566227), Branko Luki{\'{c}} (Nikola, Blagaj Rijeka, 1915, 7507436), Mile Matija{\v{s}} (Risto, Gornji Orlovci, 1911, 7566215), Cvijo Samac (Vaso, Petrinja, 1901, 7507917), Ilija Stankovi{\'{c}} (Bo{\v{z}}o, Svodna, 1910, 7567097), Draginja Stijak (Vaso, Petrinja, 1912, 7507923), Du{\v{s}}an Stijak (Milan, Tavija, 1906, 7508070), Simo Stijak (Mi{\'{c}}o, Tavija, 1914, 7508069), {\DJ}uro Zgonjanin (Vaso, Velika {\v{Z}}uljevica, 1894, 7508103), Gojko {\v{S}}urlan (Aleksa, Gornje Vodi{\v{c}}evo, 1906, 7507715), Nikola Marinkovi{\'{c}} ({\DJ}uro, Kuljani, 1868, 7570876), Rajko Stojakovi{\'{c}} (Ratko, Omarska, 1924, 7566243), Novak Duki{\'{c}} (Tomo, Omarska, 1873, 7566246), Vojkan Vu{\v{c}}en (Mile, Donji Jelovac, 1887, 7556509), Uro{\v{s}} Pilipovi{\'{c}} (Mihajlo, Donje Vodi{\v{c}}evo, 1925, 7507640), Milo{\v{s}} Relji{\'{c}} (Vlado, Gornji Jelovac, 1929, 7566187), Mile Stijak (Mihajlo, Petrinja, 1888, 7507920), Cvijo {\DJ}uri{\'{c}} (Mile, Gornje Vodi{\v{c}}evo, 1896, 7565768), Milan Stankovi{\'{c}} (Bo{\v{z}}o, Aginci, 1920, 7497016), Milan Kosti{\'{c}} ({\DJ}uro, Jaruge, 1892, 7566228), Milovan Male{\v{s}}evi{\'{c}} (Vaso, Donji Podgradci, 1912, 7501856), Milinko Resan (Janko, Kalenderi, 1915, 7571068), Dragoja Lazarevi{\'{c}} (Mikan, Lje{\v{s}}ljani, 1920, 7570884), Mihajlo Peli{\'{c}} (Pavle, Kuljani, 1905, 7570878), Radoja Toma{\v{s}} (-, Pobr{\dj}ani, 1917, 7558264), Dragoja Zlokapa (Gligo, Ba{\v{c}}vani, 1924, 7497097), Stanko Vujinovi{\'{c}} (Marko, Dobrljin, 1906, 7507607). 

\textbf{\underline{KOZARA JARUGE}}: Stevo Tomi{\'{c}} (-, {\v{Z}}ivaja, 1920, 7572972). 

\textbf{\underline{KO{\v{C}}I{\'{C}}EVO}}: Draginja Vasi{\'{c}} (Savo, Seferovci, 1928, 7505823). 

\textbf{\underline{KO{\v{S}}UTARICA}}: Kosta Peji{\'{c}} (Vaso, Gornji Podgradci, 1940, 7561912). 

\textbf{\underline{KRAGUJEVAC}}: Milka Despotovi{\'{c}} (-, Beograd, -, 7493407). 

\textbf{\underline{KRALJEVO POLJE}}: Svetozar Pavlovi{\'{c}} (-, Polom, 1922, 7568803). 

\textbf{\underline{KRE{\v{C}}.DUBICA HR.}}: Ljubica Milkovi{\'{c}} (-, {\v{Z}}ivaja, -, 7572932). 

\textbf{\underline{KRIVAJA}}: {\DJ}or{\dj}o Mari{\'{c}} (Mitar, Podvolujak, 1886, 7575712). 

\textbf{\underline{KRNETE}}: Veljko Opa{\v{c}}i{\'{c}} ({\DJ}u{\dj}o, Krnete, 1913, 7508603). 

\textbf{\underline{KRSTINJA}}: Mila Dragi{\'{c}} (Stevan, Mracelj, 1887, 7560261). 

\textbf{\underline{KRUPANJ}}: Milo{\v{s}} Stojanovi{\'{c}} (Todor, Jelanjska, 1908, 7575180). 

\textbf{\underline{KR{\v{C}}EVINE}}: {\DJ}or{\dj}o Prnji{\'{c}} (Pero, Volari, 1902, 7575072). 

\textbf{\underline{KUKUJEVAC}}: Julka Radakovi{\'{c}} (Teodor, Kukunjevac, 1888, 7526125), Jula Radakovi{\'{c}} (-, Kukunjevac, 1888, 7564551), Nikola Milanovi{\'{c}} (Stevo, Kukunjevac, 1904, 7526101). 

\textbf{\underline{KUKUNJAC}}: {\DJ}uro Kosi{\'{c}} (Te{\v{s}}o, Kukunjevac, 1890, 7561213). 

\textbf{\underline{KUKUNJEVAC}}: Stevan Dra{\v{z}}i{\'{c}} (Jovan, Kukunjevac, 1883, 7561198), Jelena Grubni{\'{c}} (-, Bair, 1912, 7564242), Ljuban Grubni{\'{c}} (-, Kukunjevac, 1928, 7564244), Ostoja Rado{\v{s}}evi{\'{c}} (-, Kukunjevac, 1932, 7574389), Lazo Jovanovi{\'{c}} (Petar, Kukunjevac, 1886, 7561216), Mihajlo Vuleti{\'{c}} (Jovo, Kukunjevac, 1914, 7561262). 

\textbf{\underline{KUSONJE}}: Luka Kuli{\'{c}} (Jovan, Kusonje, 1904, 7575995), Dragan Radoj{\v{c}}i{\'{c}} (-, Kusonje, 1929, 7574376), Anica Vezmar (-, Kusonje, 1889, 7526298). 

\textbf{\underline{KUTINA}}: Nikola Pucarin (Ilija, Stupova{\v{c}}a, 1904, 7571450), Marko {\v{C}}enagd{\v{z}}ija (Mirko, Kadin Jelovac, 1920, 7557452). 

\textbf{\underline{LAKTA{\v{S}}I}}: Milo{\v{s}} Deli{\'{c}} (-, Lakta{\v{s}}i, 1908, 7565344), Milan Davidovi{\'{c}} (Simo, Lakta{\v{s}}i, 1909, 7508617), Milan Janjo{\v{s}} (Ilija, Glamo{\v{c}}ani, 1921, 7565240). 

\textbf{\underline{LAMOVITA}}: Stevan Grahovac-Savi{\'{c}} ({\DJ}ura{\dj}, Lamovita, 1898, 7566350). 

\textbf{\underline{LASINJA}}: Janko Parapati{\'{c}} (Janko, Desni {\v{S}}tefanki, 1915, 7531001), Marko Parapati{\'{c}} (Lovro, Desni {\v{S}}tefanki, 1886, 7530904). 

\textbf{\underline{LEPOGLAVA}}: {\DJ}uro Bernfest (Viktor, Petrinja, 1922, 7564074), Zvonimir Bi{\v{s}}{\'{c}}an (Ivan, Zagreb, 1925, 7568272), Ivan Domitrovi{\'{c}} (Mirko, Ferdinandovac, 1922, 7569116), Zvonko Feketi{\'{c}} (Ivan, Zagreb, 1921, 7569138), Josip Fo{\v{c}}i{\'{c}} (Mijo, Zagreb, 1909, 7569150), Nikola Futa{\v{c}} (Mirko, Zagreb, 1902, 7569162), Petar Grubor (Stevo, Split, 1930, 7551160), Ivan Habei{\'{c}} (Stjepan, Mala Ostrna, 1912, 7566731), Ivan Habri{\'{c}} (Andrija, Zagreb, 1917, 7569192), Mustafa Had{\v{z}}iefendi{\'{c}} (Mustafa, Grada{\v{c}}ac, 1914, 7562321), Marijan Herceg (Janko, Samobor, 1911, 7561082), Alojz Kolar (Stjepan, Zabok, 1913, 7559549), Miljenko Kova{\v{c}}evi{\'{c}} (Ante, Trebinje, 1922, 7568445), Tomo Kuzmec (Mate, Zagreb, 1917, 7568399), Josip Malekovi{\'{c}} (Nikola, Zapre{\v{s}}i{\'{c}}, 1914, 7568346), Davor Mar{\v{c}}i{\'{c}} (Lucijan, Zadar, 1921, 7568351), Ferdo Mati{\'{c}} (Liberat, Pula, 1920, 7568363), Nasiha Mehmedba{\v{s}}i{\'{c}} (Mahmut, Sarajevo, 1924, 7562676), Vitomir Suki{\'{c}} (Josip, Osijek, 1908, 7555155), Stjepan Vavra (Adolf, Banja Luka, 1908, 7571220), Ivo Vican (Stjepan, Split, 1906, 7571588), Stjepan {\v{S}}o{\v{s}}tar (Dragutin, Klanjec, 1912, 7552201), Andrija {\v{S}}vager (Mijo, Nu{\v{s}}tar, 1909, 7561507), Kristina Dan{\v{c}}evi{\'{c}} (-, {\DJ}akovo, 1899, 7569109), Antun Jelan{\v{c}}i{\'{c}} (-, Bani{\'{c}}evac, 1919, 7554764), Josip Kaved{\v{z}}ija (-, Slavonski Brod, 1906, 7537302), Milan Mili{\'{c}} (-, Slano, 1891, 7568314), Stjepan Radeni{\'{c}} (-, Zagreb, 1906, 7571719), Petar Mili{\'{c}} (Stjepan, Slobodnica, 1925, 7562781), Gustav Pfeiffer (Sigmund, Sisak, 1900, 7563600), Muhamed D{\v{z}}ud{\v{z}}a (Osman, Mostar, 1908, 7575231), Dragica Stani{\'{c}} (Ivan, Zagreb, 1916, 7553998), Gabrijel Hess (Ivan, {\DJ}akovo, 1901, 7531874), Ru{\v{z}}ica Hess (Gabrijel, {\DJ}akovo, 1920, 7531873), Maks Golob (Franjo, Herceg Novi, 1900, 7519996), Nisim Pardo ({\v{S}}abetaj, Sarajevo, 1896, 7571684), Antun Hajdarovi{\'{c}} ({\DJ}uro, Gardinovec, 1920, 7569195), Zvonko Grgi{\'{c}} (Marko, Zagreb, 1930, 7569184), Pavao-Ribin Pavlin (Ivan, Pr{\v{c}}anj, 1913, 7571689), Franjo-{\v{C}}eki{\'{c}} Pintar ({\DJ}uro, Bukevje, 1910, 7571702), Marija Starina (August, Lovran, 1907, 7561278), Salih Vojvodi{\'{c}} (Ibro, Zenica, -, 7562823), Franjo Bedekovi{\'{c}} (Mirko, Zagreb, 1909, 7569094), Ignac Bun{\v{c}}i{\'{c}} (Franjo, {\v{Z}}itomir, 1905, 7569058), Zvonimir Me{\v{z}}nari{\'{c}} (Josip, Zagreb, 1915, 7568322), Dragutin Pijelik (Franjo, Zagreb, 1924, 7553845), Slavko Bira{\v{c}}-Dru{\v{z}}i{\'{c}} (Juro, Zagreb, 1919, 7568269), Ivanka Brachtel (Rudolf, Glina, 1902, 7570411), Julijana Degregori (-, Osijek, 1925, 7534914), Milivoj Gavran{\v{c}}i{\'{c}} (Oton, Virovitica, 1905, 7569171), Emilija-Milka Kleinhaus (Nikola, Vara{\v{z}}din, 1915, 7551641), Anka Kolar (-, Zagreb, -, 7568426). 

\textbf{\underline{LESKOVAC}}: Stojka Jeki{\'{c}} (Mikan, Sreflije, 1925, 7558476). 

\textbf{\underline{LIJE{\v{S}}ANJ}}: Maksim Maksimovi{\'{c}} (Lazar, Lije{\v{s}}anj, 1934, 7568887), Radojka Maksimovi{\'{c}} (-, Lije{\v{s}}anj, 1937, 7568877), Ljubinka Simi{\'{c}} (Ljubo, Lije{\v{s}}anj, 1929, 7568903), Stana Maksimovi{\'{c}} (-, Lije{\v{s}}anj, 1934, 7568879). 

\textbf{\underline{LIJE{\v{S}}{\'{C}}E}}: Uro{\v{s}} Mitri{\'{c}} (Simeun, Lije{\v{s}}{\'{c}}e, 1903, 7567077), Stanoje Kova{\v{c}}evi{\'{c}} (-, Lije{\v{s}}{\'{c}}e, 1896, 7566002), Milan Gluvak (Dimitrije, Lije{\v{s}}{\'{c}}e, 1912, 7565972), Marko Teodosi{\'{c}} (Simo, Lije{\v{s}}{\'{c}}e, 1901, 7565872), Branko {\v{Z}}ari{\'{c}} (Jovan, Lije{\v{s}}{\'{c}}e, 1903, 7566051), Branko Dujani{\'{c}} (Ljubosav, Lije{\v{s}}{\'{c}}e, 1912, 7565991). 

\textbf{\underline{LIKA}}: Dragica {\DJ}akovi{\'{c}} (Mile, Mracelj, 1900, 7560270), Gojko {\DJ}akovi{\'{c}} (Ilija, Mracelj, 1934, 7560272), {\DJ}uro {\DJ}akovi{\'{c}} (Ilija, Mracelj, 1937, 7560271), Elis Danon (Avram, Sarajevo, 1910, 7516995), Vaso Ka{\v{c}}ilovi{\'{c}} (-, Virje, -, 7523123). 

\textbf{\underline{LIKA GOSPI{\'{C}}}}: Salamon Gaon (Avram, Travnik, 1905, 7575524). 

\textbf{\underline{LIPIK}}: Mirko Cvetkovi{\'{c}} (Rade, Bair, 1902, 7554452), Stevo Kova{\v{c}}evi{\'{c}} (Gli{\v{s}}o, Lovska, 1920, 7547056), Marko Prodanovi{\'{c}} ({\DJ}uro, Bujavica, 1902, 7565382), Teodor Valenti{\'{c}} (Nikola, Bujavica, 1889, 7565385), Milan Grbi{\'{c}} (-, Kusonje, 1910, 7564214), {\DJ}uro Pribi{\'{c}} (Mi{\'{c}}o, Demirovac, 1900, 7556104), Nikola Miokovi{\'{c}} (Jovo, Bujavica, 1924, 7565375), Teodor Gaji{\'{c}} ({\DJ}uro, Kukunjevac, 1909, 7526041). 

\textbf{\underline{LIPOGLAVA}}: Alojz Veber (-, Ra{\v{c}}e, -, 7562993). 

\textbf{\underline{LI{\v{S}}TICA}}: Pero Radulovi{\'{c}} (Todor, Trebinje, 1897, 7563024). 

\textbf{\underline{LJE{\v{S}}LJANI}}: Petar Vujanovi{\'{c}} (Stevan, Lje{\v{s}}ljani, 1876, 7508097). 

\textbf{\underline{LJUBIJA}}: Ilija Stojanovi{\'{c}} (Mile, Ljubija, 1900, 7509329). 

\textbf{\underline{LOBOGRAD}}: Mia Kolari{\'{c}}-Kohn (Leon, Zagreb, 1927, 7553620). 

\textbf{\underline{LOBOR GRAD}}: Mila Deutsch (Eugen, Osijek, 1921, 7564000), Teodor Pollak ({\v{Z}}iga, Bjelovar, 1906, 7520534), Irena Spitzer (-, {\v{C}}akovec, 1900, 7551291). 

\textbf{\underline{LOBORGRAD}}: Jakob Kampos (Jozef, Sarajevo, 1931, 7517883), Helga Spitzer (Maks, Glina, 1930, 7544179). 

\textbf{\underline{LOGOR LEPOGLAVA}}: Franjo Rumbak (Franjo, Krapinske Toplice, 1897, 7571737), Jelena-Helena Visinger (Ivan, Trnovec, 1901, 7571595). 

\textbf{\underline{LOGOR NEPOZNAT NJEMA{\v{C}}KA}}: Sidonija Hirschl (Hinko, Zagreb, 1901, 7553516). 

\textbf{\underline{LOGOR PAG}}: Milan Berl (Eugen, Rijeka, 1903, 7569101). 

\textbf{\underline{LOV{\v{C}}A}}: Stevo Savurdi{\'{c}} (Petar, Strmen, 1883, 7555332), Ana Grnovi{\'{c}} (Adam, Lov{\v{c}}a, 1900, 7576670), Petar Grnovi{\'{c}} (Tanasije, Lov{\v{c}}a, 1893, 7576671), Stojan Maljkovi{\'{c}} (Ilija, Lov{\v{c}}a, 1890, 7576680), Ana Rajkovi{\'{c}} (Jovan, Lov{\v{c}}a, 1903, 7576681), Petar Trnini{\'{c}} (Gligo, Lov{\v{c}}a, 1897, 7576689), Kata {\v{C}}izmi{\'{c}} (Milo{\v{s}}, Lov{\v{c}}a, 1937, 7545328). 

\textbf{\underline{LUKA}}: Franjo Bartoli{\'{c}} (Jakob, Donji Mihaljevec, 1920, 7551306). 

\textbf{\underline{LUKAVAC}}: Ne{\dj}o Markovi{\'{c}} (Jovo, Tr{\v{s}}i{\'{c}}, -, 7569021). 

\textbf{\underline{M GR{\DJ}EVAC}}: Ignjatija Mujadin (Nikola, Mali Gr{\dj}evac, 1898, 7576461). 

\textbf{\underline{MAGLAJ}}: Milan Tripi{\'{c}} (Tripo, Krsno Polje, 1920, 7575593). 

\textbf{\underline{MAGLJANI}}: Petar {\v{S}}ipka (Milan, Kobatovci, 1928, 7508559). 

\textbf{\underline{MAJEVICA}}: Boro Popovi{\'{c}} (Vaso, Orahova, 1926, 7573813). 

\textbf{\underline{MALA BARNA}}: Du{\v{s}}an Domitrovi{\'{c}} (Nikola, Mala Barna, 1928, 7576438). 

\textbf{\underline{MALA PERATOVICA}}: Ja{\'{c}}im Kova{\v{c}}evi{\'{c}} (Tomo, Mala Peratovica, 1907, 7523595). 

\textbf{\underline{MALI GRDJEVAC}}: Stevo Grubi{\'{c}} (Nikola, Mali Gr{\dj}evac, 1899, 7576498). 

\textbf{\underline{MALI GR{\DJ}EVAC}}: Aleksa Bajevi{\'{c}} (Milovan, Mali Gr{\dj}evac, 1878, 7576513), Jovo Keseri{\'{c}} (Lazo, Mali Gr{\dj}evac, 1940, 7576485), Lazo Keseri{\'{c}} (Jandrija, Mali Gr{\dj}evac, 1899, 7576483), Rade Keseri{\'{c}} (Jovo, Mali Gr{\dj}evac, 1920, 7576482), Rade Milekovi{\'{c}} (Stevan, Mali Gr{\dj}evac, 1870, 7576489), Soka Milekovi{\'{c}} (Bajo, Mali Gr{\dj}evac, 1903, 7576490), Rade Mujadin (Mile, Mali Gr{\dj}evac, 1892, 7576459), Jovo {\v{C}}izmi{\'{c}} (Stevo, Mali Gr{\dj}evac, 1923, 7576497), Lazo Rastovi{\'{c}} (-, Gornja Kova{\v{c}}ica, -, 7576405), Ljuba Trbojevi{\'{c}} (-, Mali Gr{\dj}evac, -, 7576476), Kosta Milekovi{\'{c}} (Akso, Mali Gr{\dj}evac, 1899, 7576488), Mile Mujadin (Akso, Mali Gr{\dj}evac, 1907, 7576462), Milka Belanovi{\'{c}} (-, Mali Gr{\dj}evac, -, 7576503), Mile Belanovi{\'{c}} (-, Mali Gr{\dj}evac, -, 7576506). 

\textbf{\underline{MALO BLA{\v{S}}KO}}: Bo{\v{s}}ko Popovi{\'{c}} (Risto, Malo Bla{\v{s}}ko, 1907, 7565228). 

\textbf{\underline{MALO PALAN{\v{C}}I{\v{S}}TE}}: An{\dj}a Kos (Petar, Me{\dj}uvo{\dj}e, 1886, 7499456). 

\textbf{\underline{MARTINCI}}: Mirjana Joveli{\'{c}} (Stanko, Martinci, 1928, 7560548). 

\textbf{\underline{MARTINSKA {\v{S}}UMA}}: Ilija Bugarinovi{\'{c}} (-, Gornja {\v{S}}u{\v{s}}njara, -, 7521284). 

\textbf{\underline{MATHAUSEN}}: Ljubica Paji{\'{c}} (Mile, Svinica Krstinjska, 1928, 7560335), Janja Vujaklija (Te{\v{s}}o, Strmen, 1889, 7555317), Gojko {\v{S}}inik (Ostoja, Cimiroti, 1924, 7573957), Stevo Babi{\'{c}} (Stevan, Prevr{\v{s}}ac, 1898, 7576706), Muhamed Hod{\v{z}}i{\'{c}} (-, Te{\v{s}}anj, -, 7562712), Milko Ikoni{\'{c}} (Jevto, {\v{C}}elopek, -, 7568984), Stanoje Simi{\'{c}} (-, Polom, 1900, 7568807), {\DJ}uro Janji{\'{c}} (Krstivoj, Prevr{\v{s}}ac, 1901, 7576703), Stevo Jankovi{\'{c}} (Spasoje, Boljani{\'{c}}, 1913, 7575806), Adam Ognjanovi{\'{c}} ({\DJ}oka, Ada{\v{s}}evci, 1920, 7554977), Lazar Ognjanovi{\'{c}} ({\DJ}oka, Ada{\v{s}}evci, 1922, 7495485), Branko Vuj{\v{c}}i{\'{c}} (Milo{\v{s}}, Ada{\v{s}}evci, 1896, 7495525). 

\textbf{\underline{MATIJEVI{\'{C}}I}}: Nikola Majki{\'{c}} (-, Matijevi{\'{c}}i, 1924, 7557760). 

\textbf{\underline{MA{\v{S}}I{\'{C}}}}: Ilija Dabi{\'{c}} (-, Ma{\v{s}}i{\'{c}}, -, 7567381). 

\textbf{\underline{MEHINO STANJE}}: Milivoj Uzelac (Milan, Maljevac, 1939, 7557418). 

\textbf{\underline{MEZGRAJA}}: Rado Deli{\'{c}} (Mijo, Mezgraja, 1894, 7561011). 

\textbf{\underline{ME{\DJ}E{\DJ}A}}: Gojko Goga (Milorad, Me{\dj}e{\dj}a, 1940, 7499259). 

\textbf{\underline{MILISAVCI}}: Luka Sibin{\v{c}}i{\'{c}} (Simo, Milisavci, 1885, 7555795). 

\textbf{\underline{MILO{\v{S}}EVO BRDO}}: Zdravko Mandi{\'{c}} (Stoko, Ga{\v{s}}nica, 1936, 7573701). 

\textbf{\underline{MOSTAR}}: Nedeljko Bukvi{\'{c}} (Trifko, Mostar, 1921, 7563143), Samija Feji{\'{c}} (Muhamed, Mostar, 1910, 7512004), Tomo Milivojevi{\'{c}} ({\'{C}}etko, Mostar, 1906, 7563061), {\v{S}}piro To{\v{s}}i{\'{c}} (Pero, Mostar, 1903, 7576009), Albina Veli{\v{c}}an (Matija, Sarajevo, 1907, 7575228), Samija Krajina (Muhamed, Mostar, 1910, 7562571), Mirko {\'{C}}orluka (Drago, Stolac, 1911, 7563132), Ilija Matkovi{\'{c}} ({\v{S}}{\'{c}}epan, Bogodol, 1905, 7511937), Milenko Ili{\'{c}} (Vaso, Mostar, 1920, 7563097), Lazar Gambeli{\'{c}} (Ilija, Nevesinje, 1908, 7555112), Mehmed Omeragi{\'{c}} (Avdo, Mostar, 1917, 7555094), Vasilije {\v{S}}paravalo ({\v{Z}}arko, Mostar, 1925, 7562985). 

\textbf{\underline{MOTIKE}}: Vaskrsije Todi{\'{c}} (Simeun, Motike, 1911, 7575280). 

\textbf{\underline{MRA{\v{C}}AJ}}: Dragan Torbica (-, {\v{Z}}ivaja, -, 7572976). 

\textbf{\underline{MR{\v{C}}EVCI}}: Vidosava Balaban (-, Mr{\v{c}}evci, 1902, 7565301). 

\textbf{\underline{MUDRIKE}}: Todor Vujinovi{\'{c}} (Marko, Mudrike, 1924, 7575417). 

\textbf{\underline{NA DRVARU}}: Mihael Papo (Moise, Sarajevo, 1912, 7519278). 

\textbf{\underline{NA IGMANU}}: Aron Albahari ({\v{S}}abetaj, Sarajevo, 1921, 7516156), Jakov Levi (Izidor, Sarajevo, 1914, 7576627). 

\textbf{\underline{NA PUTU ZA LOGOR}}: Salamon Alkalaj (Mojsije, Travnik, 1903, 7564098). 

\textbf{\underline{NA SUTJESCI}}: Simo Mili{\v{c}}i{\'{c}} (Nikola, Podosoje, 1923, 7575027), Jakov Albahari ({\v{S}}abetaj, Te{\v{s}}anj, 1872, 7516157). 

\textbf{\underline{NEMA{\v{C}}KA}}: Bogdan Gavri{\'{c}} ({\DJ}uro, Donja Paklenica, 1910, 7575580), {\v{Z}}ivan Krsti{\'{c}} (Risto, Male{\v{s}}evci, 1913, 7561014), Ljubo Vidovi{\'{c}} (Pero, Ma{\v{s}}i{\'{c}}i, 1903, 7572808), Voislav {\DJ}ur{\dj}evi{\'{c}} (Jovo, Kijevci, 1924, 7504382), Stojan Cvjetkovi{\'{c}} (Milan, Ku{\'{c}}ice, 1921, 7575659), Nedeljko Lazarevi{\'{c}} (Simo, Gornja Paklenica, 1901, 7575581), Rajko Borjanovi{\'{c}} (Milutin, Sovjak, 1940, 7561980), Pavo Luki{\'{c}} (Mile, Jablanica, 1919, 7569298), Mikajilo {\v{S}}orga (Marko, Gornji Jelovac, 1920, 7569399), Simo Bro{\'{c}}ilo (Jovo, Klakar Donji, 1900, 7565709). 

\textbf{\underline{NEMA{\v{C}}KA-STALAG}}: Jovica Nal{\v{c}}i{\'{c}} (-, Ada{\v{s}}evci, -, 7554690), Risto Sekuli{\'{c}} (Tanasije, Ljeljen{\v{c}}a, 1907, 7561001). 

\textbf{\underline{NEMA{\v{C}}KI LOGOR}}: Josip Capek ({\DJ}uro, Zagreb, 1910, 7569063), Emil Hirtweil (Adolf, Vinkovci, 1911, 7540260), Savo Kaurinovi{\'{c}} (Ilija, Bukvik Donji, 1907, 7575870), Ana Miljatovi{\'{c}} (Ostoja, {\v{S}}evarlije, 1926, 7558663), Du{\v{s}}an Petrovi{\'{c}} (Petar, Svinica Krstinjska, 1923, 7560338), Branko Vrani{\'{c}} (Nikola, Lu{\v{z}}ani, 1925, 7574022), Nikola {\v{Z}}ivanovi{\'{c}} ({\v{Z}}ivko, Budimci, 1913, 7532943), Ferdo Berliner (-, Osijek, 1885, 7534883), Vlajko Eri{\'{c}} (-, {\v{C}}elopek, -, 7568976), Du{\v{s}}an Lazarevi{\'{c}} (Risto, {\v{C}}elopek, -, 7568981), Stanko Lazarevi{\'{c}} (Risto, {\v{C}}elopek, -, 7568980), {\DJ}or{\dj}o Milo{\v{s}}evi{\'{c}} (Jovan, {\v{C}}elopek, -, 7568986), Bogdan Stojki{\'{c}} (-, {\v{C}}elopek, -, 7568979), Stojan Zeti{\'{c}} (Jovo, Vrbovljani, -, 7571466), Nedeljko Sikanovi{\'{c}} (Marinko, Bakoti{\'{c}}, 1892, 7575596), Branko Todorovi{\'{c}} ({\DJ}or{\dj}o, Donji Rakovac, 1918, 7575592), Vlado {\DJ}uki{\'{c}} ({\DJ}or{\dj}e, Han Plo{\v{c}}a, 1904, 7575392), Emilija Grgur (Lazar, {\DJ}ala, 1927, 7554934), Zdravko Gvozdenac (Mihajlo, Cimiroti, 1921, 7572462), Simo {\v{S}}a{\v{s}}i{\'{c}} (Vasilj, Kirin, 1920, 7558625), Tereza Hirtweil (-, Vinkovci, 1878, 7540259), Jozef Montiljo (Mo{\v{s}}o, Bijeljina, 1912, 7513560), Milivoj Savi{\'{c}} (Krsto, Veliko Nabr{\dj}e, 1902, 7532753), Ljubica Grgur-Kuzmanov (Milan, {\DJ}ala, 1907, 7554933). 

\textbf{\underline{NEVESINJE}}: Gavrilo Kova{\v{c}} ({\DJ}uro, Nevesinje, 1896, 7555128). 

\textbf{\underline{NI{\v{S}} CRVENI KRST}}: Berta Albahari (-, Zagreb, 1906, 7553148), Marija Naran{\v{c}}i{\'{c}} (-, Smrti{\'{c}}, -, 7554504). 

\textbf{\underline{NORVE{\v{S}}KA}}: Milovan Bili{\'{c}} (Milo{\v{s}}, Brejakovi{\'{c}}i, 1897, 7575264), Vinko Bora{\v{s}} (Petar, Ov{\v{c}}ari, 1920, 7574835), Milo{\v{s}} Bo{\v{z}}i{\'{c}} (Milan, Ostoji{\'{c}}evo, 1895, 7567588), Milan Bu{\'{c}}an (Petar, Svinica Krstinjska, 1924, 7560326), Petar Bu{\'{c}}an (Nikola, Svinica Krstinjska, 1892, 7560329), Salem Mahi{\'{c}} (Meho, Ljubu{\v{s}}ki, 1919, 7562394), Zdravko Pani{\'{c}} (Cvijan, Vranjak, 1904, 7565731), {\v{Z}}ivko Pani{\'{c}} (Cvijan, Vranjak, 1920, 7575155), Ratko Papaz (Simo, Gornji Kotorac, 1915, 7574757), {\DJ}uro Radovi{\'{c}} (Marko, Ostro{\v{z}}in, 1916, 7558667), Marko Rudan (Stevan, Donji Sjeni{\v{c}}ak, 1895, 7562537), Jevto Stojanovi{\'{c}} (Simo, Gornji Lokanj, 1922, 7569005), Stani{\v{s}}a Te{\v{s}}i{\'{c}} (Stojan, Vranjak, 1920, 7575156), Sreten Vukovi{\'{c}} (Jovan, Jasenovac, 1914, 7557289), Milan {\DJ}oki{\'{c}} ({\v{Z}}ivan, Velika Obarska, 1900, 7567574), Milo{\v{s}} {\v{S}}ormaz (Nikola, Gornji Jelovac, 1909, 7569400), Ivan Dragomanovi{\'{c}} (-, Komarnica, -, 7571518), Janko Savi{\'{c}} (-, Voji{\v{s}}nica, 1902, 7561569), Jovo {\v{S}}tekovi{\'{c}} (-, Mali Zdenci, -, 7557051), Simo Despotovi{\'{c}} (Tanasije, Trije{\v{s}}nica, 1901, 7576032), Lazo Klincov (Nikola, Drakseni{\'{c}}, 1896, 7498087), Mile Rudan (Stojan, Donji Sjeni{\v{c}}ak, 1885, 7558578), Milo{\v{s}} {\v{Z}}ivanovi{\'{c}} (Rajko, Trije{\v{s}}nica, 1915, 7576033), Jovo Raj{\v{c}}evi{\'{c}} (Savo, Jasenovac, 1911, 7557197), Zvonimir Kri{\v{s}}kovi{\'{c}} (Tomo, Staro Petrovo Selo, 1918, 7533867), Jula Bala{\'{c}} (Mile, Mlaka, 1903, 7547319). 

\textbf{\underline{NOVA GRADI{\v{S}}KA}}: Milan Nikoli{\'{c}} (Savo, Slobo{\v{s}}tina, 1920, 7537031), Albert Papo (Santo, Sarajevo, 1898, 7518929), Milan Zuber (Marko, Borovac, 1905, 7554455), Jakov Bogojevi{\'{c}} (-, Jurkovac, -, 7562451), Bo{\v{z}}o Drpi{\'{c}} (-, Stoj{\v{c}}inovac, -, 7562490), Ilija Drpi{\'{c}} (-, Stoj{\v{c}}inovac, -, 7562492), Ana Kodor (-, Jurkovac, -, 7562459), Milka Mila{\v{s}}inovi{\'{c}} (-, {\v{S}}e{\v{s}}kovci, -, 7564622), Albert Papo (Santo, Sarajevo, 1898, 7519201), {\DJ}or{\dj}e Drpi{\'{c}} (-, Stoj{\v{c}}inovac, -, 7562491), Brana Bujak (Mile, Ljubija, 1930, 7509327), Milka Korica-Lali{\'{c}} (-, Rogo{\v{z}}a, 1915, 7567920). 

\textbf{\underline{NOVO SELO}}: Cvijan Bjelopetrovi{\'{c}} (Petar, Gornji Poloj, 1920, 7559620), Krsto Ili{\'{c}} (-, Novo Selo, -, 7568823), Stevo Ili{\'{c}} (-, Novo Selo, -, 7568826), Milutin Markovi{\'{c}} (-, Novo Selo, -, 7568821), Stana Tomi{\'{c}} (-, Novo Selo, -, 7568816). 

\textbf{\underline{NOVO SELO GARE{\v{S}}}}: Ilija Krupljan (Stevan, Velika Br{\v{s}}ljanica, 1903, 7567867). 

\textbf{\underline{NOVO VES}}: Josip {\v{Z}}malec ({\DJ}uro, Zagreb, 1926, 7571554). 

\textbf{\underline{NOVSKA}}: Bo{\v{z}}o Zori{\'{c}} (Nikola, Gornji Jelovac, 1939, 7569369), Mileva Zori{\'{c}} (Nikola, Gornji Jelovac, 1941, 7569370), {\DJ}uro Zori{\'{c}} (Nikola, Gornji Jelovac, 1924, 7566391), Simo Komleni{\'{c}} (-, Novska, 1895, 7530837). 

\textbf{\underline{NOVSKA,BRO{\v{C}}ICE}}: Ana Vuka{\v{s}}inovi{\'{c}} (-, Jasenovac, 1879, 7555587). 

\textbf{\underline{OBOROVO}}: Vinko Dore{\v{s}}i{\'{c}} (Stjepan, Lonja Ivani{\v{c}}ka, 1922, 7573392). 

\textbf{\underline{OD STRANE USTA{\v{S}}A}}: Ana Grubi{\'{c}} (Stojan, Crkveni Bok, 1902, 7555340). 

\textbf{\underline{OD USTA{\v{S}}A}}: In{\dj}ija Vujanovi{\'{c}} (Ilija, Donja {\v{C}}emernica, 1912, 7558520). 

\textbf{\underline{OD VR. DIVIZIJE}}: Stanoje Strigi{\'{c}} (Stjepan, Gornja Paklenica, 1908, 7575584). 

\textbf{\underline{ODVE.IZ BOLNICE}}: Dragica Leskovar (Ivan, Zagreb, 1924, 7565053). 

\textbf{\underline{OKRUGLICA}}: Slavko Bunji{\'{c}}evi{\'{c}} (Nikola, Visoko, 1914, 7575382), Vito Bakarac (-, Kula Banjer, -, 7563680). 

\textbf{\underline{OKU{\v{C}}ANI}}: Ljubica {\v{Z}}ivkovi{\'{c}} (Vidoje, Veliko Nabr{\dj}e, 1942, 7565694), Milica {\v{Z}}ivkovi{\'{c}} (Vidoje, Veliko Nabr{\dj}e, 1942, 7565695), Borislav Krmar (-, Oku{\v{c}}ani, -, 7571519), Du{\v{s}}an Mrkonji{\'{c}} (-, Oku{\v{c}}ani, -, 7564902). 

\textbf{\underline{OLOVO}}: Danko Staki{\'{c}} ({\DJ}or{\dj}e, Vukovine, 1901, 7575629). 

\textbf{\underline{OMARSKA}}: Mira Mari{\'{c}} (Stoji{\'{c}}, Omarska, 1913, 7566244), Danica Stojakovi{\'{c}} (Stojan, Omarska, 1933, 7566242), Marko Stojakovi{\'{c}} (Stojan, Omarska, 1937, 7509404). 

\textbf{\underline{ORAHOVA}}: Kosa Marinkovi{\'{c}} (Lazo, Ga{\v{s}}nica, 1930, 7573713), Marko Martinovi{\'{c}} ({\DJ}uro, Orahova, 1938, 7505381). 

\textbf{\underline{ORAHOVICA}}: Milko {\DJ}uri{\'{c}} (Luka, Orahovica, 1900, 7575172). 

\textbf{\underline{OSIJEK}}: Lazo Sabli{\'{c}} (Arso, Cremu{\v{s}}ina, 1890, 7576330), Du{\v{s}}an Vurdelja (Mile, Donja Gata, 1905, 7574894), {\DJ}or{\dj}e Vukovi{\'{c}} (Sreten, Erdut, 1892, 7534827). 

\textbf{\underline{OSMACI}}: Mirko Astarevi{\'{c}} (Mato, Osmaci, 1890, 7568918). 

\textbf{\underline{OTERAN OD KU{\'{C}}E}}: Bjelan Avramovi{\'{c}} (-, Zelinje, 1910, 7568850), Stojan Radulovi{\'{c}} (Savo, Prisjeka Donja, 1890, 7574942), Mile Kerkez (Luka, Zavolje, 1900, 7574925). 

\textbf{\underline{OTO{\v{C}}AC}}: Gertruda Friedfeld (Lavoslav, Zagreb, 1920, 7553365), Tereza Friedfeld (-, Zagreb, 1892, 7553364). 

\textbf{\underline{OZERKOVI{\'{C}}I}}: Bogdan Zoranovi{\'{c}} (Bo{\v{s}}ko, Ozerkovi{\'{c}}i, 1909, 7575265). 

\textbf{\underline{O{\'{C}}ENOVI{\'{C}}I, BRATUNAC}}: Milojka Stevanovi{\'{c}} (-, Lije{\v{s}}anj, 1941, 7568893). 

\textbf{\underline{P. GORA}}: Mile Baji{\'{c}} ({\DJ}uro, Donja Brusova{\v{c}}a, 1905, 7560062). 

\textbf{\underline{PAG}}: Klaudija Kraus (Josip, Zagreb, 1891, 7563757), Salamon Levi (David, Sarajevo, 1906, 7518280), Moric Levi (Rafael, Sarajevo, 1903, 7518505). 

\textbf{\underline{PAKRAC}}: Milja Pralica (Stevo, {\v{C}}itluk, 1938, 7556092), Gojko Trbojevi{\'{c}} (Marko, Pucari, 1939, 7558305). 

\textbf{\underline{PALE}}: Darinka Nenadi{\'{c}} (Vojin, Grkovci, -, 7567067). 

\textbf{\underline{PAPUK}}: Vilko Mermelstein (Izidor, Na{\v{s}}ice, -, 7574647), Petar Filip{\v{c}}uk (Pavao, Lipovljani, 1922, 7555381). 

\textbf{\underline{PAU{\v{C}}JE}}: Slavka Buletinac (-, Pau{\v{c}}je, -, 7559842), Kosta Mihaljevi{\'{c}} (-, Pau{\v{c}}je, -, 7559855). 

\textbf{\underline{PA{\v{S}}IN POTOK}}: Milka Kukuruzovi{\'{c}} (Dragi{\'{c}}, Pa{\v{s}}in Potok, 1941, 7529908). 

\textbf{\underline{PETRINJA}}: Milan Andrija{\v{s}}evi{\'{c}} (Nikola, Umeti{\'{c}}, 1924, 7576723), Petar Bira{\v{c}} (Stevan, Umeti{\'{c}}, 1898, 7576724), Damjan Milojevi{\'{c}} (Stevan, Umeti{\'{c}}, 1916, 7576728), Du{\v{s}}an Preradovi{\'{c}} (Stanko, Umeti{\'{c}}, 1920, 7576725). 

\textbf{\underline{PETRINJA BANIJA}}: Dmitar Lemi{\'{c}} (-, {\v{Z}}ivaja, 1909, 7572977). 

\textbf{\underline{PETROVA GORA}}: Dragica Dragi{\'{c}} (Du{\v{s}}an, Mracelj, 1935, 7560259), Danica Kotaranin (Milan, Mracelj, 1941, 7560276), Milorad {\v{C}}u{\v{c}}kovi{\'{c}} (Petar, Staro Selo Topusko, 1926, 7558694), Milka Basara (-, Gornja Brusova{\v{c}}a, 1892, 7530578). 

\textbf{\underline{PE{\v{C}}UH          MA{\DJ}ARSKA}}: Hugo Stern (David, Zagreb, 1898, 7554010). 

\textbf{\underline{PLITVI{\v{C}}KA JEZERA}}: Dragi{\'{c}} Dudukovi{\'{c}} (Janko, Mracelj, 1941, 7560264). 

\textbf{\underline{PODGRADCI}}: Murat Bajri{\'{c}} (Ma{\v{s}}o, Kozaru{\v{s}}a, 1920, 7562621). 

\textbf{\underline{PODMILO{\v{C}}JE}}: Marko {\v{S}}ari{\'{c}} ({\DJ}or{\dj}e, {\'{C}}usine, 1880, 7574997), Spasoja Daki{\'{c}} (Marko, {\'{C}}usine, 1908, 7574996). 

\textbf{\underline{PODRAVSKA SLATINA}}: Milja Bijeli{\'{c}} (-, Sjeverovci, -, 7569301). 

\textbf{\underline{PODVOLIJAK}}: Cvijetin Kenji{\'{c}} (Mihajlo, Podvolujak, 1900, 7575713). 

\textbf{\underline{PODVOLUJAK}}: Rade Kenji{\'{c}} (Mitar, Podvolujak, 1922, 7575717). 

\textbf{\underline{POGLA{\DJ}EVO BOSNA}}: Petar Milkovi{\'{c}} (-, {\v{Z}}ivaja, 1897, 7572926). 

\textbf{\underline{POGLE{\DJ}EO BOSNA}}: Milan Joki{\'{c}} (-, {\v{Z}}ivaja, 1923, 7572917). 

\textbf{\underline{POLJSKA}}: Armin Greif (-, Zagreb, -, 7493418), Lavoslav Neumann (Aleksandar, Zagreb, 1869, 7553807). 

\textbf{\underline{POSLEDICE IZ LOGORA}}: Kata Dokman (Nikola, Strmen, 1897, 7555289). 

\textbf{\underline{POSLEDICE LOGORA}}: Stevo Savurdi{\'{c}} (Petar, Strmen, 1905, 7555331). 

\textbf{\underline{POTOK RAKOV}}: Mustafa Bjelavac (Avdo, Mostar, 1897, 7555084). 

\textbf{\underline{PO{\v{Z}}ARNICA}}: Te{\v{s}}o Stojanovi{\'{c}} (Pero, Kova{\v{c}}ica, 1915, 7561022). 

\textbf{\underline{PRAVCU}}: Jela Petrovi{\'{c}} (-, Irig, -, 7568611), Lenka Petrovi{\'{c}} (-, Irig, -, 7568609), Milan Jovanovi{\'{c}} (-, Irig, -, 7568594), Leposava Petrovi{\'{c}} (-, Irig, -, 7568619), Mirjana Petrovi{\'{c}} (-, Irig, -, 7568613), Jovan Jovanovi{\'{c}}-Petrovi{\'{c}} (-, Irig, -, 7568595), Sava Jovanovi{\'{c}}-Petrovi{\'{c}} (-, Irig, -, 7568596), Tina Jovanovi{\'{c}}-Petrovi{\'{c}} (-, Irig, -, 7568593), Trifun Jovanovi{\'{c}}-Petrovi{\'{c}} (-, Irig, -, 7568592). 

\textbf{\underline{PREKOPA, GLINA}}: Danica Pavlica (Mile, Majske Poljane, 1935, 7544249). 

\textbf{\underline{PRIJEDOR}}: Branko Babi{\'{c}} (Pero, Vitasovci, 1905, 7571020), Mile Kecman ({\DJ}oko, Gornji Orlovci, 1910, 7566217), Spasoja Koji{\'{c}} (Stevan, Bistrica, 1895, 7566294), Jovan Kuki{\'{c}} (Ilija, Bistrica, 1882, 7566324), Kosta Kuki{\'{c}} (Stojak, Bistrica, 1883, 7566331), Ostoja Kuki{\'{c}} (Zekan, Bistrica, 1897, 7566271), Rade Kuki{\'{c}} (Jovan, Bistrica, 1890, 7566323), Svetko Kuki{\'{c}} (Mijat, Bistrica, 1902, 7566274), Du{\v{s}}an Lazarevi{\'{c}} (Mikan, Lje{\v{s}}ljani, 1923, 7570885), Ilija Mami{\'{c}} (Ostoja, Bistrica, 1913, 7566329), Jovan Mami{\'{c}} (Ostoja, Bistrica, 1914, 7566314), Vaskrsija Mami{\'{c}} (Ostoja, Bistrica, 1901, 7566313), Risto Mili{\v{c}}evi{\'{c}} (Gavro, Bistrica, 1885, 7566282), Radivoj Nikoli{\'{c}} (Nikola, Bistrica, 1928, 7566278), Du{\v{s}}an Nini{\'{c}} (Pejo, Bistrica, 1906, 7566302), Pane Nini{\'{c}} (Stojko, Bistrica, 1894, 7566300), Gavro Pani{\'{c}} (Petar, Trgovi{\v{s}}te, 1889, 7570999), Ilija Pani{\'{c}} (Petar, Trgovi{\v{s}}te, 1881, 7570996), Stevo Pani{\'{c}} (Jovan, Trgovi{\v{s}}te, 1921, 7570997), Stevo Paspalj (Kostadin, Bistrica, 1926, 7566326), Branko Tadi{\'{c}} (Jovo, Bistrica, 1912, 7566307), Luka Teji{\'{c}} (Marko, Lamovita, 1890, 7566341), Du{\v{s}}an Vu{\v{c}}kovac ({\DJ}uro, Bistrica, 1900, 7566322), Ljuban Zori{\'{c}} (Mile, Gornji Jelovac, 1913, 7569368), Vaskrsija Zrni{\'{c}} (Milo{\v{s}}, Bistrica, 1888, 7566283), {\DJ}ura{\dj} Zrni{\'{c}} (Milo{\v{s}}, Bistrica, 1885, 7566284), Vuka{\v{s}}in {\v{S}}arac (Jovo, Svodna, 1923, 7570977), Stevo {\v{S}}obot (Ilija, Gornji Orlovci, 1910, 7566413), Du{\v{s}}an {\v{Z}}ivkovi{\'{c}} (Simo, Trgovi{\v{s}}te, 1903, 7570991), Milan Babi{\'{c}} ({\DJ}ura{\dj}, Radomirovac, 1925, 7507971), {\DJ}ura{\dj} Babi{\'{c}} (Mile, Radomirovac, 1889, 7507973), Jovo Dra{\v{z}}i{\'{c}} (Stojan, Radomirovac, 1913, 7507975), Milan Stankovi{\'{c}} (Jovo, Radomirovac, 1922, 7507981), Marko Tomi{\'{c}} (Trivo, Radomirovac, 1886, 7507983), Lazar Vujasin (Jovan, Radomirovac, 1922, 7507984), Ostoja Nini{\'{c}} (Pajo, Bistrica, 1904, 7566301), Milan Kne{\v{z}}evi{\'{c}} (Aleksa, Bistrica, 1910, 7566308), Jovan Radanovi{\'{c}} (Trivo, Radomirovac, 1887, 7507979), Luka Radanovi{\'{c}} (Trivo, Radomirovac, 1892, 7507980), Nikola Jankovi{\'{c}} (Trivun, Dera, 1890, 7566238), Blagoja Mudrini{\'{c}} (Stojan, Gornji Jelovac, 1900, 7566191), Milja Rodi{\'{c}} (Vaso, Ko{\v{s}}u{\'{c}}a, 1920, 7557243). 

\textbf{\underline{PRNJAVOR}}: Matija Poropati{\'{c}} (Ivan, Grabovac Banski, 1900, 7566970), Du{\v{s}}an Te{\v{s}}i{\'{c}} (Savo, Donja Ilova, 1921, 7567542), Radomir Simi{\'{c}} (Cvijan, Grabik Ilova, 1921, 7567539). 

\textbf{\underline{PSUNJ}}: Draginja Vlajsavljevi{\'{c}} (Branko, Medari, 1923, 7533823). 

\textbf{\underline{PUCARI}}: Milka Dimi{\'{c}} (Petar, Pucari, 1872, 7558287), Rosa Misira{\v{c}}a (-, Demirovac, -, 7556290). 

\textbf{\underline{RAB}}: Cadik Salom (Salamon, Sarajevo, 1895, 7519752), Gabriel Finci (Jahiel, Sarajevo, 1928, 7517298). 

\textbf{\underline{RADINA SLAVONIJ}}: Veljko Babi{\'{c}} (Mihajlo, Donja Lepenica, 1900, 7575119). 

\textbf{\underline{RASTO{\v{S}}NICA}}: Lako {\DJ}oki{\'{c}} (Jovan, Rasto{\v{s}}nica, 1921, 7560990), Branko Miljanovi{\'{c}} (Miko, Rasto{\v{s}}nica, 1924, 7560992). 

\textbf{\underline{RAVNI GAJ BOSNA}}: Pero Goji{\'{c}} (-, {\v{Z}}ivaja, 1918, 7572914). 

\textbf{\underline{RAVNICE}}: Jovica Majki{\'{c}} ({\DJ}ika, Ravnice, 1883, 7508005). 

\textbf{\underline{RAZBOJ}}: Stojan Kasagi{\'{c}} (Lazo, Razboj Ljev{\v{c}}anski, 1914, 7575125). 

\textbf{\underline{RA{\DJ}ENOVCI}}: Lazar Zastavnikovi{\'{c}} (Gli{\v{s}}o, Raj{\v{c}}i{\'{c}}i, 1934, 7555999), Saja Ra{\dj}enovi{\'{c}} (Savo, Ra{\dj}enovci, 1924, 7555904). 

\textbf{\underline{RELJEVO-RALJOVA}}: Todor Sokolovi{\'{c}} ({\DJ}or{\dj}e, Sarajevo, 1907, 7519804). 

\textbf{\underline{RILA{\'{C}}E}}: Dragoljub Konjevi{\'{c}} (Petar, Crkveni Bok, 1911, 7555341). 

\textbf{\underline{RIPA{\v{C}}"ZABORE"}}: Desa Durakovi{\'{c}} (Marko, Gorjevac, 1926, 7574899). 

\textbf{\underline{RIPO{\v{C}}KI KLANAC}}: Rajko Crnojevi{\'{c}} ({\DJ}uro, Veliko Kr{\v{c}}evo, 1926, 7545720). 

\textbf{\underline{ROD{\v{Z}}ANIK}}: Petar Mir{\v{c}}i{\'{c}} (-, Vrbovljani, 1890, 7554629). 

\textbf{\underline{ROGATICA}}: Ljubo Zari{\'{c}} (Jovo, Prisoje, 1916, 7574776). 

\textbf{\underline{ROMANIJA}}: Nikola Popovi{\'{c}} (Vaso, Orahova, 1922, 7573812). 

\textbf{\underline{RUMA}}: Rade Dragi{\v{s}}i{\'{c}} (Pero, Br{\v{s}}adin, 1922, 7540807). 

\textbf{\underline{SA. MOST}}: Milan Kurjak (Stevan, Podlug, -, 7574664). 

\textbf{\underline{SACHSENHAUSEN}}: Ljubica {\v{S}}alindrija (Ostoja, Mlaka, 1927, 7547492). 

\textbf{\underline{SANICA}}: Jovo Vu{\v{c}}kovi{\'{c}} (Ostoja, Sanica, 1902, 7574946), Jovan Petrovi{\'{c}} (Mile, Sanica, 1896, 7574915). 

\textbf{\underline{SANSKI MOST}}: Meho {\v{S}}ljivar (Omer, Sanica Donja, 1893, 7574952). 

\textbf{\underline{SARAJEVO}}: {\v{S}}alom Albahari (Salamon, Sarajevo, 1914, 7516164), Meho Dedovi{\'{c}} (Selim, Turovi, 1910, 7562731), Nusret Had{\v{z}}i{\'{c}} (Mehmed, Mostar, 1915, 7562562), Danilo Krstovi{\'{c}} (Mitar, Kijevo, 1907, 7575358), Branko Miljanovi{\'{c}} (Filip, Sarajevo, 1923, 7518742), Alija Mu{\v{s}}i{\'{c}} (Hasan, Ljubu{\v{s}}ki, 1906, 7562395), Borislav Poluga (Todor, Sarajevo, 1921, 7576635), Vaso Vitkovi{\'{c}} (Petar, Trnovo, 1912, 7575359), An{\dj}a Kne{\v{z}}evi{\'{c}} (Jozo, Prozor, 1910, 7563159), Ljubica Riba{\v{s}} (Josip, Mostar, 1926, 7555140), Milena Grubanovi{\'{c}} ({\DJ}or{\dj}e, Mokro, 1924, 7512541), Branko Sila{\dj}i (Aco, Sokolac, 1922, 7576147), Du{\v{s}}an Petrovi{\'{c}} (Konstantin, Sarajevo, 1887, 7576111), Sadik Sunagi{\'{c}} (Hilmo, Konjic, 1927, 7511824), Luka Malenica (Jure, Posu{\v{s}}je, 1906, 7575411), Aleksandar Savi{\'{c}} (Radovan, Travnik, 1923, 7575255), Drago Savari{\'{c}} (Ferdo, Sarajevo, 1920, 7576275), Sead Hromi{\'{c}} (Hilmija, Sarajevo, 1896, 7562692), Ivan Tvrtkovi{\'{c}} (Aleksandar, Kre{\v{s}}evo, 1906, 7576639), Faik Pinjo (Omer, Sarajevo, 1919, 7562716), Radoslav Buha (Simo, Fojnica, 1903, 7575361), Mehmed Ferhatovi{\'{c}} (Bajro, Sarajevo, 1900, 7562665), Rade Kova{\v{c}}evi{\'{c}} (Aleksa, Novi Grad, 1908, 7576645), Du{\v{s}}an Savi{\'{c}} (Radovan, Travnik, 1925, 7575257), Zorica {\v{C}}erkez (Ljubomir, Sarajevo, 1926, 7575213), Borislav {\v{S}}arovi{\'{c}} (Vaso, Sarajevo, 1915, 7576155), Sulejman Bajraktarevi{\'{c}} (Ibrahim, Sarajevo, 1919, 7516770), {\v{C}}edomir Savi{\'{c}} (Radovan, Travnik, 1921, 7575256), D{\v{z}}emal Tanovi{\'{c}} (Huso, Gacko, 1913, 7576272), Smail {\DJ}iho (Mehmed, Ostro{\v{z}}ac, 1921, 7574215), Osman {\DJ}umhur (-, Konjic, -, 7576252), Dragutin Japec (Ignac, Jak{\v{s}}inec, 1920, 7559546), Jahiel-Musehaj Kamhi (-, Sarajevo, -, 7568225), Hasan Nal{\v{c}}ad{\v{z}}i{\'{c}} (Avdo, Sarajevo, 1919, 7574749). 

\textbf{\underline{SELO MALE{\v{S}}I{\'{C}}}}: Blago Luki{\'{c}} (Stevan, Tr{\v{s}}i{\'{c}}, -, 7569023). 

\textbf{\underline{SENJ}}: {\DJ}uro Kne{\v{z}}evi{\'{c}} (Ilija, {\v{Z}}uta Lokva, 1924, 7557564). 

\textbf{\underline{SINJE}}: Julika Adribo{\v{z}}i{\'{c}} (Petar, Svinica, 1882, 7576709). 

\textbf{\underline{SISAK}}: Golubica Babi{\'{c}} (Golub, Gornja Jurkovica, 1938, 7502704), Kata Babi{\'{c}} (Nikola, Topolovica, 1895, 7576527), Savka Babi{\'{c}} (Golub, Gornja Jurkovica, 1941, 7502705), Ana Basari{\'{c}} (Tanasije, Gornja Ra{\v{s}}enica, 1941, 7576416), Joco Ba{\v{c}}i{\'{c}} (Du{\v{s}}an, Kosovac, 1938, 7533721), Mira Bjelajac (Marko, Ga{\v{s}}nica, 1940, 7502186), Milan Blesi{\'{c}} (Marko, Krnete, 1925, 7575954), Stanko Bojad{\v{z}}ija (Milan, Dereza, 1936, 7564420), Stojan Bojad{\v{z}}ija (Milan, Dereza, 1937, 7564421), Nevenka Cviki{\'{c}} (Mirko, Bakinci, 1940, 7565275), Smilja Dardi{\'{c}} (Branko, Velika Peratovica, 1940, 7576542), Stoja Dragi{\v{s}}i{\'{c}} (Milan, Trebovljani, 1941, 7506181), Draginja Gligi{\'{c}} (Milan, Klekovci, 1919, 7498822), Borislav Gojkovi{\'{c}} (Lazo, Ga{\v{s}}nica, 1938, 7502276), Vojislav Gojkovi{\'{c}} (Lazo, Ga{\v{s}}nica, 1940, 7502278), Mirko Grbi{\'{c}} (Kosta, Raj{\v{c}}i{\'{c}}i, 1930, 7555934), Janja Ivanovi{\'{c}} ({\DJ}uka, Gornji Podgradci, 1936, 7503043), Milica Ivanovi{\'{c}} (Cvijo, Gornja Obrije{\v{z}}, 1935, 7525829), Mladen Ivanovi{\'{c}} ({\DJ}uka, Gornji Podgradci, 1941, 7503044), Bosa Jankovi{\'{c}} (Rade, Ga{\v{s}}nica, 1941, 7502330), Mila Korica (Savo, Kukunjevac, 1914, 7561185), Saveta Kosi{\'{c}} ({\DJ}uro, Kukunjevac, 1897, 7561214), Nada Kova{\v{c}}evi{\'{c}} ({\DJ}or{\dj}ija, Jablanica, 1935, 7504056), Grozda Laji{\'{c}} (Pavle, Milo{\v{s}}evo Brdo, 1940, 7504868), Dragomir Ljepojevi{\'{c}} (Savo, Jablanica, 1936, 7504118), Ivan Malko{\v{c}} (Ivan, Novska, 1912, 7555694), Luka Mandi{\'{c}} (Dragoljub, Turjak, 1940, 7506742), Mom{\v{c}}ilo Mandi{\'{c}} (Stoko, Ga{\v{s}}nica, 1933, 7573692), Cvijeta Markovi{\'{c}} (Pero, Ga{\v{s}}nica, 1941, 7502460), Drago Pureta (Stoko, Ga{\v{s}}nica, 1936, 7502511), Stana Pureta (Proko, Ga{\v{s}}nica, 1930, 7502509), Du{\v{s}}anka Rajakovi{\'{c}} (Jovan, Raj{\v{c}}i{\'{c}}i, 1931, 7556013), Milan Rajakovi{\'{c}} (Jovan, Raj{\v{c}}i{\'{c}}i, 1937, 7556014), Du{\v{s}}anka Ra{\dj}enovi{\'{c}} (Petar, Ra{\dj}enovci, 1938, 7555868), Ignjo Ra{\dj}enovi{\'{c}} (Petar, Ra{\dj}enovci, 1934, 7555869), Ljuban Ra{\dj}enovi{\'{c}} (Stevo, Ra{\dj}enovci, 1935, 7555883), Milka Ra{\dj}enovi{\'{c}} (Stevo, Ra{\dj}enovci, 1939, 7555897), Nikola Ra{\dj}enovi{\'{c}} (Joco, Ra{\dj}enovci, 1942, 7562072), Zorka Ra{\dj}enovi{\'{c}} (Milan, Ra{\dj}enovci, 1931, 7555919), Dragan Ron{\v{c}}evi{\'{c}} (Stojan, Lu{\v{z}}ani, 1939, 7504679), Draginja Samard{\v{z}}ija (Dmitar, Turjak, 1935, 7505762), Nada Savanovi{\'{c}} (Savo, Srednja Jurkovica, 1938, 7506097), Milorad Savi{\'{c}} (Rajko, Ga{\v{s}}nica, 1938, 7502554), Smilja Slijep{\v{c}}evi{\'{c}} (Du{\v{s}}an, Turjak, 1939, 7572378), Stoja Sovilj (Mirko, Ga{\v{s}}nica, 1940, 7502578), Zagorka Sovilj (Mirko, Ga{\v{s}}nica, 1938, 7502579), Rosa Spasojevi{\'{c}} (Du{\v{s}}an, Gornja Jurkovica, 1942, 7502833), Darinka Stani{\v{s}}ljevi{\'{c}} (Gligo, Turjak, 1940, 7572544), Vidosava Stani{\v{s}}ljevi{\'{c}} (Pavle, Turjak, 1933, 7506845), Dragica Suzi{\'{c}} (Ilija, Dragelji, 1930, 7572612), Mirko Suzi{\'{c}} (Ilija, Dragelji, 1934, 7572611), Branka Tend{\v{z}}eri{\'{c}} (Mile, Gornji Podgradci, 1933, 7503399), Radivoj Tend{\v{z}}eri{\'{c}} (Mile, Gornji Podgradci, 1936, 7503404), Vidosava Tend{\v{z}}eri{\'{c}} (Mile, Gornji Podgradci, 1939, 7572748), Stevo Tepi{\'{c}} (Jovo, Donji Cerovljani, 1942, 7576759), Radosava Timarac (Marjan, Sovjak, 1935, 7572754), Jovo Torbica (Milan, Gornja Ra{\v{s}}enica, 1937, 7576430), Pela Trbojevi{\'{c}} (Ilija, Gakovo, 1939, 7576378), Pejo Turudija (Simo, Bistrica, 1907, 7572775), Jula Vidi{\'{c}} (Stanko, Cremu{\v{s}}ina, 1893, 7576353), Mirko Vra{\v{c}}ar (Luka, Ga{\v{s}}nica, 1940, 7502675), Milja Vukota (Du{\v{s}}an, Trebovljani, 1936, 7506418), Desa Vukoti{\'{c}} (Mirko, Milo{\v{s}}evo Brdo, 1939, 7572855), Dragoja Zrni{\'{c}} (Rade, Grbavci, 1935, 7503726), Kosa {\v{C}}eki{\'{c}} (Tomo, Grbavci, 1935, 7503611), Miladin {\v{C}}iki{\'{c}} (Lazo, Ga{\v{s}}nica, 1936, 7502217), Milo{\v{s}} {\DJ}enadija (Dragoje, Mrakodol, 1900, 7565796), Mirko {\v{S}}e{\v{s}}um (Milovan, Ga{\v{s}}nica, 1929, 7502599), {\v{Z}}ivko {\v{S}}inik (Mirko, Cimiroti, 1937, 7573964), Gojko {\v{S}}mitran (Lazo, Turjak, 1937, 7572655), Jovan {\v{S}}ok{\v{c}}evi{\'{c}} ({\DJ}ura{\dj}, Turjak, 1933, 7506905), Ru{\v{z}}a {\v{S}}ok{\v{c}}evi{\'{c}} (Uro{\v{s}}, Kijevci, 1940, 7504529), Zorka {\v{S}}ok{\v{c}}evi{\'{c}} (Uro{\v{s}}, Kijevci, 1938, 7504531), Rade {\v{S}}ukalo (Jovan, Turjak, 1940, 7572710), Omer Ahmeta{\v{s}}evi{\'{c}} (-, Banja Luka, -, 7566183), Kata Bi{\v{s}}kupovi{\'{c}} (-, Bestrma, -, 7549953), Branka Ciganovi{\'{c}} (-, Kukunjevac, 1937, 7574381), Milan Kotur (-, Ga{\v{s}}nica, 1937, 7502340), Draga Maleti{\'{c}} (-, Kukunjevac, 1939, 7526087), Pava Parmak (-, Velika Peratovica, -, 7576548), Milka Petrovi{\'{c}} (-, Cremu{\v{s}}ina, -, 7576328), Milka Ra{\dj}enovi{\'{c}} (Mirko, Ra{\dj}enovci, 1939, 7562071), Nikola Ra{\dj}enovi{\'{c}} (Vaso, Ra{\dj}enovci, 1941, 7562073), Ana Savi{\'{c}} (-, Ga{\v{s}}nica, 1939, 7502530), Saveta {\v{S}}tula (-, Kukunjevac, 1918, 7574393), Cveta Kukrika (Milo{\v{s}}, Mrakodol, 1920, 7507873), Ljeposava {\v{C}}iki{\'{c}} (Pero, Ga{\v{s}}nica, 1938, 7502237), Nikola Bojad{\v{z}}ija (Milan, Dereza, 1941, 7525664), Kosana Gon{\v{c}}in ({\DJ}or{\dj}ije, Sovjak, 1930, 7505931), Damjan Lipovac (Petar, Turjak, 1939, 7506736), Marica Ljepojevi{\'{c}} (Savo, Jablanica, 1934, 7504119), Marko Ljepojevi{\'{c}} (Savo, Jablanica, 1938, 7569321), Vidosava Majstorovi{\'{c}} (Mirko, Gornja Jurkovica, 1935, 7502813), Vasilija Miji{\'{c}} (Simo, Hrgar, 1912, 7574904), Stana Milankovi{\'{c}} (Kosta, Bo{\v{z}}i{\'{c}}i, 1936, 7497502), Vidoja Milovanovi{\'{c}} (Marko, Cerovljani, 1940, 7501415), Milan Popovi{\'{c}} (Stevo, Trebovljani, 1930, 7506301), {\DJ}uro Popovi{\'{c}} (Stevo, Trebovljani, 1933, 7506304), Ljubica Savi{\'{c}} (Bo{\v{s}}ko, Milo{\v{s}}evo Brdo, 1940, 7504936), Bojana Simi{\'{c}} (Jovo, Dragovi{\'{c}}, 1931, 7556924), Marko Vra{\v{c}}ar (Luka, Ga{\v{s}}nica, 1934, 7502672), Marko {\v{C}}iki{\'{c}} (Pero, Ga{\v{s}}nica, 1934, 7502238), Mirko {\v{C}}iki{\'{c}} (Pero, Ga{\v{s}}nica, 1934, 7502239), Savo {\v{S}}inik (Risto, Gornji Podgradci, 1924, 7503379), Mladen {\v{S}}obot (Mile, Jaruge, 1938, 7509271), Veselka {\v{S}}ok{\v{c}}evi{\'{c}} (Tedo, Kijevci, 1930, 7504520), Vukosava {\v{S}}ok{\v{c}}evi{\'{c}} (Tedo, Kijevci, 1938, 7504521), Stojan Trifunovi{\'{c}} (Vasilije, Vinska, 1925, 7565877), Draginja Kukavica (Pero, Gornje Vodi{\v{c}}evo, 1921, 7570863), Milka Kukavica (Pero, Gornje Vodi{\v{c}}evo, 1920, 7570862), Mirko Mandi{\'{c}} (Stoko, Ga{\v{s}}nica, 1930, 7502442), Ljuba Koji{\'{c}} (Gavrilo, Raj{\v{c}}i{\'{c}}i, 1938, 7555961), Rosa Toma{\v{s}} (Miloja, Brekinja, 1923, 7574240), Savka Toma{\v{s}} (Miloja, Brekinja, 1924, 7574241), Gojko {\DJ}uki{\'{c}} (Stevan, Gornje Jame, 1912, 7562031), {\DJ}uja {\v{Z}}ivkovi{\'{c}} (Ljuboja, Kukunjevac, 1901, 7561269), Mira Devi{\'{c}} (-, Velika Barna, 1940, 7523838), Branko Savi{\'{c}} (Milo{\v{s}}, Raj{\v{c}}i{\'{c}}i, 1939, 7556042), Mileva Jefteni{\'{c}} (Mile, Grbavci, 1938, 7503640), Mikajlo Slijep{\v{c}}evi{\'{c}} (Du{\v{s}}an, Turjak, 1935, 7572377), Radovan Vurin (-, Verija, 1942, 7558791), Mira Aralica (Rajko, Ga{\v{s}}nica, 1939, 7502163), Zora Petrovi{\'{c}} (Blagoje, Gornja Jurkovica, 1939, 7572211), Vuka Sredojevi{\'{c}} (Uro{\v{s}}, Sovjak, 1937, 7572385), Mihajlo Ciganovi{\'{c}} (-, Kukunjevac, 1939, 7574382), Zdravko Grube{\v{s}}i{\'{c}} (Nikola, Trnovac, 1936, 7506490), Ljuba Kljai{\'{c}} (-, Velika Barna, 1941, 7523878), Draga Kula{\v{s}} ({\DJ}uro, Kukunjevac, 1925, 7561187), Mihajlo Majstorovi{\'{c}} (Mirko, Gornja Jurkovica, 1928, 7502810), Niko Milo{\v{s}}evi{\'{c}} (Jakov, Sovjak, 1940, 7506005), Du{\v{s}}a Mujadin (-, Mali Gr{\dj}evac, -, 7576464), Mihajlo Risti{\'{c}} (Marko, Grbavci, 1909, 7566168), Bogdan Stojni{\'{c}} (Rajko, Turjak, 1934, 7572409), Du{\v{s}}an Stojni{\'{c}} (Stanko, Vrba{\v{s}}ka, 1936, 7507272), Mara Stojni{\'{c}} (Milutin, Turjak, 1939, 7506876), {\DJ}or{\dj}e Trifunovi{\'{c}} (Mile, Vinska, 1942, 7565881), Petar Maruni{\'{c}} (Janko, Kukunjevac, 1935, 7528365), Dragan Sarajlija (-, Batinjani Pakra{\v{c}}ki, 1938, 7564565), Marica Sarajlija (-, Batinjani Pakra{\v{c}}ki, 1937, 7574359), Mihajlo Sarajlija (-, Batinjani Pakra{\v{c}}ki, 1934, 7574360), Ljubomir Vukajlovi{\'{c}} (Petar, Jablan, 1920, 7565248), Vaso Ra{\v{s}}kovi{\'{c}} (Vaso, Krnete, 1938, 7508607), Cvijeta {\v{S}}ukalo (Jovan, Turjak, 1930, 7506924). 

\textbf{\underline{SLABINJA}}: Ljuba Bo{\v{z}}i{\'{c}} (-, Slabinja, 1928, 7557494). 

\textbf{\underline{SLAVONSKA PO{\v{Z}}EGA}}: Spasoje Aleksi{\'{c}} (Petar, Lu{\v{z}}ani Novi, 1914, 7511014), Savo Desan{\v{c}}i{\'{c}} (Nikola, Donji Podgradci, 1938, 7501730), Vaso Desan{\v{c}}i{\'{c}} (Nikola, Donji Podgradci, 1934, 7501732), Milan Gali{\'{c}} (Ostoja, Gornja Jurkovica, 1938, 7502744), Jefto Gligi{\'{c}} (Milutin, Milo{\v{s}}evo Brdo, 1939, 7504827), Janja Janjetovi{\'{c}} (Stanoje, Donji Podgradci, 1937, 7501792), Du{\v{s}}anka Klincov (Stevo, Elezagi{\'{c}}i, 1932, 7502144), Milenko Klincov (Milo{\v{s}}, Elezagi{\'{c}}i, 1939, 7502143), Nikola Kotur (Stojan, Ga{\v{s}}nica, 1940, 7502349), Gojko Lalo{\v{s}} (Vaso, Gornja Jurkovica, 1935, 7502781), Milan Lalo{\v{s}} (Vaso, Gornja Jurkovica, 1937, 7502784), Mirko Lalo{\v{s}} (Milan, Gornja Jurkovica, 1941, 7502764), Stevka Milakovi{\'{c}} (Du{\v{s}}an, Elezagi{\'{c}}i, 1938, 7502150), Dmitra Risti{\'{c}} (Branko, Miljevi{\'{c}}i, 1934, 7505062), Bo{\v{z}}o Suboti{\'{c}} (Boro, Vrba{\v{s}}ka, 1931, 7507278), Jela Suboti{\'{c}} (Ostoja, Vrba{\v{s}}ka, 1938, 7507305), Miladin Vidovi{\'{c}} (Tomo, Ga{\v{s}}nica, 1933, 7502659), Ranko Vidovi{\'{c}} (Lazar, Ga{\v{s}}nica, 1939, 7502654), Ilija {\v{Z}}ivkovi{\'{c}} (Risto, Elezagi{\'{c}}i, 1900, 7572474), Vaso Romi{\'{c}} (-, Kru{\v{s}}evo, 1872, 7564826), Kosa Desan{\v{c}}i{\'{c}} (Nikola, Donji Podgradci, 1938, 7501729), Mile Milakovi{\'{c}} (Du{\v{s}}an, Elezagi{\'{c}}i, 1940, 7502149), Stojan {\DJ}uki{\'{c}} (Milan, Elezagi{\'{c}}i, 1932, 7502136), Du{\v{s}}an {\v{S}}inik (Stojan, Milo{\v{s}}evo Brdo, 1937, 7504984), Bosiljka Kuki{\'{c}} (Simo, Sunja, 1914, 7550651), Milutin Janjetovi{\'{c}} (Nikola, Donji Podgradci, 1936, 7501786), Radovan Jerini{\'{c}} (Risto, Barica, 1924, 7510153), Vida Lalo{\v{s}} (Vlado, Gornja Jurkovica, 1937, 7573662), Vuka Lalo{\v{s}} (Vlado, Gornja Jurkovica, 1939, 7573661), Mira Suboti{\'{c}} (Ostoja, Vrba{\v{s}}ka, 1937, 7507306), Stanoje Samard{\v{z}}ija (Milo{\v{s}}, Turjak, 1942, 7505775). 

\textbf{\underline{SLAVONSKI BROD}}: {\v{S}}imo {\DJ}akovi{\'{c}} (Franjo, Vrhovina, 1900, 7562795). 

\textbf{\underline{SLAVSKO POLJE}}: Milo{\v{s}} Gabri{\'{c}} (-, Slavsko Polje, -, 7571501). 

\textbf{\underline{SMEDEREVSKA PALANKA}}: Milo{\v{s}} Vukojevi{\'{c}} (-, Visoko, -, 7567603). 

\textbf{\underline{SMILJANSKO POLJE}}: Marija Lemai{\'{c}} (Stevo, Smiljan, 1870, 7528443). 

\textbf{\underline{SMREKA}}: Bo{\v{z}}o Gudalo (Nikola, Donja Trebeu{\v{s}}a, 1920, 7575451). 

\textbf{\underline{SMREKE}}: Marko Gudalo (Simo, Donja Trebeu{\v{s}}a, 1912, 7575452), Mi{\'{c}}o Gudalo (Jovo, Donja Trebeu{\v{s}}a, 1919, 7575449), Pero Gudalo (Pero, Donja Trebeu{\v{s}}a, 1890, 7575450), Simo Gudalo (Simo, Donja Trebeu{\v{s}}a, 1910, 7575453), Pero Kezija (Stanko, Donja Trebeu{\v{s}}a, 1901, 7575459), Simo Kezija (Stanko, Donja Trebeu{\v{s}}a, 1910, 7575457), Stojan Kezija (Vaso, Donja Trebeu{\v{s}}a, 1907, 7575458), {\v{Z}}ivko Kezija (Janko, Donja Trebeu{\v{s}}a, 1920, 7575454), Ilija Poletan (Jovo, Donja Trebeu{\v{s}}a, 1907, 7575466), Jovo Poletan (Ilija, Donja Trebeu{\v{s}}a, 1875, 7575468), Ljubo Poletan (Jovo, Donja Trebeu{\v{s}}a, 1912, 7575467), Mile Poletan ({\DJ}uro, Donja Trebeu{\v{s}}a, 1906, 7575465), Mi{\'{c}}o Poletan (Bo{\v{z}}o, Donja Trebeu{\v{s}}a, 1918, 7575463), Stojan Poletan (Bo{\v{z}}o, Donja Trebeu{\v{s}}a, 1908, 7575464), Jovo Tegeltija (Jovo, Donja Trebeu{\v{s}}a, 1916, 7575462), Nedeljko {\DJ}uri{\'{c}} (Jovo, Donja Trebeu{\v{s}}a, 1913, 7575445), Pane {\DJ}uri{\'{c}} (Jovo, Donja Trebeu{\v{s}}a, 1916, 7575446), Aleksa Kezija ({\DJ}or{\dj}e, Donja Trebeu{\v{s}}a, 1919, 7575456), {\DJ}or{\dj}ija Tegeltija (Nikola, Donja Trebeu{\v{s}}a, 1917, 7575461). 

\textbf{\underline{SMUKE-SLIMENA}}: Nedeljko Male{\v{s}} (Cvijo, Gornje Kr{\v{c}}evine, 1898, 7575471). 

\textbf{\underline{SOKOLI{\v{S}}TE}}: Mirko Stjepanovi{\'{c}} (Ilija, Sokoli{\v{s}}te, 1915, 7508032). 

\textbf{\underline{SONICA}}: Milan Kurid{\v{z}}a (Mile, Prisjeka Donja, 1895, 7574940). 

\textbf{\underline{SPLIT}}: Menahem Papo (Moise, Sarajevo, 1915, 7519277). 

\textbf{\underline{SREFLIJE}}: Jelka {\v{S}}taki{\'{c}} (Blagoje, Sreflije, 1918, 7558413). 

\textbf{\underline{SREM}}: Armando Paparela (Vjekoslav, Split, 1912, 7562926). 

\textbf{\underline{SREMSKA MITROVICA}}: Antun Duha{\v{c}} (Josip, Prekopakra, 1908, 7526568), Jelena Me{\dj}e{\v{s}}ki (-, Irig, 1909, 7568599), Bojana Novakovi{\'{c}} (Milovan, Ne{\v{s}}tin, 1887, 7493862), {\v{Z}}ivojin Jovanovi{\'{c}} (Milivoj, Divo{\v{s}}, 1919, 7494639), Branko Mari{\'{c}} (Ivan, Vi{\v{s}}nji{\'{c}}evo, 1930, 7496470), Bo{\v{z}}idar Bedeni{\v{c}}ar (-, Zagreb, 1922, 7569095), Stefanija-Mila Mehand{\v{z}}i{\'{c}}-Stoj{\v{s}}i{\'{c}} (Jovan, Be{\v{s}}ka, 1906, 7568067). 

\textbf{\underline{SREMSKA RA{\v{C}}A}}: Bogdan Obradovi{\'{c}} (Panta, Bosut, 1936, 7494464). 

\textbf{\underline{STAMBOL{\v{C}}I{\'{C}}}}: Mladen Prodanovi{\'{c}} (Vojislav, Sarajevo, 1928, 7575252). 

\textbf{\underline{STARA}}: Teodor Srbljanin (-, Donji Hrastovac, 1898, 7573127). 

\textbf{\underline{STARI MAJDAN}}: Fevzija Muji{\'{c}} (Fehim, Stari Majdan, 1917, 7509690). 

\textbf{\underline{STAZA}}: Jovan Dragojevi{\'{c}} (-, Donji Hrastovac, 1910, 7573176). 

\textbf{\underline{STRELJAN}}: Ivan {\DJ}ermanovi{\'{c}} (Ilija, Rajevo Selo, 1904, 7575863). 

\textbf{\underline{STRIGOVA}}: Cvijo Marin (Mikana, Zovik, 1930, 7508114), {\DJ}or{\dj}o Mandi{\'{c}} (Stojan, Strigova, 1890, 7558586). 

\textbf{\underline{STRMEN}}: Petra Arbutina (Ilija, Strmen, 1884, 7555276), Stevan La{\dj}evi{\'{c}} (Jovan, Strmen, 1895, 7555308), {\DJ}uro Maslovara (Jovan, Strmen, 1903, 7555312), Matija Radovanovi{\'{c}} (Stevan, Strmen, 1890, 7555319), Jelica La{\dj}evi{\'{c}} (Stevan, Strmen, 1863, 7555309), Milo{\v{s}} Arbutina (-, Strmen, -, 7555275). 

\textbf{\underline{STRUG}}: Ilija Laji{\'{c}} (Milan, Bistrica, 1938, 7561682), Branka Zrni{\'{c}} (Radovan, Gornji Podgradci, 1930, 7503540), Savka Zrni{\'{c}} (Radovan, Gornji Podgradci, 1933, 7503544), Mirko Zvijerac (Milan, Bistrica, 1932, 7501035), Anka Zvijerac (Milan, Bistrica, 1938, 7501033), Sava Zvijerac (Milan, Bistrica, 1939, 7501036). 

\textbf{\underline{STRUG-SLAVONIJA}}: Stoja Mijatovi{\'{c}} (Mihajlo, Jablanica, 1922, 7504157). 

\textbf{\underline{STRUGE}}: Mithat Muminovi{\'{c}} (Nail, Struge, 1932, 7559927). 

\textbf{\underline{STUPOVA{\v{C}}A}}: Petar Pucarin (Ilija, Stupova{\v{c}}a, 1900, 7565353). 

\textbf{\underline{SUNJA}}: Milan Kukavica (-, Crkveni Bok, 1940, 7549453). 

\textbf{\underline{SUNJA SISAK}}: Milka Radojevi{\'{c}} (-, {\v{C}}etvrtkovac, 1924, 7564895). 

\textbf{\underline{SUNJE}}: Simo Adribo{\v{z}}i{\'{c}} (Ilija, Svinica, 1880, 7576708). 

\textbf{\underline{SUTJESKA}}: Jozef Levi (Benjamin, Sarajevo, 1924, 7518256), Ibrahim Raljevi{\'{c}} (Salih, Mostar, 1918, 7562586). 

\textbf{\underline{SUVA{\v{C}}A BOSNA}}: Stevo Paprika (-, {\v{Z}}ivaja, 1920, 7572940). 

\textbf{\underline{SVETOBLA{\v{Z}}JE}}: Bo{\v{z}}o Veselinovi{\'{c}} (Mile, Golubi{\'{c}}, 1912, 7551080). 

\textbf{\underline{SVINICA}}: Nikola Adribo{\v{z}}i{\'{c}} (Ilija, Svinica, 1904, 7576712), Vladimir Brki{\'{c}} (Nikola, Svinica Krstinjska, 1925, 7560323), Mara Adribo{\v{z}}i{\'{c}} (Nikola, Svinica, 1937, 7576713), Milja Adribo{\v{z}}i{\'{c}} (Nikola, Svinica, 1936, 7576714), Vukosava Adribo{\v{z}}i{\'{c}} (Milan, Svinica, 1937, 7576710), Stojan Borota (Kuzman, Veliko Kr{\v{c}}evo, 1869, 7576717). 

\textbf{\underline{TOBOLI{\'{C}}}}: Mile Sekuli{\'{c}} (-, Toboli{\'{c}}, -, 7571266). 

\textbf{\underline{TOPOLOVICA}}: Nikola Babi{\'{c}} (-, Topolovica, -, 7576528). 

\textbf{\underline{TOPUSKO}}: Slavko Hirschl (-, Zagreb, 1879, 7563859). 

\textbf{\underline{TRAVNIK}}: Milan Drmi{\'{c}} (Stanko, {\v{C}}osi{\'{c}}i, 1920, 7575422), Simo Drmi{\'{c}} (Mitar, {\v{C}}osi{\'{c}}i, 1900, 7575424), Danilo Komljenovi{\'{c}} (Risto, {\v{C}}osi{\'{c}}i, 1915, 7575429), Aleksa Mali{\'{c}} (Simo, {\v{C}}osi{\'{c}}i, 1905, 7575434), Milan Mali{\'{c}} ({\DJ}uro, {\v{C}}osi{\'{c}}i, 1902, 7575432), Risto Mali{\'{c}} ({\DJ}uro, {\v{C}}osi{\'{c}}i, 1905, 7575433), Milo{\v{s}} Vukovi{\'{c}} ({\DJ}or{\dj}e, Bugojno, 1900, 7575534), Pero {\DJ}uri{\'{c}} (Mitar, Karaula, 1892, 7575472), Nenad Vaski{\'{c}} (Simo, {\v{C}}osi{\'{c}}i, 1920, 7575438), Petar Topalovi{\'{c}} (Jovo, {\v{C}}osi{\'{c}}i, 1900, 7575437), Risto Topalovi{\'{c}} (Jovo, {\v{C}}osi{\'{c}}i, 1908, 7575436), Nikola Maksimovi{\'{c}} (Niko, {\v{C}}osi{\'{c}}i, 1915, 7575430), Pero Maksimovi{\'{c}} (Stanko, {\v{C}}osi{\'{c}}i, 1905, 7575435). 

\textbf{\underline{TRGOVI}}: Mika Joka (-, Trgovi, -, 7558040). 

\textbf{\underline{TRNOVA}}: Branko Milovanovi{\'{c}} (Ilija, Crnjelovo Gornje, 1923, 7567558). 

\textbf{\underline{TRNOVO}}: Nikola {\'{C}}ur{\v{c}}i{\'{c}} (Danilo, Brodac Donji, 1919, 7575866). 

\textbf{\underline{TRNOVO         POLJSKA}}: Milan Latin{\v{c}}i{\'{c}} ({\DJ}uka, Krnete, 1913, 7575951). 

\textbf{\underline{TRST}}: Milo{\v{s}} Pepi{\'{c}} (Rade, Ma{\v{s}}i{\'{c}}i, 1923, 7572205). 

\textbf{\underline{TR{\v{S}}I{\'{C}}}}: Mi{\'{c}}o Simi{\'{c}} (Lazo, Tr{\v{s}}i{\'{c}}, -, 7569014), Nikola Zeki{\'{c}} (Marko, Tr{\v{s}}i{\'{c}}, -, 7569026), Krsto Joksimovi{\'{c}} (Mihajlo, Tr{\v{s}}i{\'{c}}, -, 7569025), Bojo Boji{\'{c}} (Tomo, Tr{\v{s}}i{\'{c}}, -, 7569029). 

\textbf{\underline{TURBE}}: Mile Leki{\'{c}} (Aleksa, Jazovac, 1925, 7573671). 

\textbf{\underline{TURJAK}}: Silvo Me{\v{z}}nari{\v{c}} (-, Metlika, 1913, 7563427). 

\textbf{\underline{TUTNJEVAC}}: Cvijetin Blagojevi{\'{c}} (Pero, Tutnjevac, 1914, 7560975), Lazar Gavri{\'{c}} (Gavro, Tutnjevac, 1923, 7560977), Stanko Milovanovi{\'{c}} (Cvijan, Tutnjevac, 1924, 7560980). 

\textbf{\underline{TUZLA}}: Blagoje Novi{\v{c}}i{\'{c}} (Savo, Kitovnice, 1909, 7575888), Milan Pajkanovi{\'{c}} (Jovan, Tuzla, 1911, 7574481), {\DJ}or{\dj}o Gotovac (Stevo, Kitovnice, 1919, 7575890), Trivko Novi{\v{c}}i{\'{c}} (Savo, Kitovnice, 1913, 7575889), Andrija Lebeni{\v{c}}ki (Andrija, Tuzla, 1876, 7575871). 

\textbf{\underline{U BORBAMA-NOV}}: Halid Lokvan{\v{c}}i{\'{c}} (Ahmet, Hrasnica, 1922, 7562352), Zaim Lokvan{\v{c}}i{\'{c}} (Ahmet, Hrasnica, 1920, 7562351). 

\textbf{\underline{U SARAJEVU}}: Red{\v{z}}o Pand{\v{z}}i{\'{c}} (-, Sarajevo, -, 7562719). 

\textbf{\underline{U SLAVONIJI}}: Nikola Grozdani{\'{c}} (-, Morovi{\'{c}}, -, 7568663). 

\textbf{\underline{UBIJEN OD STRANE UST}}: Krsto Markovi{\'{c}} (Marko, Predra{\v{z}}i{\'{c}}i, 1906, 7575642), Milivoje Markovi{\'{c}} (Krsto, Predra{\v{z}}i{\'{c}}i, 1923, 7575644), Jovo Pejkanovi{\'{c}} (Ilija, Predra{\v{z}}i{\'{c}}i, 1909, 7575635), Mitar Todorovi{\'{c}} (Savo, Predra{\v{z}}i{\'{c}}i, 1913, 7575646). 

\textbf{\underline{UBILE USTA{\v{S}}E}}: Jovan Popovi{\'{c}} (Zarija, Vlasenica, 1894, 7575378). 

\textbf{\underline{USKOCI}}: Evica Grubi{\v{s}}i{\'{c}} (-, Grubi{\v{s}}no Polje, 1885, 7572987), Ana {\v{S}}u{\'{c}}ur (Nikola, Grubi{\v{s}}no Polje, 1912, 7523510), Du{\v{s}}an Kalender (Savo, Uskoci, 1927, 7554869). 

\textbf{\underline{USTIKOLINA}}: Salih Delali{\'{c}} (Ibrahim, Ustikolina, 1908, 7562265). 

\textbf{\underline{UTOLICA}}: Du{\v{s}}an Kova{\v{c}}evi{\'{c}} (Adam, Lov{\v{c}}a, 1913, 7576678). 

\textbf{\underline{VARAD{\v{Z}}IN}}: Ivan Kranj{\v{c}}evi{\'{c}} (Marko, Rijeka, 1911, 7568437), Vladimir {\v{C}}eba{\v{s}}ek (-, Zvornik, 1907, 7576302), Jovo {\v{S}}ajatovi{\'{c}} (-, Mala Rijeka, -, 7565453). 

\textbf{\underline{VARE{\v{S}}}}: Oskar Hola{\v{c}}ek (Adolf, Vare{\v{s}}, 1925, 7512929). 

\textbf{\underline{VEL. PERATOVICA}}: Lazo Radakovi{\'{c}} (-, Gakovo, -, 7576386). 

\textbf{\underline{VEL. PISANICA}}: Tomo Paravina (Rade, Velika Pisanica, 1903, 7521182). 

\textbf{\underline{VELA LUKA}}: Leon Romano (Jozef, Sarajevo, 1908, 7519629). 

\textbf{\underline{VELIKA BARNA NA}}: Vlajko Divjak (Tomo, Velika Barna, 1929, 7576606). 

\textbf{\underline{VELIKA GRADUSA}}: Bo{\v{s}}ko {\v{Z}}ica (-, Velika Gradusa, 1916, 7550766). 

\textbf{\underline{VEZKOSE        KEZKOSE}}: Petar Tomi{\'{c}} (Miko, Crnjelovo Donje, 1910, 7567568). 

\textbf{\underline{VINKOVCI}}: Marko Bari{\v{s}}i{\'{c}} ({\DJ}uro, Vranovci, 1906, 7562418), Stevan Utje{\v{s}}inovi{\'{c}} (Dmitar, Novi Grabovac, 1904, 7555639), Marija Pa{\v{s}}kuljevi{\'{c}} (Janko, {\v{Z}}upanja, 1887, 7567685), Katica Male{\v{s}}evi{\'{c}}-{\v{S}}undukovi{\'{c}} (-, Irig, 1901, 7568598). 

\textbf{\underline{VIROVITICA}}: Milica Koji{\'{c}} (-, Klekovci, 1930, 7557124), Sava Velagi{\'{c}} (-, Cremu{\v{s}}ina, 1914, 7564758), Savo Velagi{\'{c}} (-, Cremu{\v{s}}ina, -, 7576352). 

\textbf{\underline{VITEZ}}: Stjepan Radoj{\v{c}}i{\'{c}} (Jovica, Boljani{\'{c}}, 1913, 7575801). 

\textbf{\underline{VITLOVSKA,}}: Kosta Balaban (-, Mr{\v{c}}evci, 1908, 7575939). 

\textbf{\underline{VOJSKOVA}}: {\DJ}uja {\DJ}uri{\'{c}} (Vaso, Dobrljin, 1911, 7567878), Nikola Popovi{\'{c}} (Marko, Vojskova, 1865, 7558967). 

\textbf{\underline{VOLARI}}: {\DJ}uro Ni{\v{s}}i{\'{c}} (Vaso, Volari, 1900, 7575084), Jovo Prnji{\'{c}} (Mitar, Volari, 1911, 7575074), Mitar Prnji{\'{c}} ({\DJ}uro, Volari, 1881, 7575075), Milan Prnji{\'{c}} (Mitar, Volari, 1909, 7575076). 

\textbf{\underline{VONJI{\'{C}}}}: Ljubica Jagrovi{\'{c}} (Du{\v{s}}an, Gornja Brusova{\v{c}}a, 1934, 7530167), Sofija Ma{\'{c}}e{\v{s}}i{\'{c}} (Tomo, Mracelj, 1883, 7560279), Mila Bu{\'{c}}an (-, Gejkovac, 1940, 7530081), Milan Paji{\'{c}} (Milovan, Jagrovac, 1900, 7560217). 

\textbf{\underline{VORIN}}: Milan Kolund{\v{z}}i{\'{c}} (Stevan, Stara Krivaja, 1920, 7561789). 

\textbf{\underline{VOZU{\'{C}}A}}: Simo Petrovi{\'{c}} (Vaskrsije, Vukovine, 1905, 7575672), Stevo Blagojevi{\'{c}} (Blagoje, Vozu{\'{c}}a, 1908, 7575775), Savo Jeli{\'{c}} (Jovo, Vukovine, 1921, 7511530), Milo{\v{s}} Njegomirovi{\'{c}} (Vaskrsije, Vukovine, 1920, 7575631), Branko Pavlovi{\'{c}} (Sarafijan, Vozu{\'{c}}a, 1898, 7575687), Branko Staki{\'{c}} (Jovo, Vukovine, 1924, 7575628), Milo{\v{s}} Vra{\v{c}}evi{\'{c}} (Mitar, Vukovine, 1911, 7575684), Milovan Smilji{\'{c}} (Mihajlo, Hrge, 1898, 7575633), Savo Petrovi{\'{c}} (Josip, Vukovine, 1900, 7575680). 

\textbf{\underline{VRACA SARAJEVO}}: Jovo Mihajlovi{\'{c}} (Ilija, Sarajevo, 1909, 7518733). 

\textbf{\underline{VRA{\v{Z}}I{\'{C}}}}: Ilija Kalaba (Lazo, Vra{\v{z}}i{\'{c}}, 1890, 7575041). 

\textbf{\underline{VRA{\v{Z}}JA DIVIZIJA}}: Dujko {\v{Z}}ivkovi{\'{c}} ({\DJ}uro, Donja Paklenica, 1914, 7575587). 

\textbf{\underline{VRELA DONJA}}: Vid {\DJ}uki{\'{c}} (-, Klakar Gornji, 1938, 7560189). 

\textbf{\underline{VRPOLJE}}: Damjan Kopanja (Mile, Lubovo, 1917, 7575018). 

\textbf{\underline{VRTLJINOVEC}}: Nikola Serti{\'{c}} (Ilija, Sinac, -, 7557576). 

\textbf{\underline{VUKOVAR}}: Vera Mr{\dj}enov (Du{\v{s}}an, Sremska Mitrovica, 1922, 7564949), Milo{\v{s}} Timi{\'{c}} ({\DJ}oka, Sremska Mitrovica, 1933, 7495026), Verica Vuj{\v{c}}i{\'{c}} (Branko, Ada{\v{s}}evci, 1936, 7495527), Zdenko Pollak (Ljudevit, Vukovar, 1920, 7560757). 

\textbf{\underline{ZAGORSKA SELA}}: Rudolf Lon{\v{c}}ar (-, Razvor, -, 7568377). 

\textbf{\underline{ZAGREB}}: Bosiljka Blanu{\v{s}}a (Du{\v{s}}an, Kalenderi, 1942, 7507763), Milovan Borjanovi{\'{c}} (Pavle, Sovjak, 1921, 7573328), Savo Brkanac ({\DJ}uka, Kukunjevac, 1935, 7526006), Ana Brkanovi{\'{c}} ({\DJ}uro, Gornja Obrije{\v{z}}, 1939, 7525817), Milica Brki{\'{c}} (Petar, Gejkovac, 1938, 7530071), {\v{Z}}ivko Buletinac (Srban, Pau{\v{c}}je, 1940, 7532083), Ljuba Daki{\'{c}} (Ilija, Kusonje, 1938, 7556445), Ljuban Daki{\'{c}} (Ilija, Kusonje, 1938, 7556446), Petra Daki{\'{c}} (Ilija, Kusonje, 1935, 7556447), {\DJ}uro Damjanovi{\'{c}} ({\DJ}uro, Kusonje, 1942, 7556450), Bogdan Gaji{\'{c}} (Milo{\v{s}}, Kusonje, 1938, 7556451), Pero Gavranovi{\'{c}} (Simo, Klakar Donji, 1939, 7565957), Svetozar Gligi{\'{c}} (Luka, Kusonje, 1941, 7556452), Rade Jovi{\'{c}} (Stanko, Zrinska, 1902, 7574533), Milan Lazi{\'{c}} (Spasoje, Nova Krivaja, 1940, 7561834), {\v{Z}}ivko Marjanovi{\'{c}} (Dragi{\'{c}}, Maljevac, 1930, 7529810), Bosiljka Markovi{\'{c}} (Mile, Stara Krivaja, 1940, 7522607), Jovanka Mika{\v{c}} (Pero, Klakar Gornji, 1940, 7566097), Stana Mika{\v{c}} (Pero, Klakar Gornji, 1942, 7566099), {\v{Z}}ivko Milijevi{\'{c}} (Mi{\'{c}}o, Pobr{\dj}ani, 1937, 7507937), Nevenka Mujadin (Mitar, Mali Gr{\dj}evac, 1933, 7576460), Du{\v{s}}an Novkovi{\'{c}} (Zorka, Lakta{\v{s}}i, 1920, 7575948), Branko Ostarija{\v{s}} ({\DJ}uro, Dugo Selo, 1909, 7571540), Adam Pe{\v{s}}ut (Vaso, Petrinja, 1932, 7549014), Zdravko Popovi{\'{c}} (Mile, Stara Krivaja, 1939, 7522627), Stevo Radovi{\'{c}} (Nikola, Mala Paukova, 1930, 7550052), Mileva Reljanovi{\'{c}} (Savo, Vojskova, 1941, 7500582), Petar Sarapa (Ostoja, La{\dj}evac, 1917, 7571741), {\DJ}uro Saro{\v{s}}evi{\'{c}} ({\DJ}uro, Ivankovo, 1917, 7568237), Branko Savatovi{\'{c}} (Jovan, Ostrvica, 1920, 7568411), Mustafa Skaka (Uzeir, Sarajevo, 1918, 7562700), {\DJ}ura{\dj} Trkulja (Vid, Vakuf, 1925, 7574014), Du{\v{s}}an Uji{\'{c}} (Ilija, Mali Gr{\dj}evac, 1936, 7576468), Milan Uji{\'{c}} (Ilija, Mali Gr{\dj}evac, 1938, 7576470), Uro{\v{s}} Vasiljevi{\'{c}} (Mitar, Rogolji, 1914, 7572791), Nada Vidakovi{\'{c}} (Zdravko, Jasenovac, 1941, 7546782), Dobrila Vidi{\'{c}} (Luka, Brusnica Mala, 1940, 7565738), Mara Vukman (Bogdan, Klakar Gornji, 1940, 7510638), Ilija {\v{C}}elica (Pero, Novoselci, 1928, 7499657), Milica {\v{S}}a{\v{s}}i{\'{c}} (Ilija, Me{\dj}e{\dj}a, 1923, 7499364), Dragica Bo{\v{z}}i{\'{c}} (-, Jurkovac, -, 7562462), Vaso Dra{\v{z}}i{\'{c}} (-, Pau{\v{c}}je, -, 7565512), Josip Han{\v{z}}ek (-, Karivaro{\v{s}}, 1902, 7559589), Milan Naran{\v{c}}i{\'{c}} (-, Veliko Nabr{\dj}e, 1942, 7565628), Milena Vidakovi{\'{c}} (-, Pau{\v{c}}je, -, 7565458), Simo Vrane{\v{s}} (Gligorije, Tu{\v{z}}evi{\'{c}}, -, 7557534), Miladin Vukomanovi{\'{c}} (-, {\v{C}}enkovo, -, 7565494), Vlajko {\v{Z}}ivkovi{\'{c}} (-, {\v{C}}enkovo, -, 7565508), Du{\v{s}}anka Uji{\'{c}} (Cvjetko, Mali Gr{\dj}evac, 1939, 7576469), Filip Bajevi{\'{c}} ({\DJ}uro, Mali Gr{\dj}evac, 1936, 7576511), Gojko Brki{\'{c}} (Milo{\v{s}}, Svinica Krstinjska, 1931, 7530581), Marica Crnobrnja (Vaso, Stara Krivaja, 1941, 7522589), Berta Heinrich (Ignatz, Bjelovar, 1886, 7520496), {\v{Z}}ivko To{\v{s}}i{\'{c}} (Ostoja, Milo{\v{s}}evo Brdo, 1929, 7502650), Ljuban Vidovi{\'{c}} (Pero, Gunjevci, 1938, 7556889), Muhamed Bilalbegovi{\'{c}} (Ahmet, Sanski Most, -, 7573389), {\'{C}}iro Dunki{\'{c}} (Pa{\v{s}}ko, {\v{S}}ibenik, 1921, 7558363), Ljuba {\v{S}}kori{\'{c}} (Ljuboja, Nova Krivaja, 1940, 7522369), Nada Crnobrnja (Mirko, Stara Krivaja, 1940, 7522588), Radmila Radovanovi{\'{c}} (Marko, Brusnica Mala, 1942, 7565950), Josip Fleischer (Josip, Brestovac Daruvarski, 1885, 7571557), Jovan Biljanovi{\'{c}} (Rajko, Bukovica Mala, 1942, 7510922), Bosa Damjanovi{\'{c}} (-, Kusonje, 1940, 7564433), Jovo Dodo{\v{s}} (Milo{\v{s}}, Hajti{\'{c}}, 1908, 7544210), Ru{\v{z}}a Radovanovi{\'{c}} (-, Jurkovac, -, 7562457), Bosa {\DJ}uri{\'{c}} (Aleksa, Brusnica Mala, 1941, 7510200), Spaso Ljutica (Jovan, Kalebovac, 1920, 7529956), Ljuba Pe{\v{s}}ut ({\DJ}uro, Lu{\v{s}}{\v{c}}ani, 1930, 7548807), Milorad {\v{S}}etki{\'{c}} (Dmitar, Trebovljani, 1940, 7506365), {\DJ}uro Banjeglav (Petar, Prokike, 1922, 7557566), Tomo Benakovi{\'{c}} (-, Gundinci, 1922, 7571449), Miroslav Juhn (Mavro, Podgora{\v{c}}, 1897, 7571267), Geza Oblath (Jakob, Vrbovec, 1882, 7563624), {\v{Z}}iga Krämer (Ljudevit, Zagreb, 1901, 7553649). 

\textbf{\underline{ZAJA{\v{C}}A}}: Stanoje Simi{\'{c}} (-, Lije{\v{s}}anj, 1919, 7568869). 

\textbf{\underline{ZAKOPA}}: {\DJ}uro Resanovi{\'{c}} (-, Zakopa, -, 7558057). 

\textbf{\underline{ZAVIDOVI{\'{C}}}}: Pero Bojani{\'{c}} (Milan, Zavidovi{\'{c}}i, 1920, 7575750). 

\textbf{\underline{ZAVIDOVI{\'{C}}I}}: Bukica Pilc (-, Zavidovi{\'{c}}i, -, 7567949). 

\textbf{\underline{ZELENGORA}}: Stojan Karan (Mitar, Ra{\v{s}}{\'{c}}ani, 1921, 7574777). 

\textbf{\underline{ZELICE}}: Stevo Mijatovi{\'{c}} (-, Drinja{\v{c}}a, 1920, 7568704). 

\textbf{\underline{ZELINA}}: Mitar Gavri{\'{c}} (Gajo, Zelina, 1880, 7514815). 

\textbf{\underline{ZELINJE}}: Petko Mijatovi{\'{c}} ({\DJ}oko, Tr{\v{s}}i{\'{c}}, -, 7569018), Risto Vukadinovi{\'{c}} (-, Zelinje, 1916, 7568847), Rado-Dobro Gvozdi{\'{c}} (-, Zelinje, 1908, 7568846). 

\textbf{\underline{ZEMUN}}: Petar Latinovi{\'{c}} (Jovan, Janjila, 1880, 7574676), Milan {\'{C}}aji{\'{c}} (Marko, Nepravdi{\'{c}}i, 1919, 7575267), Manojlo {\'{C}}aji{\'{c}} (Marko, Nepravdi{\'{c}}i, 1914, 7575266). 

\textbf{\underline{ZEMUN SAJMI{\v{S}}TE}}: Pajo Babi{\'{c}} (Jandrija, Bijakovac, 1881, 7555539), Stevo Babi{\'{c}} (Jovan, Rakovica, 1920, 7558311), Dragutin Baki{\'{c}} (Risto, Aginci, 1912, 7496970), Radivoj Barud{\v{z}}ija (Nikola, Mlaka, 1908, 7558411), Lazo Bates (Jovo, Petkovac, 1898, 7507885), Du{\v{s}}an Beri{\'{c}} (Teodor, Lamovita, 1914, 7566332), Nikola Bijeli{\'{c}} (Pero, Petrinja, 1880, 7507897), Ilija Bori{\'{c}} (Marko, Gornji Garevci, 1880, 7566234), Dragi{\'{c}} Brki{\'{c}} (Dragi{\'{c}}, Radmanovac, 1938, 7530477), Ilija Dragi{\v{s}}i{\'{c}} (Stevo, Ga{\v{s}}nica, 1900, 7573437), Milo{\v{s}} Gojkovi{\'{c}} (Petar, Ga{\v{s}}nica, 1898, 7573491), Stanoja Grahovac (Stojko, Lamovita, 1900, 7566813), Du{\v{s}}an Grgi{\'{c}} (Rade, Ga{\v{s}}nica, 1902, 7573507), Branko Gvozden (Vaso, Jo{\v{s}}ava, 1923, 7570870), Milan Ivanovi{\'{c}} (Zdravko, Beograd, 1908, 7493424), Mile Jovani{\'{c}} (Marko, Mala {\v{Z}}uljevica, 1887, 7507838), Salamon Kampos (Isak, Sarajevo, 1901, 7574608), Mile Katana (Petar, Miska Glava, 1906, 7566355), Dragan Kne{\v{z}}evi{\'{c}} (Mile, Gornje Vodi{\v{c}}evo, 1897, 7570857), Nikola Kne{\v{z}}evi{\'{c}} ({\DJ}uro, Donje Vodi{\v{c}}evo, 1889, 7507637), Jovan Koji{\'{c}} ({\DJ}ura{\dj}, Bistrica, 1895, 7566298), Marko Kova{\v{c}}evi{\'{c}} (Nikola, Sokoli{\v{s}}te, 1897, 7508025), Stanko Kukavica (Ilija, Ga{\v{s}}nica, 1909, 7573645), {\DJ}oko Kukavica (Milan, Ga{\v{s}}nica, 1922, 7502359), Oste Laji{\'{c}} (Mla{\dj}o, Lamovita, 1927, 7509310), Aleksa Lubura (Vukan, Krupac, 1907, 7575336), Du{\v{s}}an Lubura (Marko, Krupac, 1904, 7575335), Milka Luki{\'{c}} ({\DJ}okan, Vojskova, 1905, 7558950), Milo{\v{s}} Luki{\'{c}} (Mirko, Kalenderi, 1928, 7565839), Du{\v{s}}an Mandi{\'{c}} (Jovo, Svodna, 1900, 7570971), Gojko Marin (Du{\v{s}}an, Pobr{\dj}ani, 1899, 7507933), Milan Marin (Mile, Pobr{\dj}ani, 1890, 7507935), Petar Marin (Milan, Pobr{\dj}ani, 1905, 7507934), Lazo Matija{\v{s}} (Mile, Gornji Orlovci, 1926, 7566209), Ahmet Mehinovi{\'{c}} (Ibrahim, Blagaj Rijeka, 1920, 7507438), Stevo Milankovi{\'{c}} (Marko, Vla{\v{s}}kovci, 1898, 7558875), Simo Miloj{\v{c}}i{\'{c}} (Jovan, Sreflije, 1880, 7558503), Ilija Mirosav (Stanko, {\v{S}}evarlije, 1893, 7558676), Milan Muti{\'{c}} ({\DJ}or{\dj}e, Me{\dj}e{\dj}a, 1918, 7557925), Milo{\v{s}} Muti{\'{c}} ({\DJ}or{\dj}e, Me{\dj}e{\dj}a, 1922, 7557927), Marija Papi{\'{c}} (Kojo, Demirovac, 1940, 7556226), Maksim Pele{\v{s}} (Stojan, Gornji Sjeni{\v{c}}ak, 1882, 7558612), Jovo Pepi{\'{c}} (Pero, Donje Vodi{\v{c}}evo, 1896, 7507639), Lazar Pureta (Jovan, Ga{\v{s}}nica, 1906, 7573820), Savo Roki{\'{c}} (Vaso, Jasenovac, 1916, 7557200), Radojka Sekuli{\'{c}} (Vaso, Vojskova, 1939, 7558971), Mirko Sikima (Aleksa, Vojkovi{\'{c}}i, 1914, 7575332), Du{\v{s}}an Skrobi{\'{c}} (Stanko, Tuklju{\v{c}}ani, 1911, 7558738), Ostoja Sovilj (Stoko, Ga{\v{s}}nica, 1900, 7573863), {\v{Z}}arko Srdi{\'{c}} (Jovo, Crna Dolina, 1925, 7566433), Ostoja Stojakovi{\'{c}} ({\DJ}uro, Gornja Dragotinja, 1910, 7509242), Stojan Toma{\v{s}} (Jovan, Pobr{\dj}ani, 1923, 7558263), Cvijo Tomi{\'{c}} (Mile, Pobr{\dj}ani, 1894, 7507942), Petar Vrani{\'{c}} (Stanko, Donja Brusova{\v{c}}a, 1898, 7560068), Stojan Vujasin (Trivo, Sokoli{\v{s}}te, 1906, 7508039), Milan Vukovi{\'{c}} (Nikola, Gornje Vodi{\v{c}}evo, 1915, 7507731), Stevo Vuleti{\'{c}} (Petar, Gornji Jelovac, 1920, 7569356), Savo Zec (Jovan, U{\v{s}}ivac, 1889, 7558757), Milan Zori{\'{c}} (Stojan, Gornji Jelovac, 1925, 7566385), Jovan {\'{C}}uri{\'{c}} (Aleksa, U{\v{s}}ivac, 1881, 7558771), Stojan {\v{C}}i{\v{c}}i{\'{c}} (Du{\v{s}}an, Gornji Jelovac, 1928, 7569397), Bogdan {\DJ}akovi{\'{c}} (Obrad, Vitasovci, 1896, 7571022), Bo{\v{z}}o {\DJ}uki{\'{c}} (Mile, Klekovci, 1924, 7498821), Mile {\v{S}}ari{\'{c}} (Stevan, Gornja Gradina, 1902, 7556759), Rade {\v{S}}ormaz ({\DJ}or{\dj}e, Gornji Jelovac, 1923, 7566188), Ana {\v{S}}panji{\'{c}} (Mi{\'{c}}o, Jablanica, 1926, 7572686), Dragan {\v{S}}urlan (Du{\v{s}}an, Gornje Vodi{\v{c}}evo, 1924, 7565781), Milo{\v{s}} {\v{S}}urlan (Vid, Gornje Vodi{\v{c}}evo, 1940, 7507721), Pero Babi{\'{c}} (Stojan, Veliko Palan{\v{c}}i{\v{s}}te, 1892, 7566378), Mile Kukavica (-, Ga{\v{s}}nica, 1920, 7502379), Milo{\v{s}} Kukavica (Stevan, Ga{\v{s}}nica, -, 7573637), Bo{\v{s}}ko Milju{\v{s}} (-, Sreflije, 1924, 7558528), Branko Vladisavljevi{\'{c}} (-, Slobodna Vlast, -, 7570202), Mile Vuleti{\'{c}} (-, Donja Brusova{\v{c}}a, 1904, 7560065), Savo Ani{\'{c}} (Luka, Boljani{\'{c}}, 1922, 7575812), Jovo Babi{\'{c}} (Simo, Sokoli{\v{s}}te, 1892, 7508019), Cvijo Beri{\'{c}} ({\DJ}uro, Lamovita, 1895, 7566336), Krstan Beri{\'{c}} (Stojan, Lamovita, 1925, 7566335), Petar Beri{\'{c}} (Nikola, Lamovita, 1896, 7566339), Stojan Beri{\'{c}} (Jovica, Lamovita, 1893, 7566334), Ostoja Egi{\'{c}} (Pero, Brezi{\v{c}}ani, 1900, 7508931), Gavro Goji{\'{c}} (Rade, Velika {\v{Z}}uljevica, 1900, 7508081), Irma Goldberger (Herman, Uljanik, 1891, 7522667), Stevan Grahovac (Risto, Lamovita, 1898, 7566345), Ljuban Kne{\v{z}}evi{\'{c}} (Dragan, Gornje Vodi{\v{c}}evo, 1925, 7565770), Nikola Kova{\v{c}}evi{\'{c}} (Mile, Sokoli{\v{s}}te, 1900, 7508024), {\DJ}uro Kuni{\'{c}} (Nikola, Gornja Jutrogo{\v{s}}ta, 1906, 7509446), {\v{Z}}ivko Kuni{\'{c}} (Trivun, Gornja Jutrogo{\v{s}}ta, 1927, 7509123), Vladimir Lubura (Jovo, Krupac, 1921, 7575352), Nikola Lu{\v{c}}i{\'{c}} (Milan, Vojkovi{\'{c}}i, 1912, 7575333), Vlado Lu{\v{c}}i{\'{c}} (Jefto, Grlica, 1912, 7575334), Lazo Matija{\v{s}} (Dragan, Gornji Orlovci, 1930, 7509239), Ramo Mehinovi{\'{c}} (Ahmet, Blagaj Rijeka, 1890, 7507437), Mile Popovi{\'{c}} (Petar, Poljavnice, 1916, 7507957), Milan Pro{\v{s}}i{\'{c}} (Lazo, Dobrljin, 1904, 7507596), Mirko Rakita (Stanko, Grbavica, 1904, 7575916), Mile Sani{\v{c}}anin (Simo, Svodna, 1907, 7508049), {\DJ}or{\dj}o Sikima (Aleksa, Vojkovi{\'{c}}i, 1910, 7575331), Vasilije Star{\v{c}}evi{\'{c}} (Vaso, Poljavnice, 1886, 7507960), Branko Vaskovi{\'{c}} (Jovo, Vojkovi{\'{c}}i, 1920, 7575329), Du{\v{s}}an Vrane{\v{s}}evi{\'{c}} (Tanasije, Donje Vodi{\v{c}}evo, 1910, 7507659), {\DJ}ura{\dj} Vukmir (Te{\v{s}}o, Vitasovci, 1909, 7508110), Rade {\v{C}}enagd{\v{z}}ija (Dragoja, Kadin Jelovac, 1911, 7498693), Cvijo {\DJ}uri{\'{c}} (Stanko, Gornje Vodi{\v{c}}evo, 1924, 7565767), Dmitar {\DJ}uri{\'{c}} (Milan, Gornje Vodi{\v{c}}evo, 1905, 7565754), Milan {\DJ}uri{\'{c}} (Mikan, Gornje Vodi{\v{c}}evo, 1890, 7565758), Nikola {\DJ}uri{\'{c}} (Ilija, Gornje Vodi{\v{c}}evo, 1890, 7565761), Stanko {\DJ}uri{\'{c}} (Marko, Gornje Vodi{\v{c}}evo, 1899, 7565765), Nikola {\v{S}}ormaz (Risto, Gornji Jelovac, 1890, 7566393), Stanko {\v{S}}ormaz (Simo, Gornji Jelovac, 1913, 7566394), Rade {\v{S}}urlan (Ilija, Devetaci, 1899, 7507537), Milan Laji{\'{c}} (Andrija, Gornja Jutrogo{\v{s}}ta, 1910, 7509124), Petar Kotur (Teodor, U{\v{s}}tica, 1888, 7555442), Mile Kuki{\'{c}} (Jefto, Bistrica, 1890, 7566289), Vaskrsija Kuki{\'{c}} (Jefto, Bistrica, 1900, 7566290), Ljuban Miki{\'{c}} (Mihajlo, Gornji Jelovac, 1924, 7569378), {\v{Z}}ivko Kos ({\DJ}okan, Bo{\v{z}}i{\'{c}}i, 1925, 7497487), Rajko Mom{\v{c}}ilovi{\'{c}} ({\DJ}uro, Jablanica, 1932, 7569455), {\v{Z}}arko Mom{\v{c}}ilovi{\'{c}} ({\DJ}uro, Jablanica, 1934, 7569456), Marko Taborin (Jovan, Svodna, 1891, 7508053), Uro{\v{s}} Vicanovi{\'{c}} (Jovo, Romanovci, 1892, 7505687), Ostoja Grbi{\'{c}} (Mihajlo, Marini, 1920, 7509332), Ilija Kne{\v{z}}evi{\'{c}} (Ignjatije, Vidovo Selo, 1911, 7510113), Ostoja Trkulja (Mikajlo, Kadin Jelovac, 1911, 7498767), Ljubomir Vu{\v{c}}kovac (Dragica, Bistrica, 1926, 7566320), Vaskrsija Zrni{\'{c}} (Dragica, Bistrica, 1926, 7566317), Stanko Vojnovi{\'{c}} (Nikola, Velika {\v{Z}}uljevica, 1885, 7508096), {\DJ}uka Banjac (Svetozar, Vojskova, 1919, 7500510), Milan Janjuz (Ante, Sklju{\v{c}}ani, 1895, 7499977), Mihajlo Ke{\v{c}}a ({\DJ}ura{\dj}, {\v{S}}evarlije, 1926, 7500220), Vojno Lubura (Marko, Krupac, 1908, 7575339), Pero Predojevi{\'{c}} (-, Brezi{\v{c}}ani, 1900, 7566405), Te{\v{s}}o Radi{\'{c}} (Jovo, Donje Vodi{\v{c}}evo, 1898, 7570848), Jovo Jaj{\v{c}}anin (Stevo, Gornja Jutrogo{\v{s}}ta, 1886, 7566442), Borislav Kuki{\'{c}} (Vaskrsije, Bistrica, 1926, 7566292), Blago Lubura (Ilija, Krupac, 1914, 7575338), {\DJ}ure{\dj} Milanovi{\'{c}} (Vaso, Me{\dj}e{\dj}a, 1911, 7499317), Vasilj Vidi{\'{c}} (Milan, Miholjsko, 1914, 7560251), Rade Zori{\'{c}} (Dragoja, Gornji Jelovac, 1929, 7509222), Dragica Bjelivuk (Du{\v{s}}an, Gornja Brusova{\v{c}}a, 1939, 7530169), Dragoja Gojkovi{\'{c}} (Mile, Ga{\v{s}}nica, 1910, 7573496), Ljuban Jugovi{\'{c}} (Stevo, Pucari, 1925, 7574650), Ljubo Mami{\'{c}} (Milan, Bistrica, 1920, 7566319), Svetko Mirosav (Ilija, {\v{S}}evarlije, 1924, 7558677), Mihajlo {\DJ}uri{\'{c}} (Milan, Gornje Vodi{\v{c}}evo, 1925, 7565759), Mla{\dj}o {\DJ}ur{\dj}evi{\'{c}} (Stanko, Lamovita, 1910, 7566349), Ostoja Babi{\'{c}} (Milan, Jablanica, 1940, 7503768), Milo{\v{s}} Janji{\'{c}} (Trivun, Gornje Vodi{\v{c}}evo, 1885, 7570856), Nikola Stojni{\'{c}} (Simo, Pobr{\dj}ani, 1887, 7507941), Pavle Dmitra{\v{s}}novi{\'{c}} (Petar, Petkovac, 1896, 7507887), Milan Dugajli{\'{c}} ({\DJ}ura{\dj}, Gornje Vodi{\v{c}}evo, 1925, 7570854), Marko Grahovac-Savi{\'{c}} (Milo{\v{s}}, Lamovita, 1925, 7566346), {\v{Z}}ivko Pa{\v{s}}ajli{\'{c}} (Stevo, Bistrica, 1911, 7500893), Radosava Rapaji{\'{c}} (-, Klekovci, 1918, 7557175). 

\textbf{\underline{ZENICA}}: Du{\v{s}}ko Mamula (Lazo, Top{\v{c}}i{\'{c}} Polje, 1909, 7575400), Ilija Mamula (Lazo, Top{\v{c}}i{\'{c}} Polje, 1903, 7575399), Milo{\v{s}} Mamula (Lazo, Top{\v{c}}i{\'{c}} Polje, 1901, 7575398), Jozo Primorac (Anto, Bija{\v{c}}a, 1890, 7574853), Simo Stojanovi{\'{c}} (Niko, Top{\v{c}}i{\'{c}} Polje, 1914, 7575401), {\v{Z}}eljko Josi{\'{c}} (Stipo, {\v{S}}urkovac, 1913, 7566358), Vojo Dabar{\v{c}}i{\'{c}} (-, Kifino Selo, 1913, 7512074). 

\textbf{\underline{ZOLJE}}: Stanko {\'{C}}irkovi{\'{c}} (-, Zolje, 1910, 7567621). 

\textbf{\underline{ZRIN}}: Stanko Rude{\v{z}} (Marko, Rude{\v{z}}i, 1863, 7557996). 

\textbf{\underline{ZRINSKA}}: Toma Vitanovi{\'{c}} (Gajo, Zrinska, 1902, 7574529). 

\textbf{\underline{ZVORNIK}}: Ilija Mla{\dj}enovi{\'{c}} (Jakov, Drinja{\v{c}}a, 1906, 7568709), Milo{\v{s}} Markovi{\'{c}} (-, Novo Selo, -, 7568822). 

\textbf{\underline{{\v{C}}APRGINCI}}: Du{\v{s}}an Arbutina (-, {\v{C}}aprginci, -, 7564361), {\DJ}uja Deliba{\v{s}}i{\'{c}} (-, {\v{C}}aprginci, -, 7564360). 

\textbf{\underline{{\v{C}}ARDAK KOD JAJC}}: {\v{C}}edo Savi{\v{c}}i{\'{c}} (Stevan, Lubovo, 1903, 7575004). 

\textbf{\underline{{\v{C}}AZMA}}: Isak Poljokan (Rafael, Banja Luka, 1911, 7496838). 

\textbf{\underline{{\v{C}}ETVRTKOVAC}}: Mile Baji{\'{c}} (Stevan, {\v{C}}etvrtkovac, -, 7571455). 

\textbf{\underline{{\v{C}}IFLUK}}: Petar Bala{\'{c}} (Todo, {\v{C}}ifluk, 1919, 7575031), Stevo Bala{\'{c}} (Todo, {\v{C}}ifluk, 1924, 7575029). 

\textbf{\underline{{\v{C}}ITLUK}}: Milja Mijuk (Ljuban, {\v{C}}itluk, 1938, 7556086), Dosta Serdar (Dragan, {\v{C}}itluk, 1933, 7556095). 

\textbf{\underline{{\v{C}}OVAC}}: Ostoja Krnjai{\'{c}} (-, La{\dj}evac, 1904, 7554765). 

\textbf{\underline{{\v{C}}OVI{\'{C}}I}}: Pantelija Kuki{\'{c}} (Arsenije, Gori{\v{c}}ka, 1905, 7543771). 

\textbf{\underline{{\DJ}AKOVO}}: Klara Abinun (Juda, Banja Luka, 1897, 7575506), Ernica Altarac (Isak, Travnik, 1937, 7575511), Ginica Altarac (Avram, Travnik, 1936, 7575516), Isak Altarac (Avram, Travnik, 1934, 7575513), Tilda Altarac (Isak, Travnik, 1908, 7575515), Sara Daniti (Mordehaj, Sarajevo, 1899, 7574205), Santo Gaon (Jakob, Sarajevo, 1939, 7574263), {\v{S}}emaja Hazan (Juda, Sarajevo, 1930, 7574277), Blanka Kabiljo (Menahem, Sarajevo, 1930, 7563494), Rahela Kabiljo (David, Sarajevo, 1924, 7574587), Erna Kajon ({\v{S}}im{\v{s}}on, Sarajevo, 1917, 7517778), Rifka Kajon ({\v{S}}im{\v{s}}on, Sarajevo, 1919, 7517780), Elvira Kalderon (Elijas, Travnik, 1910, 7575529), Erdonja Kunorti ({\v{S}}abetaj, Sarajevo, 1898, 7574636), Jela Lap{\v{c}}evi{\'{c}} (Ostoja, U{\v{s}}tica, 1874, 7555445), Bea Levi (David, Sarajevo, 1882, 7574682), Blanka Levi (Isak, Sarajevo, 1929, 7574684), Estera Levi (A{\v{s}}er, Sarajevo, 1941, 7518210), Nehama Levi (Rafael, Sarajevo, 1933, 7574706), Simha Levi (Jakob, Sarajevo, 1920, 7568149), Vita Levi (Izidor, Sarajevo, 1933, 7518359), Blanka Maestro (Mo{\v{s}}o, Travnik, 1926, 7518699), Moric Maestro (Jozef, Sarajevo, 1931, 7518685), Rena Maestro (Avram, Sarajevo, 1920, 7518645), Blanka Montiljo (Jakob, Sarajevo, 1904, 7574759), Rena Montiljo (Naftali, Sarajevo, 1894, 7518832), Dragan Paji{\'{c}} (Milo{\v{s}}, Vojni{\'{c}}, 1939, 7560430), Marko Paji{\'{c}} (Milo{\v{s}}, Vojni{\'{c}}, 1933, 7560431), Milka Paji{\'{c}} (Milo{\v{s}}, Vojni{\'{c}}, 1936, 7560433), Laura Papo (Rafael, Sarajevo, 1912, 7519320), Rena Papo (Jakob, Sarajevo, 1937, 7568178), Sara Papo (Jakob, Sarajevo, 1909, 7519155), Sarina Papo (Juda, Sarajevo, 1933, 7568176), Sara Pesah (Isidor, Derventa, -, 7561559), Zlata Popovi{\'{c}} (Isak, Sarajevo, 1905, 7576114), Blanka Romano (Mojsije, Sarajevo, 1900, 7519637), Albert Salom (Sado, Zenica, 1936, 7513298), Veljko Sibin{\v{c}}i{\'{c}} (Rade, Milisavci, 1941, 7547073), Zora Singer (Eduard, {\v{C}}azma, 1901, 7553973), Sonja Spiegel (Hinko, Zagreb, 1920, 7563418), Petra Suboti{\'{c}} (Pane, Milisavci, 1939, 7555813), Melita Ungar (Moric, Derventa, 1903, 7576174), Regina {\v{S}}ajer (Adolf, Gacko, 1915, 7576152), Rahela Alkalaj (-, Sarajevo, -, 7568228), Mazalta Altarac (-, Sarajevo, 1885, 7574978), Moric Altarac (Isak, Travnik, 1939, 7575512), Rahela Altarac (-, Sarajevo, 1882, 7516274), Mira Jakab (-, Zagreb, 1920, 7568462), Lea Levi (-, Sarajevo, 1880, 7574699), Regina Levi (Izidor, Sarajevo, -, 7576628), Rifka Levi (-, Sarajevo, 1894, 7518200), Estera Ma{\v{c}}oro (-, Sarajevo, 1890, 7568502), Bar Kohba Papo (-, Sarajevo, 1914, 7562387), Estera Papo (-, Sarajevo, 1883, 7519083), Simha Papo (Jakob, Sarajevo, -, 7519156), Jozef Romano (-, Sarajevo, -, 7519599), Erna Alkalaj (Moise, Sarajevo, 1904, 7568116), Roza Alkalaj-Altarac (-, Sarajevo, -, 7568227), Sadik Altarac (Avram, Sarajevo, 1930, 7516318), Salamon Baruch (Mojsije, Sarajevo, 1933, 7516832), Dona Koen (Isak, Sarajevo, 1890, 7574617), Josef Konforti (Zadik, Travnik, 1912, 7512809), Jozefina Schwamm (Bero, Sarajevo, 1940, 7561857), Irena Spitzer (Samuel, Osijek, 1893, 7565099), Zora Hochstädter (Josip, Zagreb, 1910, 7563845), Blanka Weisser-Konforti (Mordehaj, Sarajevo, 1910, 7568508), Olga Bauer-Fuchs (Leopold, Nova Gradi{\v{s}}ka, 1883, 7533844), Cipora Adi{\v{z}}es (Mojsije, Sarajevo, 1908, 7516078), Albert Altarac (Rafael, Sarajevo, 1921, 7516550), Avram Altarac (-, Doboj, 1880, 7511061), Estera Altarac (Avram, {\v{Z}}ep{\v{c}}e, 1892, 7511634), Lenka Atijas (David, Travnik, 1916, 7512728), Berta Engel (Jakob, Tuzla, 1883, 7563970), Vida Eskenazi (Salamon, Sarajevo, 1896, 7517182), Rifka Romano (Isak, Sarajevo, 1923, 7519616), Blanka Salom (-, Banja Luka, 1882, 7496852), Armin Stein (Karl, {\v{C}}erevi{\'{c}}, 1878, 7563397), Simha Danon-Salom (Jehuda, Sarajevo, 1889, 7574224), Rikica Konforti (Ezer, Travnik, 1927, 7575532), Sara Kamhi (Merjam, Sarajevo, 1924, 7517813), Sara Ozmo (Haim, Sarajevo, 1903, 7574756), Ermoza Tuvi (Avi{\v{s}}alom, Sarajevo, 1898, 7576173), Juda Altarac (Salom, Sarajevo, 1912, 7574133), Eli{\v{s}}a Kabiljo (Benjamin, Sarajevo, 1896, 7517591), Flora Kabiljo (Salamon, Sarajevo, 1887, 7517715), Luna Kabiljo (Eli{\v{s}}a, Sarajevo, 1910, 7517620), Hana Konforti (Salamon, Travnik, 1886, 7575531), Sara Levi (Mo{\v{s}}o, Sarajevo, 1906, 7568148), Hilda Maestro (Mo{\v{s}}o, Travnik, 1930, 7575533), Estera Papo (Salamon, Sarajevo, 1924, 7519337), Elizabeta Blum (Mihajlo, Sarajevo, 1916, 7574175), Sultana Gaon (Naftali, Sarajevo, 1894, 7517439), Rahela Kabiljo (Isak, Sarajevo, 1898, 7517543), Luna Kamhi (Moise, Sarajevo, 1899, 7517846), Estera Papo (Jakob, Sarajevo, 1904, 7519292), Luna Papo (Jakob, Sarajevo, 1926, 7519143), Rahela Papo (Jakov, Sarajevo, 1907, 7576222), Blanka Salom (Jakob, Jajce, 1898, 7508286), Betti Spitzer (Ignaz, Zagreb, 1913, 7565130), Aja Kohn ({\DJ}uka, Zagreb, 1932, 7563768), Itta Salom (Sadik-Sado, Zenica, 1932, 7576288), Rezika Konforti (Zadik, Travnik, 1929, 7512805), Gracija Abinun (-, Sarajevo, 1913, 7515941), Gracija Alkalaj (Izodor, Sarajevo, 1927, 7516219), Lun{\v{c}}ika Alkalaj (-, Sarajevo, -, 7576259), Hanzi Cenover (-, Zagreb, 1922, 7569068), Lun{\v{c}}ika Levi (Salamon, Sarajevo, 1928, 7518540), Sarika Montiljo (Mojsije, Sarajevo, 1914, 7568155), Hana Papo (Jozef, Sarajevo, 1930, 7519173), Rozika Papo (Zadik, Sarajevo, 1914, 7576226), Sarika Papo (Jozef, Sarajevo, 1937, 7576227), Je{\v{s}}ua Perera (Avram, Sarajevo, 1923, 7519413), Regina-Ren{\v{c}}ika Albahari (Salamon, Sarajevo, 1900, 7574095), Eli{\v{s}}ah Kabiljo (Salamon, Sarajevo, 1903, 7574575), Erdonja Katan (Majer, Sarajevo, 1908, 7574609), Sinjora Levi (Avram, Sarajevo, 1925, 7518282), Merjam Papo (Salamon, Sarajevo, 1886, 7519348), Salamon Papo (Rafael, Sarajevo, 1895, 7519323), Estera Pesah (Juda, Sarajevo, 1871, 7576186), Delisija Trinki (Rafael, Sarajevo, 1919, 7519908), Berta Gaon (Avram, Travnik, 1883, 7575526), Safira Salom-Sarafi{\'{c}} (-, Banja Luka, 1884, 7564581), Rifka Atijas (-, Sarajevo, 1910, 7516696), Sinjora Atijas-Kabiljo (-, Sarajevo, -, 7574160), Mazaltov Papo (Menahem, Sarajevo, 1915, 7519224), Flora Abinun (Mo{\v{s}}o, Sarajevo, 1886, 7574078), Gracija Albahari-Adroki (-, Sarajevo, 1903, 7574090), Debora Baruch (Kalmi, Sarajevo, 1925, 7568215), Marga Deutsch (-, Zagreb, -, 7553282), Hana Finci-Kabiljo (-, Sarajevo, 1890, 7559118), Elza Habermann (-, Vinkovci, 1894, 7540232), Vida Hausner-Lipa (-, Sarajevo, 1923, 7561530), Rut Hirschl (Ernest, Sarajevo, 1924, 7574280), Katica Hirschler (Ladislav, Koprivnica, 1881, 7563854), Ru{\v{z}}a Hirschler (Ladislav, Koprivnica, 1882, 7563853), Dana Hochstädter (Josip, Zapre{\v{s}}i{\'{c}}, 1911, 7563843), Mazalta Kabiljo-Izrael (Benjamin, Sarajevo, 1880, 7517597), Rena Kajon (Mojsije, Sarajevo, 1880, 7574596), Blanka Kajon-Sumbulovi{\'{c}} (-, Sarajevo, 1904, 7574594), Solec Kampos (David, Sarajevo, 1907, 7574607), Malvina-Hermina Löwinger-Bruckner (Simon, Sarva{\v{s}}, 1886, 7563710), Sarika Montiljo-Atijas (Juda, Sarajevo, 1898, 7574760), Anula Papo-Levi (Isak, Sarajevo, 1911, 7568158), Rahela Pesah (Avram, Sarajevo, 1902, 7519475), Hana Pinto-Papo (Isak, Sarajevo, 1885, 7576194), Ita Atijas (Salamon, Travnik, 1933, 7512738), Haja Bienenfeld (Salamon, Zagreb, 1932, 7553219), Bjanka Papo (Jakob, Sarajevo, 1922, 7574762), Buena Romano (Isak, Sarajevo, 1935, 7561827). 

\textbf{\underline{{\DJ}EVANJE}}: Radovan Koji{\'{c}} (Ne{\dj}o, Donje {\DJ}evanje, 1912, 7568793). 

\textbf{\underline{{\DJ}IMRIJE}}: Petar Mijatovi{\'{c}} (Cvijan, Suho Polje, 1924, 7514058). 

\textbf{\underline{{\v{S}}ABAC}}: Drago Simi{\'{c}} (-, Lije{\v{s}}anj, 1921, 7568896). 

\textbf{\underline{{\v{S}}AMARICA}}: Ilija Mio{\v{c}}inovi{\'{c}} ({\DJ}uro, Mio{\v{c}}inovi{\'{c}}i, 1897, 7548893). 

\textbf{\underline{{\v{S}}A{\v{S}}}}: Stevo Bla{\v{z}}evi{\'{c}} (Mijo, Utolica, 1884, 7576731). 

\textbf{\underline{{\v{S}}EOVICA}}: Stana Grubi{\v{s}}i{\'{c}} (-, {\v{S}}eovica, 1913, 7564235), Joco Prodanovi{\'{c}} (-, {\v{S}}eovica, 1923, 7564548), Mara Prodanovi{\'{c}} (-, {\v{S}}eovica, 1894, 7564539), Ljubica Kova{\v{c}}i{\'{c}} (-, {\v{S}}eovica, 1898, 7564491), Ljuba Prodanovi{\'{c}} (-, {\v{S}}eovica, 1939, 7557010). 

\textbf{\underline{{\v{S}}ILJKOVA{\v{C}}A}}: Dragica Popovi{\'{c}} ({\DJ}uro, {\v{S}}iljkova{\v{c}}a, 1938, 7574892), Milan Popovi{\'{c}} ({\DJ}uro, {\v{S}}iljkova{\v{c}}a, 1932, 7574893), Ranka Popovi{\'{c}} ({\DJ}uro, {\v{S}}iljkova{\v{c}}a, 1936, 7574891), Simo Krmar (Milivoj, {\v{S}}iljkova{\v{c}}a, 1932, 7510134), {\DJ}uro Krmar (Milivoj, {\v{S}}iljkova{\v{c}}a, 1936, 7510133). 

\textbf{\underline{{\v{S}}IPOVO}}: Du{\v{s}}an Malinovi{\'{c}} (Jovan, Lubovo, 1922, 7575016), An{\dj}elko Baji{\'{c}} (Jovo, Be{\v{s}}njevo, 1905, 7575904), Ilija Jak{\v{s}}i{\'{c}} (Maksim, Sokolac, 1879, 7575935), Milan Kalaba (Lazo, Vra{\v{z}}i{\'{c}}, 1910, 7575036), Petar Leto (Mile, {\v{C}}ifluk, 1921, 7575024), Jovan O{\v{s}}ap (Mijat, Sokolac, 1880, 7575937), Ile Savi{\v{c}}i{\'{c}} (Stojan, Lubovo, 1905, 7575002), Milovan Bala{\'{c}} (Todo, {\v{C}}ifluk, 1919, 7575030), Pavo Baji{\'{c}} (Manojlo, Be{\v{s}}njevo, 1905, 7575903), {\v{C}}edo Baji{\'{c}} (Manojlo, Be{\v{s}}njevo, 1897, 7575902), Nino Kaurin (Nikola, {\v{C}}ifluk, 1912, 7575023), Ile Baji{\'{c}} (Ilija, Be{\v{s}}njevo, 1888, 7575900), Todo Bala{\'{c}} (Pero, {\v{C}}ifluk, 1883, 7575032), Krsman Ubiparip (Vaso, Vra{\v{z}}i{\'{c}}, 1905, 7575047), Du{\v{s}}an Kopanja (Pero, Lubovo, 1915, 7575017). 

\textbf{\underline{{\v{S}}IPOVO-JAJCE}}: Milan Materi{\'{c}} (Obrad, Drvar, 1906, 7574876). 

\textbf{\underline{{\v{S}}UMA,MALA BARNA}}: Nikola Ranisavljevi{\'{c}} (Tanasije, Mala Barna, 1910, 7576435). 

\textbf{\underline{{\v{S}}U{\v{S}}UPARI}}: Milan Orelj (Du{\v{s}}an, {\v{S}}a{\v{s}}, 1940, 7550679). 

\textbf{\underline{{\v{Z}}EP{\v{C}}E}}: Josip Rausnitz (Ignac, Virovitica, 1879, 7546104). 

\textbf{\underline{{\v{Z}}IROVAC}}: Aleksa {\DJ}ustebek (-, Br{\dj}ani {\v{S}}amari{\v{c}}ki, 1910, 7557587). 

\textbf{\underline{{\v{Z}}IVINICE}}: Ilija Mijatovi{\'{c}} ({\DJ}uro, Olovci, 1921, 7575326), Rado Mijatovi{\'{c}} ({\DJ}uro, Olovci, 1920, 7575325). 

\textbf{\underline{{\v{Z}}UMBERAK}}: Dragi{\'{c}} Usorac (Petar, Gornja Trebinja, 1921, 7560029)

\subsection{Other sources}
\label{subsec:other}

The two previously mentioned sources already provided a large number of matches. Because of that and for the sake of simplicity, only a single match is given here for each of the other found sources that also contain claims about persons' fates that are contrary to the ones given in the Jasenovac list. Additionally, for the persons from each of these matches only a single such source countering the Jasenovac list is given despite the fact that quite often there are numerous such sources. For example, the Jasenovac list claims that \href{https://www.ushmm.org/online/hsv/person_view.php?PersonId=7516845}{Kalmi Baruch} born to father Salamon in Sarajevo in 1896 allegedly died in Jasenovac in 1942~\cite{ushmm2021kbaruch}, but the \href{https://zbl.lzmk.hr/?p=714}{Jewish Biography Lexicon}~\cite{zbl2021baruh}, the \href{https://www.ushmm.org/online/hsv/person_view.php?PersonId=5089869}{United States Holocaust Memorial Museum}~\cite{ushmm2021drbaruh}, \href{https://yvng.yadvashem.org/nameDetails.html?itemId=1988323}{a personal testimony} of his cousin available in the digital archive of the Yad Vashem~\cite{yv2021baruh}, and the website dedicated to the past, present, and the future of Jews of former Yugoslavia \href{http://elmundosefarad.wikidot.com/dr-kalmi-baruh-biografija}{El Mundo Sefard}~\cite{ems2021dr} show that he died in 1945 after being liberated from the German camp Bergen-Belsen. While here these sources are useful to resolve the father's name issue, for the sake of simplicity in the following examples only a single source is given.

The Jasenovac list claims that \href{https://ushmm.org/online/hsv/person_view.php?PersonId=7564031}{Mirko Breyer} born in 1863 allegedly died in Jasenovac~\cite{ushmm2021breyer}, but the Croatian Encyclopedia, among others, \href{https://www.enciklopedija.hr/Natuknica.aspx?ID=9456}{claims} that he died in 1946~\cite{he2021breyer}.

The Jasenovac list claims that \href{https://ushmm.org/online/hsv/person_view.php?PersonId=7551369}{Marko Graf} born to father Leopold in Ivanec in 1890 allegedly died in Jasenovac in 1941~\cite{ushmm2021graf}, but the The Jewish Biography Lexicon~\cite{ig2021zbl} \href{https://zbl.lzmk.hr/?p=91}{claims} that he survived the war~\cite{zbl2021graf}.

The Jasenovac list claims that \href{https://ushmm.org/online/hsv/person_view.php?PersonId=7562945}{Vojo Prnjatovi{\'{c}}} born in Sarajevo allegedly died in Jasenovac in 1942~\cite{ushmm2021vprnjatovic}. Namely, in the source for this claim i.e., the book \href{http://web.archive.org/web/20181226133453/http://www.znaci.net/00003/510.pdf}{Koncentracioni logor Jasenovac}~\cite{miletic1986koncentracioni}, it can be seen that he was in a group most of whose members were killed on January 2, 1942. However, the same book claims on April 9, 1942 he gave a testimony in Serbia after leaving Jasenovac.

The Jasenovac list claims that \href{https://ushmm.org/online/hsv/person_view.php?PersonId=7516130}{Josef Albahari} born to father Jakob in Sarajevo in 1913 allegedly died in Jasenovac in 1942~\cite{ushmm2021albahari}, but he can be found on the \href{https://www.jewishgen.org/databases/holocaust/0142_Sarajevo_survivors.html}{``List of Sarajevo survivors who went to Israel in 1948''}~\cite{jg2021sarajevo}.

The Jasenovac list claims that \href{https://ushmm.org/online/hsv/person_view.php?PersonId=7553952}{Bo{\v{z}}o Schwarz} born to father Ferdinand in Zagreb in 1920 allegedly died in Jasenovac in 1942~\cite{ushmm2021schwarz}, but he can be found on \href{https://www.jewishgen.org/databases/holocaust/0153_Zagreb_survivors.html}{the Zagreb Survivor Lists}~\cite{jg2021zagreb}.

The Jasenovac list claims that \href{https://ushmm.org/online/hsv/person_view.php?PersonId=7519035}{Isak Papo} born to father David in Sarajevo in 1911 allegedly died in Jasenovac in 1942~\cite{ushmm2021papo}, but \href{http://www.annapizzuti.it/public/dbcompleto.pdf}{the list of foreign Jews} interned in Italy during the war period~\cite{ap2021db} claims that in 1943 he was in Roana, Italy, and that in 1944 he went to Palestine.

The Jasenovac list claims that \href{https://ushmm.org/online/hsv/person_view.php?PersonId=7511544}{Josip Basch} born to father Jakov in Zavidovi{\'{c}}i in 1910 allegedly died in Jasenovac in 1942~\cite{ushmm2021basch}, but \href{https://collections.arolsen-archives.org/archive/3-2-1-4_1718000/?p=1&doc_id=80908158}{the documents} from the Arolsen Archives claim that he was applying for help in 1949~\cite{aa2021basch}.

The Jasenovac list claims that \href{https://ushmm.org/online/hsv/person_view.php?PersonId=7541001}{Re{\v{z}}inka Stern} born in Ilok in 1914 allegedly died in Jasenovac in 1942~\cite{ushmm2021stern}, but \href{https://www.myheritage.com.hr/research/collection-10460/auschwitz-death-certificates-1941-1943?itemId=54529-&action=showRecord&recordTitle=Rezinka+Stern}{one of the Auschwitz death certificates} claims that she died in Auschwitz in 1942~\cite{myheritage2021stern}.

The Jasenovac list claims that \href{https://ushmm.org/online/hsv/person_view.php?PersonId=7518712}{Jakob Maestro} born to father Salamon in Sarajevo in 1893 allegedly died in Jasenovac in 1941~\cite{ushmm2021maestro}, but \href{http://web.archive.org/web/20191023233354/http://www.znaci.net/00001/191_4.pdf}{a book} about Jewish resistance fighters claims that he was captured in 1944 and sent to Auschwitz~\cite{romano1980jevreji}.

The Jasenovac list claims that \href{https://ushmm.org/online/hsv/person_view.php?PersonId=7553106}{Ivan Komorčec} born to father Stjepan in Mihovljan in 1911 allegedly died in Jasenovac in 1944~\cite{ushmm2021komorcec}, but \href{http://web.archive.org/web/20190805082634/http://znaci.net/00003/647.pdf}{a book} about the war in Hrvatsko zagorje claims that he died as a resistance fighter in 1945~\cite{belinic1981hrvatsko}.

The Jasenovac list claims that \href{https://ushmm.org/online/hsv/person_view.php?PersonId=7537014}{Ernest Stern} born to father Filip in Po{\v{z}}ega in 1905 allegedly died in Jasenovac in 1941~\cite{ushmm2021estern}, but the genealogy site geni.com \href{https://www.geni.com/people/Ernesto-Enzi-Stern/6000000040417222921}{claims} that he died in Brazil in 1985~\cite{geni2021estern}.

On two occasions the Jasenovac list claims that \href{https://www.ushmm.org/online/hsv/person_view.php?PersonId=7553674}{Pavao L{\"{o}}w} i.e., \href{https://www.ushmm.org/online/hsv/person_view.php?PersonId=7563715}{Pavao L{\"{a}}w} born to father Alfred in Zagreb in 1910 allegedly died in Jasenovac~\cite{ushmm2021low,ushmm2021law}, but \href{https://www.reprezentacija.rs/loew-pavao/}{the website} of the history of the Serbian national football team claims that he escaped and died in 1986~\cite{rs2021low}.

It is possible to list other sources as well, but already the presented ones prove the critique wrong in its indication about the alleged lack of importance of the claim of our previous paper about the three individuals on the Jasenovac. As shown here, it turns out that they are an indication of a larger phenomenon. The fact that something similar has not been observed for other lists makes this rather significant.

\section*{Acknowledgments}
\label{sec:acknowledgements}

\textit{(Nikola Bani{\'{c}} and Neven Elezovi{\'{c}} contributed equally to this work.)} The authors would like to thank Dr. Mladen Koić for his instructions on how to use the digital archive of the Yad Vashem, Roman Leljak for providing the digitally processed version of the Belgrade museum data, Dr. Esther Gitman for collecting and providing the data about the Zagreb and Sarajevo communities and giving additional instructions on how to use some of the data, Prof. Branko Jeren and Dr. Josip Pe{\v{c}}ari{\'{c} for their suggestions, and Dr. Vladimir Geiger for providing instructions on proper comparison methodology of contradicting historical sources. Finally, the authors would like to thank Dr.~Julio Guijarro Garcia, Dr.~Josep Peguera Poch, Dr.~Jorge Ramos, Dr. Lovro Prepolec, Dr. Vuko Brigljevi{\'{c}}, and Dr.~Jure Bogdan for their kind support.

\balance
\bibliographystyle{IEEEtran}
\bibliography{literature}

\end{document}